\documentclass[12pt]{article}
\pdfoutput=1                     
\usepackage{graphicx,amssymb}
\usepackage{amsmath,amsfonts,epsfig}
\usepackage{braket}
\usepackage{cite}
\usepackage[usenames]{color}
\usepackage{float}
\usepackage{setspace}
\usepackage{here}
\usepackage[normalem]{ulem}
\usepackage{enumitem}
\usepackage{subfigure}
\usepackage{array,arydshln}
\usepackage{soul}
\newcommand{\pslash}{p\kern-1ex /}
\newcommand{\qslash}{q\kern-1ex /}
\newcommand{\lslash}{l\kern-1ex /}
\newcommand{\sslash}{s\kern-1ex /}
\newcommand{\kaslash}{k_a\kern-2ex /}
\newcommand{\kbslash}{k_b\kern-2ex /}
\newcommand{\Dslash}{{\cal D}\kern-1.5ex /}

\newcommand{\bc}{\overline{c}}
\newcommand{\tr}{{\rm tr}}

\newcommand{\beqa}{\begin{eqnarray}}
	\newcommand{\eeqa}{\end{eqnarray}}

\renewcommand{\Im}{\mathrm{Im}\,}
\newcommand{\bpm}{\begin{pmatrix}}
	\newcommand{\epm}{\end{pmatrix}}
\newcommand{\bbm}{\begin{bmatrix}}
	\newcommand{\ebm}{\end{bmatrix}}

\def\order{\ensuremath{\mathcal{O}}}
\def\del{\partial}
\newcommand{\half}{\frac{1}{2}}

\def\Tr{\text{Tr}}

\makeatletter

\@addtoreset{equation}{section}
\makeatother
\usepackage{subfig}
\begin{document}

	\voffset -0.7 true cm
	\hoffset 1.5 true cm
	\topmargin 0.0in
	\evensidemargin 0.0in
	\oddsidemargin 0.0in
	\textheight 8.6in
	\textwidth 5.4in
	\parskip 9 pt
	
	\def\Tr{\hbox{Tr}}
	\newcommand{\be}{\begin{equation}}
		\newcommand{\ee}{\end{equation}}
	\newcommand{\bea}{\begin{eqnarray}}
		\newcommand{\eea}{\end{eqnarray}}
	\newcommand{\beas}{\begin{eqnarray*}}
		\newcommand{\eeas}{\end{eqnarray*}}
	\newcommand{\nn}{\nonumber}
	\font\cmsss=cmss8
	\def\C{{\hbox{\cmsss C}}}
	\font\cmss=cmss10
	\def\bigC{{\hbox{\cmss C}}}
	\def\scriptlap{{\kern1pt\vbox{\hrule height 0.8pt\hbox{\vrule width 0.8pt
					\hskip2pt\vbox{\vskip 4pt}\hskip 2pt\vrule width 0.4pt}\hrule height 0.4pt}
			\kern1pt}}
	\def\ba{{\bar{a}}}
	\def\bb{{\bar{b}}}
	\def\bc{{\bar{c}}}
	\def\bphi{{\Phi}}
	\def\Bigggl{\mathopen\Biggg}
	\def\Bigggr{\mathclose\Biggg}
	\def\Biggg#1{{\hbox{$\left#1\vbox to 25pt{}\right.\n@space$}}}
	\def\n@space{\nulldelimiterspace=0pt \m@th}
	\def\m@th{\mathsurround = 0pt}

	\begin{titlepage}
		\begin{flushright}
			{\small OU-HET-1149}\\
			{\small TIFR/TH/23-6}
			\\
		\end{flushright}
		
		\begin{center}
			
			\vspace{1mm}
			
			{\Large \bf Sparse random matrices and }  \\[3pt] 
		                            \vspace{1mm}
		        {\Large \bf  Gaussian ensembles with varying randomness } 	 \\[3pt] 

			\vspace{5mm}
			
			\renewcommand\thefootnote{\mbox{$\fnsymbol{footnote}$}}
			Takanori Anegawa${}^{\spadesuit}$, 
			Norihiro Iizuka${}^{\spadesuit}$, 
			Arkaprava Mukherjee${}^{\clubsuit, \diamondsuit}$, \\
			\vspace{0.5mm}
			Sunil Kumar Sake${}^{\spadesuit, \diamondsuit}$, and 
			Sandip P. Trivedi${}^{\diamondsuit}$
			
			\vspace{3mm}

			${}^{\spadesuit}${\small \sl Department of Physics, Osaka University} \\ 
			{\small \sl Toyonaka, Osaka 560-0043, JAPAN} 
	
			${}^{\clubsuit}${\small \sl Department of Physics, The Ohio State University} \\
			{\small \sl   Columbus, OH 43210, USA} 
						
			${}^{\diamondsuit}${\small \sl Department of Theoretical Physics, Tata Institute of Fundamental Research} \\
			{\small \sl Colaba, Mumbai 400 005, INDIA}

		\end{center}
		
		\vspace{-1mm}
		
		\noindent
		We study a system of  $N$ qubits  with a random Hamiltonian obtained by drawing coupling constants from Gaussian distributions in various ways. This results in  a rich class of systems which include the GUE and the fixed $q$ SYK theories. Our motivation is to understand the system at large $N$. In practice most of our  calculations are carried out  using exact diagonalisation techniques (up to $N=24$). Starting with the GUE, we study the resulting behaviour  as the randomness is decreased.
While in general the system goes  from being chaotic to being more ordered as the randomness is decreased, the changes in  various properties, including the density of states,  the spectral form factor,
the    level statistics and   out-of-time-ordered correlators, reveal interesting patterns. Subject to the limitations of our analysis which is mainly numerical, we find some evidence that  the behaviour  changes  in an  abrupt  manner when the number of non-zero independent terms in the Hamiltonian is exponentially large in $N$.
We also study the opposite limit of much reduced randomness obtained in  a local version of the SYK model where the number of couplings scales linearly in $N$, and characterise its behaviour.
Our investigation suggests that a more complete theoretical analysis of this  class of systems will prove quite worthwhile.

	\end{titlepage}
	
	\setcounter{footnote}{0}
	\renewcommand\thefootnote{\mbox{\arabic{footnote}}}
	
	\newpage
	
	\setcounter{tocdepth}{2}  
	\tableofcontents
	
	\newpage

	\section{Introduction and Motivation\label{sect:intro}}


	Random Matrix theory (RMT) is well known to have interesting  connections with two-dimensional gravity. See for review \cite{DiFrancesco:1993cyw, Ginsparg:1993is}.    
	More recently an interesting and new version of this connection has been discovered in the study of two-dimensional Jackiw-Teitelboim (JT) \cite{Jackiw:1984je,Teitelboim:1983ux} gravity. It has been shown that the key features of JT gravity are correctly reproduced by the  low-energy limit of the Sachdev-Ye-Kitaev (SYK)  model, which is a quantum mechanical model of $N$ flavors of Majorana fermions with random couplings \cite{Sachdev:1992fk,Kitaevtalk,Maldacena:2016hyu,Polchinski:2016xgd}. Furthermore it was shown in  \cite{Saad:2019lba} that the partition function of JT gravity on surfaces with an arbitrary number of boundaries and handles is correctly reproduced by Random Matrix theory in a suitable double scaling limit.  
	Since JT gravity arises quite universally as a description of the low-energy dynamics for a wide class of higher dimensional near extremal black holes \cite{Nayak:2018qej,Moitra:2019bub},  these fascinating results promise to hold general lessons for the study of quantum gravity and black holes in higher dimensions as well. 
	
	The Hamiltonian of the SYK model with a $\psi^q$ coupling (with $q$ being  an even number) is given by 
	\be
	\label{H_SYK}
	H_{{\text{SYK}_q}} = i^{q/2} \sum_{1 \le i_1< i_2< \cdots i_q \le N} j_{{i_1 i_2 \cdots i_q}} \psi_{i_1} \psi_{i_2}  \cdots \psi_{i_q}   \,,
	\ee
	Here $\psi_i, i=1, \cdots N,$ are the $N$ flavors of Majorana fermions and $j_{{i_1 i_2 \cdots i_q}}$ are real random couplings drawn from a Gaussian ensemble with variance
	\be
	\label{varsykq}
	\braket{j_{{i_1 i_2 \cdots i_q}}, j_{{j_1 j_2 \cdots j_q}}}  = \frac{(q-1)! J^2}{N^{q-1}} \delta_{i_1,j_1} \delta_{i_2,j_2} \cdots \delta_{i_q,j_q}
	\ee
	We refer to this model as SYK$_q$ model. 
	Quite remarkably, this model reproduces many aspects of JT gravity, as was mentioned above. In particular, 
	the low-energy dynamics of the SYK model exhibits a characteristic pattern of symmetry breaking, tied to how time reparametrisation symmetry is realised. This gives rise to  soft Goldstone-like modes   whose dynamics is governed by the Schwarzian action. The behaviour  of this action accounts for the  thermodynamics  and low-energy  density of states and the resulting  dynamics gives rise to   Out-of-Time-Ordered Correlators (OTOCs) in the system saturating the chaos bound. These features - the pattern of  symmetry breaking, the Schwarzian action  for the time reparametrizations, the resulting thermodynamics and the behaviour of OTOCs - are all shared by  JT gravity.  
	
	As we will discuss in detail  below, the fermions $\psi^i$ can be realised as operators in an $L=2^{N/2}$ dimensional  Hilbert space obtained by taking a tensor product of  Hilbert spaces of $N/2$ qubits and the Hamiltonian $H$ can then be thought of as a $L\times L$ dimensional Hermitian matrix  acting on  this Hilbert space. Such a matrix in general has $L^2$ real independent elements. Taking all these to be independent Gaussian random variables with zero mean and equal variance, gives rise to the Gaussian Unitary Ensemble (GUE) of Random Matrix Theory (RMT). Its behaviour, even at  low-energies,
	is quite different from that of JT gravity. 
	
	In contrast, in the SYK model, even though the random variables present are  independent  Gaussian variables, as was mentioned above, their total  number is only $\order(N^q)$ which is much smaller than  $L^2$ -the number of random variables present in the GUE. 
	
%
	These observations raise several interesting questions: How much randomness is needed for agreement with gravity? What happens when we start from the GUE and begin reducing the randomness, by decreasing the number of Gaussian random variables? Is the resulting behaviour, at low-energies, dependent  on only the number of random variables or also on which variables have been retained? When do we get the behaviour at low-energies to agree with  JT gravity? Etc. 
	
	The current investigation is prompted by questions such as these  and will attempt to address some of them. 
	
	 Briefly, we follow  two lines of investigation here. Our analysis is mostly numerical. In the first investigation, starting with the GUE we  reduce  the randomness by retaining only $n<L^2$  of the matrix elements to be non-zero and taking  each of these $n$  elements to be  independent Gaussian random variables. We find   that the behaviour  changes in a fairly pronounced manner as the randomness reduces and $n$ becomes smaller.  Subject to the limitations of our numerical analysis, we also find that this change  occurs in  quite a small interval,  when $n$ reaches a value,  { $n_c\sim L^{(1.5)}$, so that the fraction of random matrix elements $n_c/L^2\sim L^{(-0.5)} \rightarrow 0$,} as $L\rightarrow \infty$.  We study     several properties of the system as $n$ is varied, including the density of states, thermodynamics, the nearest neighbour spacing distribution, spectral form factors and Out of Time Correlators  and find that these  reveal interesting difference with both the GUE and the SYK  cases. For example, initially, starting from the GUE, as $n$ is reduced, we find that the density of states is  well fitted to a rescaled Wigner semi-circle distribution. After the transition though,  this is no longer true. 

%
	

	The second line of investigation involves going to the other limit, and  now  having very little randomness. In this analysis we study  a ``local" version of the SYK model. In the Hamiltonian eq.(\ref{H_SYK}) the Hamiltonian connects all the $N$ fermions to each other through the $\psi^q$ couplings which take fluctuating values. In the altered system  we study - which we often refer to as the {\it local}\, SYK model below - only sets of $q$ nearest neighbour fermions will be coupled together by random couplings. This reduces the number of random couplings to $\order(N)$, instead of the $\order(N^q)$ in the SYK model. 
	To differentiate  our  model with the more conventional one described above, eq.(\ref{H_SYK}),   we will refer to the latter sometimes as the ``Non-local" SYK model. Once again we study several properties of the local system, along the lines of the first investigation  mentioned above,  and find these compare and  contrast in interesting ways with  other cases. 
	
%
			
	Two comments are worth making before we proceed. In  the double scaling limit considered in \cite{Saad:2019lba}, which gives agreement with JT gravity, we have to fine-tune the potential $V(M)$, and furthermore ``zoom in" on the edge of the matrix theory potential. 
		This means while all $L^2$ random variables in the Hermitian matrix $M$ are retained, its matrix elements are not independent Gaussian Random variables.
 In contrast in our study,
	as was mentioned above, we reduce the number of random variables, but those which are retained are taken to be independent Guassian random variables. 
%
%
	Second, there are other variations of the SYK model which have also been considered in the literature. These involve taking $q\rightarrow \infty$, along with $N\rightarrow \infty$. The case where $N\rightarrow \infty$ first, then $q\rightarrow \infty$, was studied in \cite{Maldacena:2016hyu}. In \cite{Berkooz:2018jqr,Lin:2022rbf} a double scaled limit was considered in which $q^2/N$ is held fixed while $q,N\rightarrow \infty$. 
	These limits are also quite revealing and we will make some  comments about  them  below. 
	
	
	{\it More on the Relation Between the SYK Model and RMT: } 
	
	It is helpful at this stage to give some more details on  how the  $N$-flavor SYK model can be realised in the Hilbert space obtained from  the tensor product of $N/2$ qubits; this will help clarify  the relation between the SYK model and RMT.

	The  Majorana fermions $\psi_i, i=1, \cdots N$ of the SYK model satisfy the Clifford algebra
	\be
	\{ \psi_i \,, \psi_j \} =  \delta_{ij}\,.\label{cliff}
	\ee
	On the Hilbert space obtained by taking the tensor product of $N/2$ qubits, this algebra can be realised by taking the following representation for the fermions in terms of tensor products of Pauli matrices:
	\begin{align}
		\psi_{2i-1}&= {\frac{1}{\sqrt{2}}}  \, \sigma^{z}_{1}\otimes\sigma^{z}_{2}\otimes \ldots \sigma^{z}_{i-1}\otimes \sigma^{x}_{i}\otimes {\bf 1} \otimes {\bf 1} \ldots \otimes {\bf 1}_{\frac{N}{2}} \,, \nonumber\\
		\psi_{2i}&=\frac{1}{\sqrt{2}}  \,  \sigma^{z}_{1}\otimes\sigma^{z}_{2}\otimes \ldots \sigma^{z}_{i-1}\otimes \sigma^{y}_{i}\otimes {\bf 1} \otimes {\bf 1} \ldots \otimes {\bf 1}_{\frac{N}{2}}  \,,\label{psiaspaul}
	\end{align}
	where $\sigma^x, \sigma^y, \sigma^z$ are the usual Pauli matrices and $\bf{1}$ is $2\times 2$ identity matrix. 
	Note that in this representation the $\psi_i$ are $L \times L$ matrices where 
	\be
	\label{defL}
	L=2^{N/2} \,.
	\ee
	
	It follows then  that a general Hermitian matrix acting on  the Hilbert space of the $N/2$ qubits  can be written as the sum over various  $\psi^q$ terms, with $q$ taking both even and odd values:
	\be
	\label{sumh}
	H= \sum_{q=0}^{N} \alpha_q\sum_{i_1<i_2<\cdots <i_q}  j_{i_1i_2\cdots i_q} \psi_{i_1}\psi_{i_2} \cdots \psi_{i_q}
	\ee
	where 
	\be
	\label{defalphaq}
	\alpha_q=e^{i \pi q(q-1)\over 4}
	\ee
	and $j_{i_1\cdots i_q}$ is real. 
	The matrix  eq.(\ref{sumh}) has 
	\be
	\label{numrandomcouplings}
	\sum_{q=0}^N \, {}_N C_q  =  2^N = 2^{N/2} \times 2^{N/2}= L^2
	\ee
	number of  real independent couplings due to binomial theorem, which is equal to the total number of independent elements in an $L\times L$ Hermitian matrix. It is also easy to see that if we now draw all the couplings $j_{i_1 \cdots i_q}$ which appear on the RHS in eq.(\ref{sumh}) from a Gaussian random ensemble with equal variance
	\be
	\label{dall}
	\langle j_{i_1i_2\cdots i_q}, j_{j_1j_2 \cdots j_q}\rangle= \sigma^2 \delta_{i_1,j_1} \delta_{i_2,j_2}, \cdots \delta_{i_q,j_q} \,,
	\ee 
	{\it i.e.}, with a variance $\sigma^2$ independent of $q$ and the coupling $j_{i_1i_2,\cdots i_q}$,  
	then the resulting ensemble for $H$ is the GUE. 
	This follows from noting that the set of matrices $\alpha_q \psi_{i_1}\psi_{i_2} \cdots \psi_{i_q}$  form  a basis in which any Hermitian matrix can be expanded. Furthermore, this basis is orthonormal with respect to the inner product, $\langle M_1,M_2\rangle= {1\over L} \Tr(M_1^\dagger M_2)$. 
	
	Restating the GUE in this way allows for a better comparison  with  the SYK model. As was mentioned above, with fixed value of $q$, the SYK model has only ${}_N C_q\sim O(N^q)$ random couplings.
	In addition we note that  in the SYK model, eq.(\ref{varsykq}) the variance (for a fixed value of $q)$ is taken  to scale as $1/N^{q-1}$,  
	 to ensure that the model has a good large $N$ limit. 
	This is very different from the GUE where, as mentioned above, besides including all powers of $q$ we also  take the variance to be independent of $q$.
	{A key difference in the resulting behaviour as was emphasised above, between the GUE and the SYK model at fixed $q$, is in the low energy dynamics, which for the $\psi^q$  case arises from the dynamics of the  Schwarzian mode.  In appendix \ref{quadbosnmodel} we show that the Schwarzian continues to arise when $q\sim N^a$, with $a<1/2$. 
	The case $a=1/2$ was discussed in \cite{Berkooz:2018jqr,Lin:2022rbf}, in which case there are $O(N^{\sqrt{N}\over 2})$ number of random couplings.  When $\lambda=q^2/N$ is small it was shown that one gets Schwarzian dynamics in this theory while  for $\lambda$ being large one gets the  GUE. This might have suggested at first glance that  GUE like behaviour would  to continue for higher values of $a$, beyond $a=1/2$, since in those cases  $\lambda\rightarrow \infty$, when $N\rightarrow \infty$. 
	 
	 We have not analysed the behaviour while changing $a$ systematically. Instead in our study  we vary the number of random couplings, $n$. 
	We find  that when  $n$ becomes sufficiently big,   $n>n_c \sim L^{1.5}\sim 2^{0.75 N}$, the behaviour is indeed of GUE type with no Schwarzian dynamics. But for $n<n_c$, while still being much bigger than $O(N^{\sqrt{N}\over 2})$,  the behaviour changes and is qualitatively different from the GUE. Unfortunately,   with the limitations of our analysis, we are not able to establish the precise nature of the low-energy behaviour, in particular whether or not a Schwarzian mode is  present,  for such values of $n$.

At the other extreme,  in terms of having less randomness, one can take the $\psi^q$ theory, for fixed $q$, as discussed in \cite{Xu:2020shn} and reduce the number of couplings by setting some of them to vanish at random. It was found, as we will also review in appendix \ref{quadbosnmodel}, that as long as the number of couplings present still grows  like $N^\alpha$, with $\alpha>1$, {\it i.e.}, the number of couplings grows faster than a linear power of $N$,   the low-energy theory will be due to the Schwarzian mode.  Together, these results  imply that the Schwarzian dynamics  at low energies is in fact quite universal - arising for any fixed value of $q$, or in double scaling limits as long as $q\le \sqrt{N}$, and persisting  in any of these cases even with vastly reduced randomness as long as the total number of couplings present are of order $N^{(1+\epsilon)}$, for $\epsilon>0$. 
What happens when the randomness is reduced even further, so that the number of couplings grows only linearly with $N$?
The local SYK model is an example of such a situation. Our analysis suggests that interesting differences set in, in  the linear case. In particular, our analysis suggests   that the Schwarzian mode is absent.

Many of our comments have focussed on the low energy dynamics and the presence of a Schwarzian theory, because they are of primary interest from the point of view of connecting with a theory of gravity. However, it is worth mentioning that the detailed spectrum, beyond the low-energy limit, and many other properties, are of course much less universal and will change. This is true as $n$, the number of non-zero matrix elements in $H$, is varied. Or as the power $q$ in the SYK model is varied, or in SYK models as  the number of non-vanishing couplings for fixed $q$ is varied.  Even when the Schwarzian mode is present, the detailed nature of the conformal fixed point at low energies changes as $q$ is varied, with the fermion having anomalous dimension $1/q$.  Interestingly though, for fixed $q$, when we reduce the number of random couplings, keeping the total number to grow faster than $O(N)$,  the anomalous dimension of the fermion fields is still given by $1/q$ and does not change.

	One more comment is in order. We had mentioned above that among the properties we study are the OTOCs and the growth of operator size, which tell us about the scrambling behaviour.  
	Using the mapping above, eq.(\ref{psiaspaul}), we see that the fermionic operators are defined in general for any system in an $L=2^{N/2}$ dimensional Hilbert space, including RMT. One can therefore use these fermionic operators, for example, to study the growth of OTOC, or the growth of operator size. In particular, we will often  consider the OTOC which  involves a sum over all flavours and is given by, 
	\cite{Kitaevtalk, Maldacena:2016hyu, Polchinski:2016xgd}, 
	\be
	\sum_{i,j=1,{i\neq j}}^N\braket{\psi_i (t) \psi_j (0) \psi_i (t) \psi_j (0)} \sim 1 - \frac{1}{N} e^{\lambda t } \,, \quad \lambda = \frac{2 \pi}{\beta}
	\label{Ft}
	\ee   
	This saturates the chaos bound with a maximal Lyapunov exponent in the fixed $q$ SYK model. In contrast, the behaviour of the same correlator in the GUE is very different, and is that of a ``hyperfast scrambler"  as discussed further below. 
	
	{\it Organisation: }
	
	The paper is organised as follows. In section 2 we review some aspects of the SYK model, the behaviour of the GUE and the connection between these  two systems.  
	In section 3 we  report on our investigation of $L\times L$ matrices with reduced randomness and in section 4 on the local version of the SYK model. We end with discussion in section 5. 
	
	Before we end this introduction section, let us  comment on some additional related literature. Our local SYK model belongs to the so-called ``sparse'' SYK models, where the number of random couplings is reduced and which have been  actively studied recently \cite{Swingletalk, Xu:2020shn, Garcia-Garcia:2020cdo, Tezukatalk}. Our local SYK model is based on the locality of fermion interactions. The question of how universal these sparse SYK models are, compared with usual SYK model, is interesting and needs to be investigated  more\footnote{We thank M. Tezuka for discussions of the sparse SYK models.}. 
	
	{
	{\it Convention in figures: }
	
	In all figures, whenever we denote $\log x$, we choose its basis $10$, {\it i.e.,} $\log x = \log_{10} x$. On the other hand, for $\ln x$, its basis is $e$, {\it i.e.,} $\ln x = \log_{e} x$.

}

	\section{The GUE and SYK$_q$ Theories: }
	\label{revsec}
	We briefly summarise the behaviour of the GUE and the SYK $\psi^q$ models  here. For concreteness in the numerical plots we take $q=4$. 
	
	\subsection{Partition functions, eigenvalue densities and thermodynamics}
	The  partition function for the GUE, $Z_{GUE}$ is given by 
	\begin{align}
		\label{GUE}
		Z_{GUE}= \frac{1}{Z_0} \int \prod_{i} dM_{ii}\prod_{i<j}d\text{R}[M_{ij}]d\text{I}[M_{ij}] \exp\left( -\frac{1}{2\sigma^2}\Tr M^2 \right) \,,
	\end{align}
	Here $M$ are $L\times L$ Hermitian Matrices, { $Z_0$ is a normalisation constant}, and 
	$\sigma$  is given by
	\begin{align}
		\sigma = \frac{1}{\sqrt{L} } \,.\label{sigrmt}
	\end{align}
	The matrix integral in eq.(\ref{GUE}) is invariant under the  $U(N)$ symmetry $M \to U^\dagger M U$. Using this, we can obtain the density of eigenvalues $\rho_L(\lambda)$, 
%
	\begin{align}
		Z_{GUE} 
		= L \int d \lambda \, \rho_L(\lambda) \,, 
	\quad \mbox{where}  \quad  
		\label{rhoLlambda}
		\rho_L(\lambda)=
		\frac{1}{\sqrt{2L}}  K_L \left(\sqrt{\frac{L}{2}} \, \lambda \right) \,, 
	\end{align}
and $K_L(x)$ is given in terms of Hermite polynomials $H_j$ as
\be
\label{fK}
K_L(x)=\frac{e^{-x^2}}{2^L (L-1)! \sqrt{\pi}}(H_L'(x)H_{L-1}(x)-H_{L-1}'(x)H_L(x)) \,.
\ee
 The expectation values of single trace operators, $\Tr(M^n)$,  can be computed using $\rho_L(\lambda)$ as 
	\be
	\label{mdens}
	  \langle \Tr(M^n)\rangle = \frac{\int d\lambda \,\rho_L(\lambda) \lambda^n }{\int d\lambda \,\rho_L(\lambda) } \,.
	\ee
	In the $L\rightarrow \infty$ limit, $K_L(\lambda)$ - the  Kernel function - becomes the famous Wigner semi-circle distribution, 
	\begin{align}
	\lim_{L \to \infty} {K_L(\lambda)} = \frac{1}{\pi}\sqrt{2L-\lambda^2}
         \end {align} 
         leading to
	\begin{align}
		\label{Wignersemicircle}
		\rho_W(\lambda) \equiv \lim_{L\to \infty} \rho_L(\lambda) = \frac{1}{2 \pi} \sqrt{4 - \lambda^2}\,,
	\end{align}
	which is normalised to unity, 
	\be
\label{nd}
\int_{-2}^{2} \,d\lambda \,\rho_W(\lambda) =1.
\ee

The { eigenvalue density} $\rho_L(\lambda)$ for the GUE with $L=4096, N=24$  and { the density of states for SYK model with} $q=4$, $J=1$, eq.~(\ref{varsykq}),  are shown in Fig \ref{finitenpic}. 
Qualitatively the eigenvalues are more ``clumped" together in the SYK model towards the centre and the density is correspondingly smaller towards the edges, compared to the GUE case. One can heuristically understand this as arising in the SYK model due to the reduced randomness which causes   smaller eigenvalue repulsion. 
\begin{figure}[tb]
\subfigure[]{\includegraphics[width=7cm,height=5.5cm]{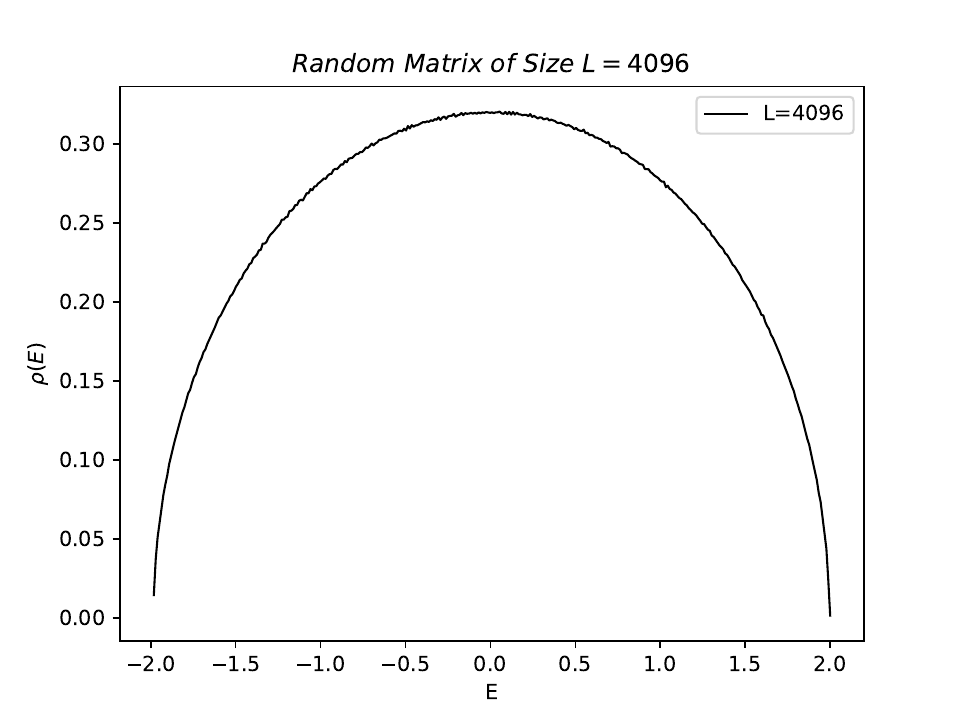}}
\subfigure[]{\includegraphics[width=7cm,height=5.5cm]{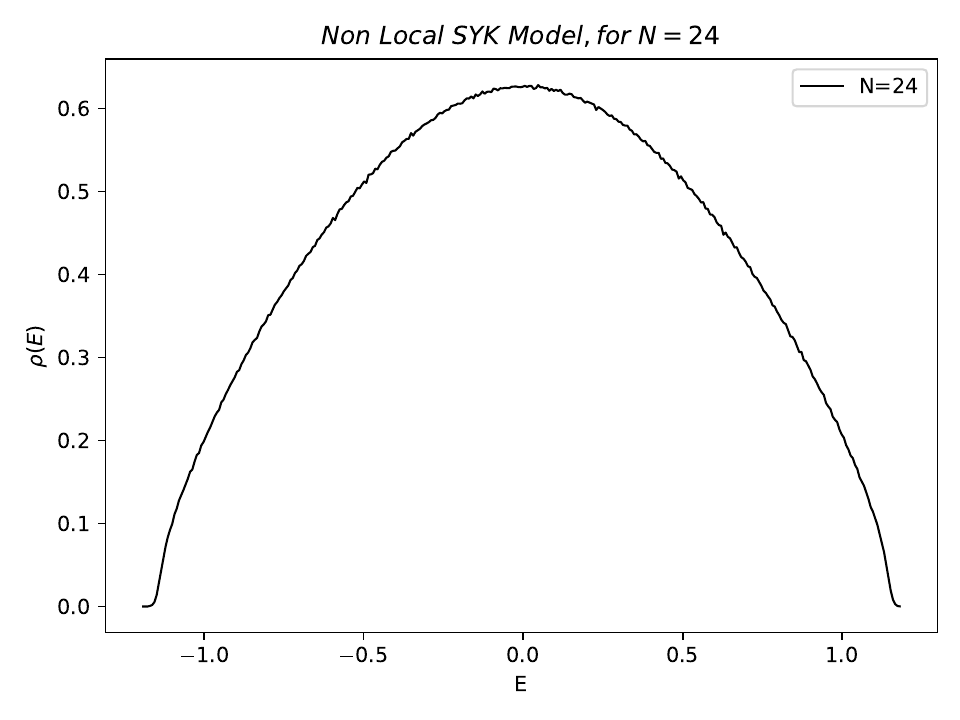}}
	\caption{(a) The density of eigenvalues for $2^{12} \times 2^{12}$ GUE  matrix. (b) The density of {states} for $q=4$, $N=24$ SYK Hamiltonian. The number of ensemble elements used is 1000.  
	}
	\label{finitenpic}
\end{figure}

Note that the { eigenvalue density} $\rho_W$ is denoted in the figure below  by $\rho(E)$ .

Reducing the rank $L$ for the GUE gives rise to oscillations,  which grow as $L$ becomes smaller.   Fig \ref{finitenpicRMT} shows the density of states  $\rho_L(E)$ for $L=10$,
which can be contrasted with the case of $L=4096$ in Fig \ref{finitenpic}.

\begin{figure}[H]
\centering
\subfigure{\includegraphics[width=7cm,height=4.5cm]{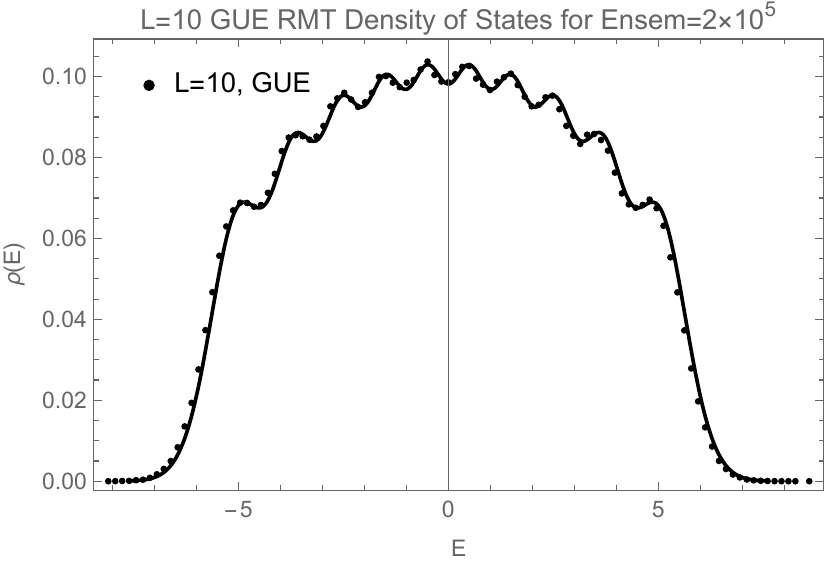}}
	\caption{The density of eigenvalues for $10 \times 10$ GUE random matrix ensemble. }
	\label{finitenpicRMT}
\end{figure}


		\noindent \underline{\bf Thermodynamics}
		
		{As we have discussed in the introduction, regarding the random matrix $M$ as Hamiltonian $H$  
	\be
	\label{RMTHamiltonian}
	H = M  \,,
	\ee 
	%
one can interpret $\lambda$ as energy eigenvalue, and $\rho_W(\lambda)$ as the density of states. 
Throughout the paper we will refer to the normalised density of states as $\rho$, where it satisfies the condition 
$\int d\lambda \rho(\lambda)=1$.} 

Then the {\it thermal} partition function\footnote{The partition function eq.~\eqref{GUE} and \eqref{rhoLlambda} are $\beta \to 0$ limit of this thermal partition function.} of the GUE in the $L\rightarrow \infty$ limit is 
	\begin{align}
			Z(T) & =L \int_{-2}^2 d\lambda \, \rho_W(\lambda) e^{- \beta \lambda} 
			= L \beta^{-1} I_1\left( 2 \beta \right)\label{pf} \\
			& \sim  L \, T^{3/2} \exp\left( {\frac{2}{T}} \right)  \left( 1 + \order(T) \right) \,,\label{lglrmtpf}
		\end{align}
		where $I_1$ is the modified Bessel function of the first kind and the second line is the $T\rightarrow 0$ behaviour, 
		and the resulting entropy at low temperature is given by 
		\be
		\label{ent}
		S =    \left( 1 +  T \frac{\partial}{\partial T}\right) \ln Z(T) \simeq  \ln L + {3\over 2} \ln (T) + \order(T^0)\,.
		\ee
	The $T\rightarrow 0$ features are due to  the density of states going like $\rho_W\rightarrow {1\over \pi} \sqrt{2+\lambda}$ near the lower edge of the spectrum at $\lambda=-2$. Note  that the temperature-dependent part of the entropy is negative  indicating that thermodynamics is not a good approximation in the very low temperature limit. 
	In fact one can show that thermodynamics is not a good approximation at any temperature in this model. 
	Thermodynamics corresponds to a saddle point in the evaluation of the partition function \eqref{pf}, which  here takes the form,
	\be
	\label{spf}
	{	Z(\beta)=L \int_{-2}^2  {d \lambda\over 2 \pi} e^{[\ln (\sqrt{4-\lambda^2}) -\beta \lambda]}. }
	\ee
	It is easy to see that there is no good saddle point approximation to this integral,  { since $L$ is an overall factor and we do not have large $N$ parameter here.}   

	In contrast, the behaviour of the SYK $\psi^q$ model is very different. Denoting the energy difference from the ground state by $E$ we have at 
	 large $N$ that the density of states  for energies $E\ll J$ is given by \cite{Maldacena:2016hyu, Cotler:2016fpe, Garcia-Garcia:2016mno, Kitaev:2017awl, Jevicki:2016ito}
	\be
	\label{rhosyk}
	\rho(E)\propto e^{S_0} \sinh\left(\sqrt{c N E\over J}\right)
	\ee
	where $c$ is a positive $\order(1)$ number ({for $ q=4$, $c\approx 0.396$}).
	Near $E\rightarrow 0$,  $\rho\propto \sqrt{E/J}$ which is analogous to the Wigner distribution near the edge $\lambda\rightarrow -2$. 
	However, in the range 
	\be
	\label{lesyk}
	{
        1  \gg {E\over J}\gg 	{1\over N} }
 	\ee
	 the partition function has a good saddle point approximation (for $N\gg 1$)
	\begin{eqnarray}
	\label{pfsyk}
	Z(\beta) & = & e^{S_0}  \int d E \rho(E) e^{-\beta E}\nonumber\\
	 & \propto & e^{S_0} {1\over \beta^{3/2}} e^{{c N \over 4\beta J}}
	\end{eqnarray}
	 and is given by the second line on the RHS above. As a result  thermodynamics is a good approximation in this regime. One can also see this from the entropy 
	 \be
	 \label{entSYK}
	 { S=S_0+   {cNT\over 2 J} {+\frac{3}{2}\ln T} }
	 \ee
	which is positive and with a  resultant positive specific heat which is linear in $T$. The constant $S_0$ scales with $N$,
	 \be
	 \label{csc}
	 S_0=c_0 N
	 \ee
	 where $c_0$ is also an $\order(1)$ constant.  The presence of this  constant  is a characteristic feature of the SYK model. Although for any finite $N$ the ground state degeneracy goes to zero, in the temperature range 
	 eq.~(\ref{lesyk}),	 the system behaves  as if,  in effect, it has a big ``ground state" degeneracy,  since eigenvalue differences are $O(e^{ - S_0})$ which is  very tiny compared to the  range eq.~\eqref{lesyk}.
	Most of the  above features arise from  the Schwarzian action. 
	 
	{ A plot of the entropy $S$ as a function of $T$ for the GUE and SYK models is shown in Fig \ref{entguenlsyk}.}
\begin{figure}[H]
\subfigure[]{\includegraphics[width=7.5cm,height=5.5cm]{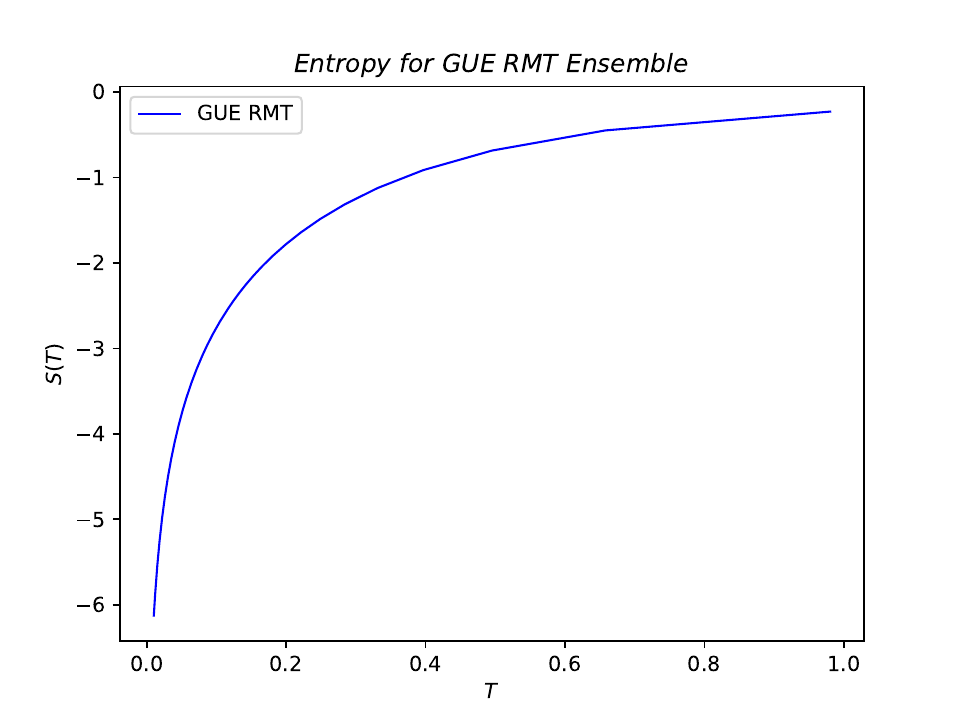}}
\subfigure[]{\includegraphics[width=7.5cm,height=5.5cm]{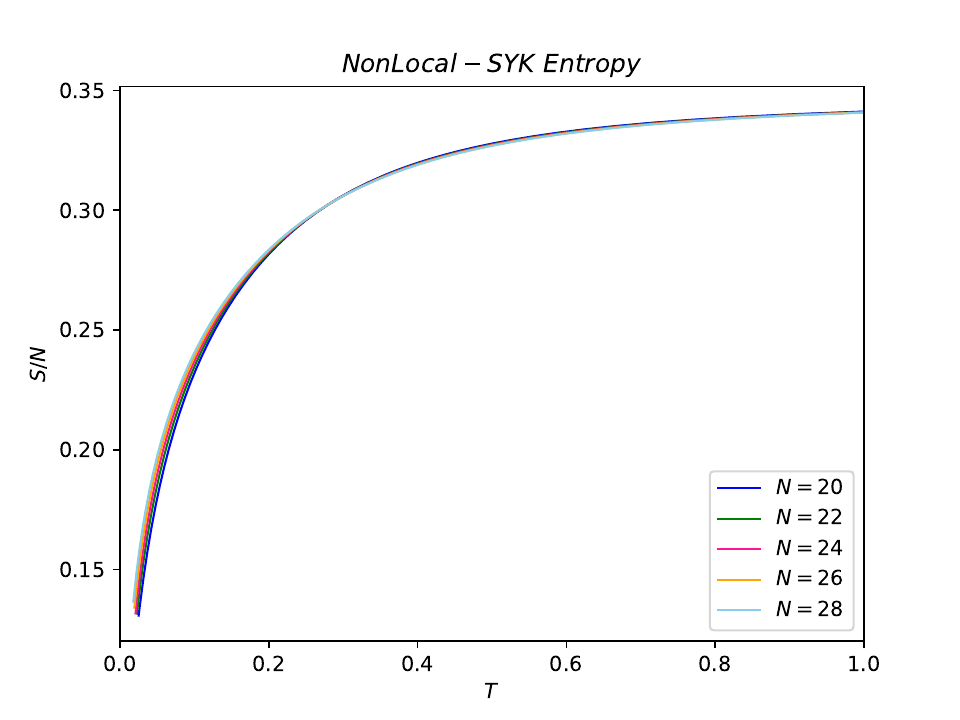}}
\caption{Fig.(a) $S(T)=S-\ln L$  vs $T$ for GUE. Fig.(b) ${S\over N}$ vs $T$ for Non-local SYK model }
\label{entguenlsyk}
\end{figure}

%
	
%
%
	\subsection{Spectral form factor (SFF) for RMT and SYK model}
	\label{subsffrmtsyk}
	This subsection briefly reviews the work of \cite{Cotler:2016fpe}. 
	
	The spectral form factor  is an important quantity to study from the perspective of chaos. 
	It is  defined to be ,	
	\be
	\label{defsff}
	{\rm SFF} \equiv g(t,\beta) \equiv {\langle Z(\beta+it)Z(\beta-it)\rangle\over \langle Z(\beta)\rangle ^2}
	\ee
	 and it can be divided into a disconnected and connected piece,
	\begin{align}
		g_d(t,\beta) &= {\langle Z(\beta,t)\rangle \langle Z(\beta,t)\rangle^* \over \langle Z(\beta)\rangle^2} \label{gd} \\
		g_c(t,\beta) &=g(t,\beta)-g_d(t,\beta)\label{gc}
	\end{align}
	where $Z(\beta,t)=Z(\beta+it)$. { In the rest of the paper, we call $g(t,\beta)$ as total SFF, $g_d(t,\beta)$ as disconnected SFF, and $g_c(t,\beta)$ as connected SFF respectively. }
	
	We note that the SFF can be expressed in terms of the two point correlator $\langle \rho(\lambda_1) \rho(\lambda_2)\rangle$
			\be
			\label{cex}
			\langle Z(\beta+i t) Z(\beta-i t)\rangle= \int d\lambda_1d\lambda_2 \,e^{-\beta(\lambda_1+\lambda_2)} 
			e^{-i t(\lambda_1-\lambda_2)}
			\langle \rho(\lambda_1) \rho(\lambda_2)\rangle
			\ee
			This shows that as $t$ increases the SFF is sensitive to the two point density correlator at smaller separations in the  eigenvalues.
			
	Interestingly the SFF of the SYK model agrees with RMT \cite{Cotler:2016fpe}. Depending on whether $N {\rm\,\,\,( mod\,\,\, 8) }$ is 2 or 6, 0, 4, respectively, the SFF of SYK model agrees with that of the GUE, GOE or GSE \cite{Cotler:2016fpe}.
	
	As a function of time $t$ there are three basic features seen in the SFF:  an initial decay, dominated by the disconnected part; a subsequent ramp due to the connected part and a final plateau. 
	The disconnected part in GUE is given by
	\begin{align}
		Z(\beta+it)\equiv \Tr\ e^{-\beta H-iHt}= {L \over \beta+ i t} I_1(2(\beta+i t)) \label{partan}
	\end{align}
	as a result, at  late times, $t\rightarrow \infty$, 
	\be
	\label{limz}
	Z\sim {L \over 2 \pi (\beta+i t)^{3/2}} e^{\beta+it} \sim {L \over t^{3/2} }e^{\beta+it}
	\ee so that both the real and imaginary part of $Z(\beta+it)$ decay with time, leading to 
\be
	\label{gddecay}
	g_d(\beta=0,t)\sim {1\over t^3}
	\ee

%
%
%
%
%
	
	In contrast to the disconnected piece, the connected part grows with time,  
	\be
	\label{gcgrow}
	g_c(\beta=0,t)\sim {t\over 2 \pi L^2}
	\ee
	for $1 \ll t <2L$ 
	The connected and disconnected contributions  therefore  become equal at a  time of order
	\be
	\label{valtd}
	t_{dip}\sim \sqrt{L}
	\ee
	which is called dip time. 
	Thereafter the connected part dominates and the SFF exhibits a ramp-like character growing linearly with $t$. 
	
	
	Finally at time $t\sim 2 L$, $g_c$ reaches a plateau attaining a constant value 
	\be
	\label{gcp}
	g_c(\beta=0,t)={1\over \pi L}
	\ee
	The plateau region arises due to nearest neighbour eigenvalues repelling each other. 
	Compared with the plateau region, the ramp region is sensitive to interactions between eigenvalues which are further apart.
	As time increases, the SFF becomes more sensitive to the eigenvalue interactions between the closer ones. 
	
	All of these  features can be seen in  Fig \ref{SFFRMT} which also show,   \cite{Cotler:2016fpe}  that the SFF in RMT and SYK are very similar. 

	\begin{figure}[H]
		\begin{minipage}[b]{0.48\linewidth}
			\centering
			\includegraphics[width=7cm,height=5.5cm]{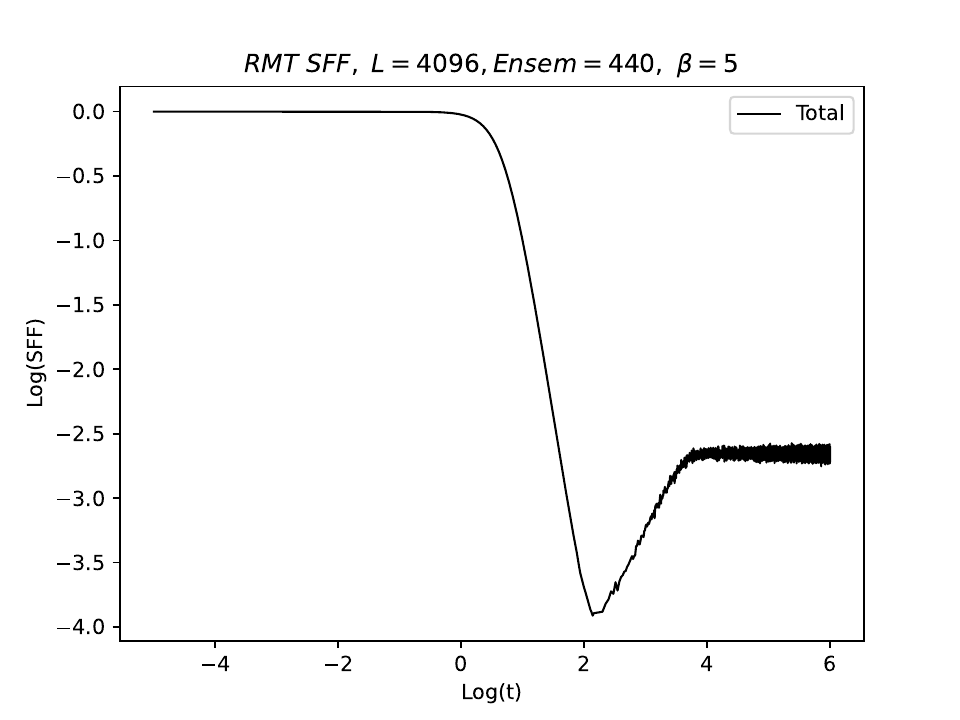}
		\end{minipage} \hspace{3mm}
		\begin{minipage}[b]{0.48\linewidth}
			\centering
			\includegraphics[width=7cm,height=5.5cm]{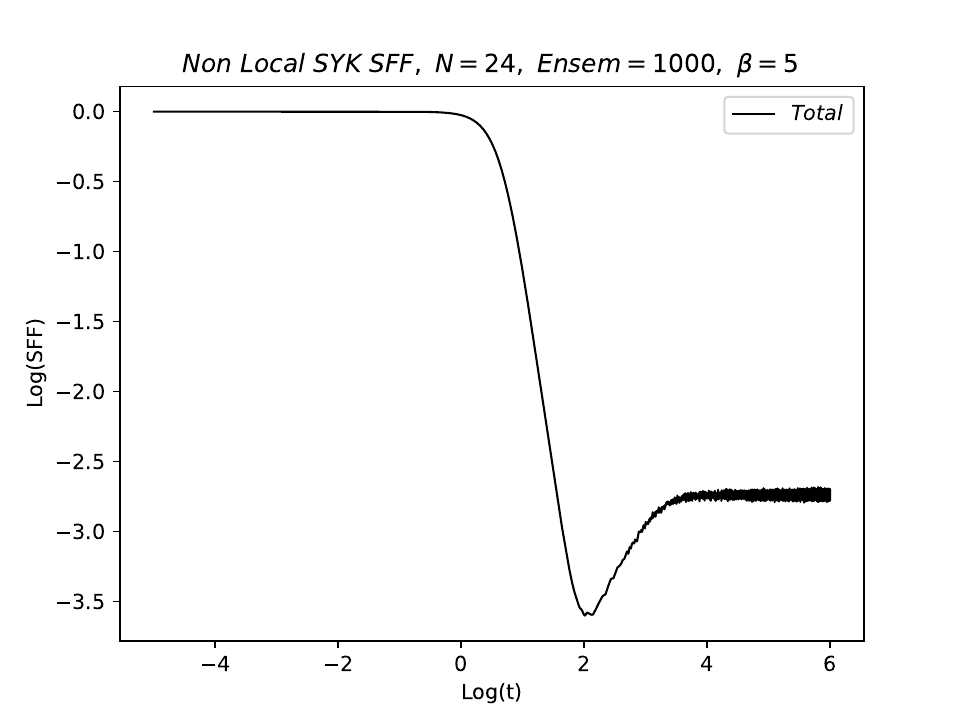}
		\end{minipage} 
		\caption{Left: The spectral form factor for the size $L= 2^{12}$ random matrices. Right: The spectral form factor for the $q=4$, $N=24$ SYK model. The size of Hilbert space are both $2^{12}=4096$. } 
			\label{SFFRMT}
	\end{figure}
	
	We also note that the behaviour of the disconnected piece of the SFF and  the  ramp region  in the connected piece, for  the SYK model, are in good agreement 
	(for sufficiently large $\beta$) 
	with the JT theory where they correspond to the behaviour of the disk and  double trumpet  partition functions, respectively \cite{Saad:2018bqo}. 
	
	\subsection{Nearest Neighbour Level Spacing for RMT and SYK model}
	\label{nlrmtsyk}
	The nearest neighbour level spacing is also well known to be a good diagnostic of chaos. 
	{For GOE this spacing takes the form of the well known Wigner surmise. 
	The probability of the spacing taking value $s$, $P(s)$ is given by 
	\be
	\label{pws}
	P(s)={\pi \over 2} s e^{-{\pi s^2\over 4}}
	\ee
}	
	 In Fig \ref{RMTspc} we plot this probability distribution for the GOE and the SYK $q=4$, $N=24$,  models. We see very good agreement with the Wigner surmise, \cite{Cotler:2016fpe, Garcia-Garcia:2016mno}. 
	  Note that  these plots have been obtained after unfolding the spectrum, as described in \cite{You:2016ldz}, see also section \ref{srmtcv} below.
	  The comparison is made with the GOE since the number of flavours,  $N=24$, is divisible by $8$, see \cite{You:2016ldz}.

	\begin{figure}[H]
		\begin{minipage}[b]{0.48\linewidth}
			\centering
			\includegraphics[width=7cm,height=5.5cm]{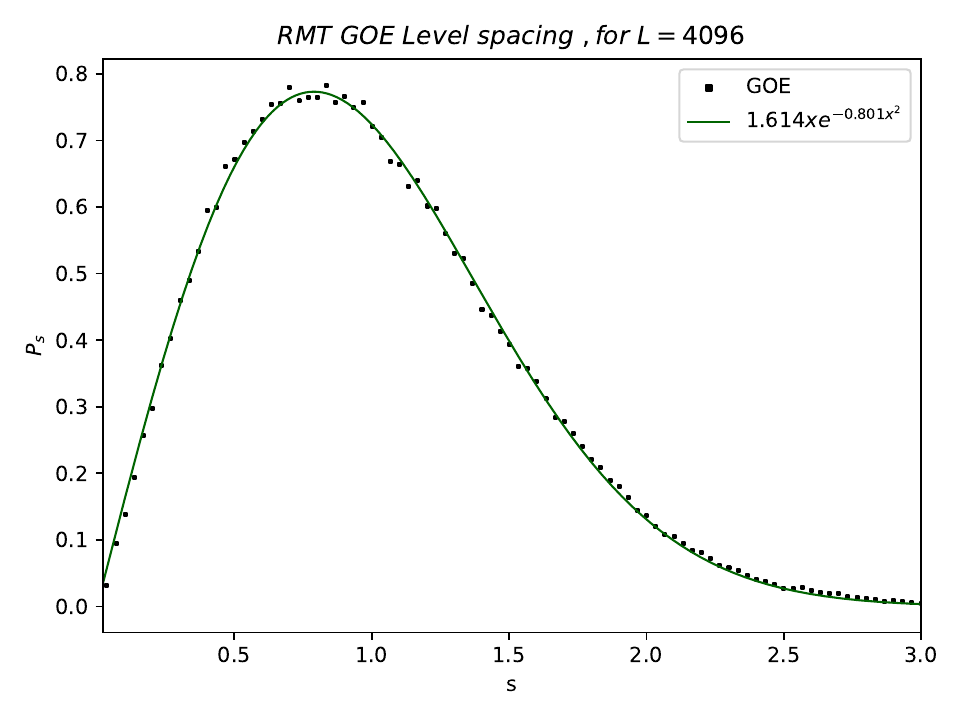}
		\end{minipage} \hspace{3mm}
		\begin{minipage}[b]{0.48\linewidth}
			\centering
			\includegraphics[width=7cm,height=5.5cm]{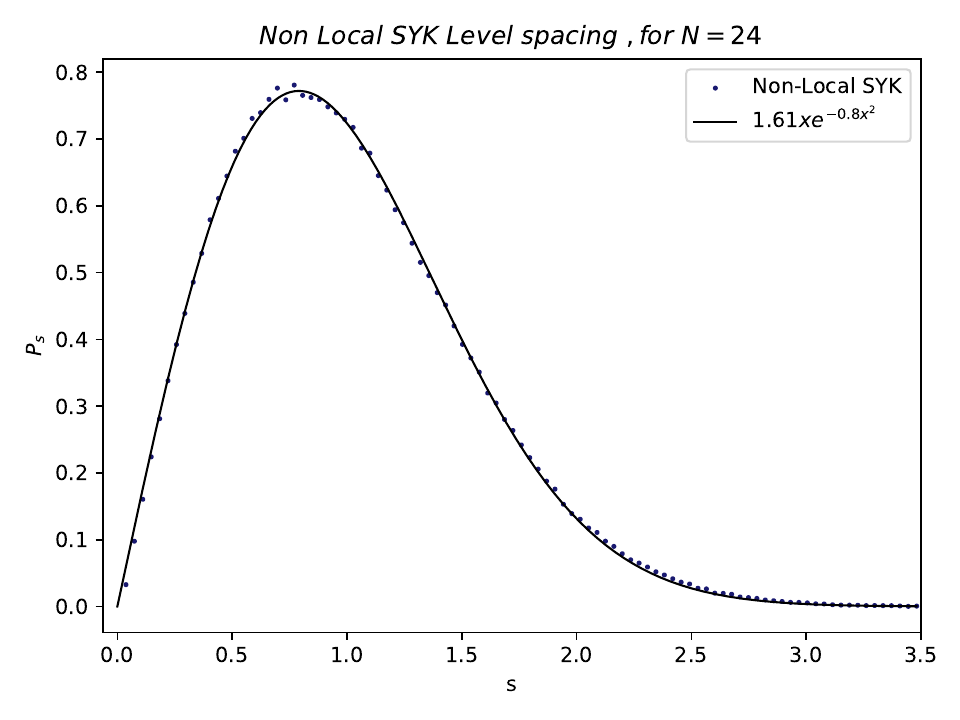}
		\end{minipage} 
		\caption{Left: Nearest Neighbour Level Spacing for the size $L= 2^{12}$ Gaussian Orthogonal random matrices. Right: Nearest Neighbour Level Spacing for the $q=4$, $N=24$ SYK model. The size of Hilbert space in both cases is  $2^{12}=4096$. } 
			\label{RMTspc}
	\end{figure}

	\subsection{OTOC}
	\label{rmtotoc}

Finally we turn to OTOCs. These are defined in eq.(\ref{Ft}) above. 
More precisely, we consider the normalized OTOCs
\be
\label{defgij}
{\tilde F}_{ij}(t)={\Tr(  y\psi_i(t) y\psi_j(0) y\psi_i(t)y \psi_j(0))\over Z  \langle\psi_j(0)\psi_j(0)\rangle\langle\psi_i(t)\psi_i(t)\rangle },\qquad y=e^{- \frac{\beta H}{ 4}}
\ee
where $Z$ is the partition function, 
{
$Z = \Tr e^{-\beta H}$ and  
\begin{align}
\langle\psi_j(0) \psi_j(0)\rangle={\Tr(e^{-\beta H} \psi_j(0) \psi_j(0))\over Z} \,, \quad \langle\psi_i(t) \psi_i(t)\rangle= {\Tr(e^{-\beta H} \psi_i(t) \psi_i(t))\over Z} \,.
\end{align}
}

In the non-local SYK model, the OTOC ${\tilde F}_{ij}(t)$ is same for all pairs as far as  $ i\neq j,$ since all the flavour of fermions are connected to each other with random couplings $j_{ijkl}$ of the same strength. Numerically, we calculate the OTOC by averaging over the pairs $(i,j)$ for $i\neq j$ and denote this quantity as  ${\tilde F}(t)$; since the OTOC is independent of the choice of flavour, this average should be  the same as that for any particular values of  $(i,j)$. Below we plot the function 
\begin{align}
	G_4(t)=1-{\tilde F}(t)\label{gtilfdef}
\end{align}
\\

\noindent \underline{\bf Non-Local SYK  OTOC, $q=4$,  $N=24$} 
\begin{figure}[H]
\hspace{-10mm}
\subfigure[]{\includegraphics[width=8cm,height=6.5cm]{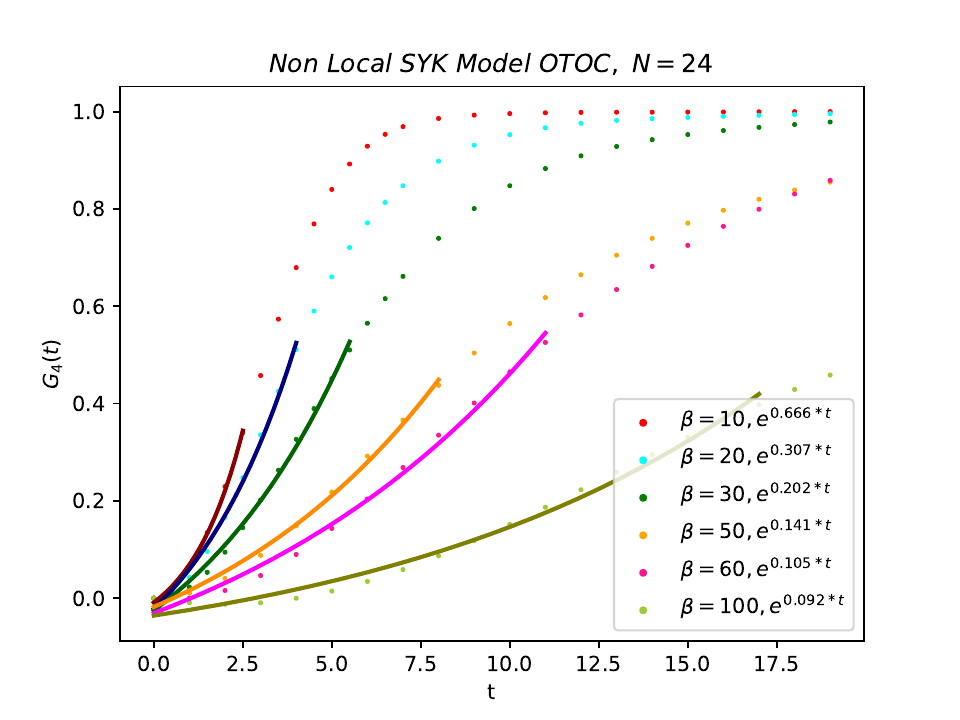}}		
\hspace{-8mm}
\subfigure[]{\includegraphics[width=8cm,height=6.5cm]{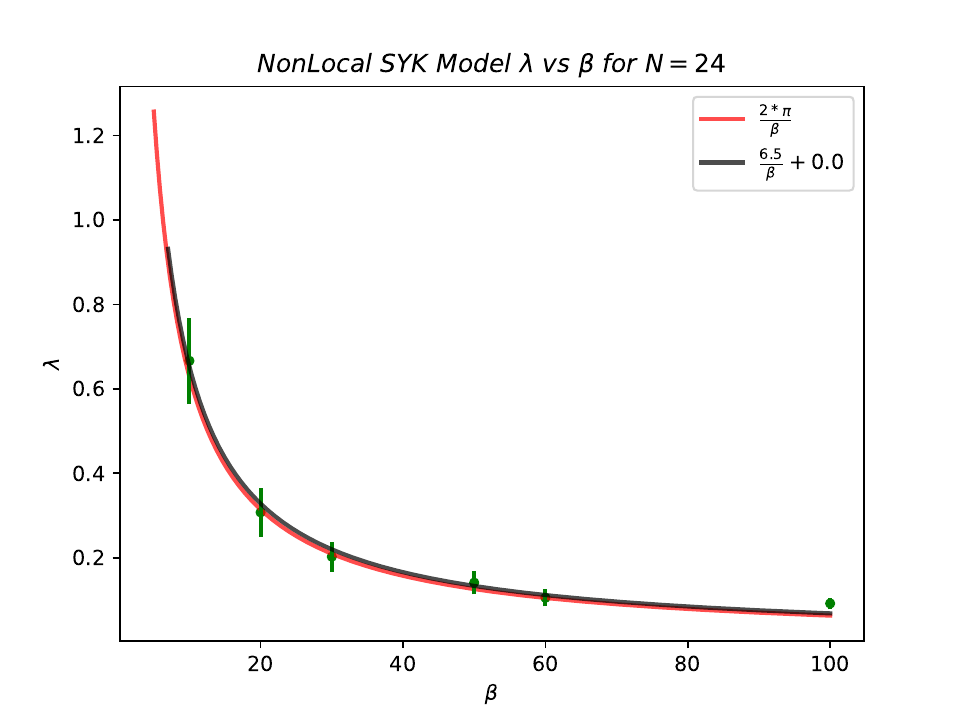}}
\caption{Non-local SYK model with $N=24$, (a) Plot of $G_{4}= 1 - {\tilde F}$ as a function of time, fitted with $A+B e^{\lambda t}$, up to $t\le t_{max}$, (b)  Lyapunov coefficient as a function of $\beta$.}
\label{nlsykotoc}
\end{figure}

The plots for the non-local SYK model are obtained as follows, see also  appendix \ref{eranl}. 
The dots in the left hand panel are the data, obtained with  an ensemble of $2$ elements. The solid curves are obtained by fitting the initial region (up to a maximum time $t_{max}$) for each $\beta$ with a curve of the form 
$A + B e^{\lambda t} $. The best fit value of $\lambda$ obtained in this manner is then plotted against $\beta$ in the right hand panel. 
The maximum time $t_{max}$ up to which we fit the OTOC with the exponential form  is obtained by starting with an initial range for $t$ where this form is a good fit and then extending  this range to the largest time, $t_{max}$, where the exponential fit continues to be a good one. This was  done by  first fitting  the data naively with ``the  eye" and then,   more systematically, by using a chi-square distribution to estimate the goodness of the fit, as discussed in appendix \ref{eranl}. 
In the right side panel the error bars for each data point corresponds to  the  $1\sigma$ errors obtained by fitting the OTOC up to $t_{max}$. 
The red curve  in this panel corresponds to $\lambda={2\pi\over \beta}$, and the black curve is obtained by fitting the data  with a curve of the form
$\lambda={A\over \beta}+B$. The best fit gives   $B=0$ and $A=6.5$, which is in reasonable agreement with the chaos bound exponent $\frac{2 \pi}{\beta}$. 

Strictly speaking, the Lyapunov exponent is defined in the large $N$ limit. Obtaining it in this limit numerically  though is quite challenging, see \cite{Kobrin:2020xms}.
From the right hand panel above we see that  while working at a moderately large value of $N$, our analysis still   allows for a fairly reliable estimate of
the Lyapunov exponent. 

Theoretically, one can show in the large $N$ limit that $G_4$ takes the form
\be
\label{fg4syk}
G_4={A\over N} e^{{2\pi t \over \beta}}
\ee
where $A$ is a coefficient of order unity determined by  the fermionic fields.
This shows that $G_4$ reaches its asymptotic value of order unity at a scrambling time
\be
\label{tscsyk}
t\sim{ \beta\over 2 \pi} \ln N.
\ee
$G_4$ is suppressed by a factor of $1/N$ because the non-local SYK model has a saddle point in the  large $N$ limit, 
and the value of the action at the saddle point scales like $N$.
We see that as a result the scrambling time  is parametrically larger than $\beta$ by $\ln N$. 

\noindent \underline{\bf  Random matrix OTOC, L=4096}\\

\vspace{-5mm}
\begin{figure}[H]
\centering
\subfigure{\includegraphics[width=8.5cm,height=7cm]{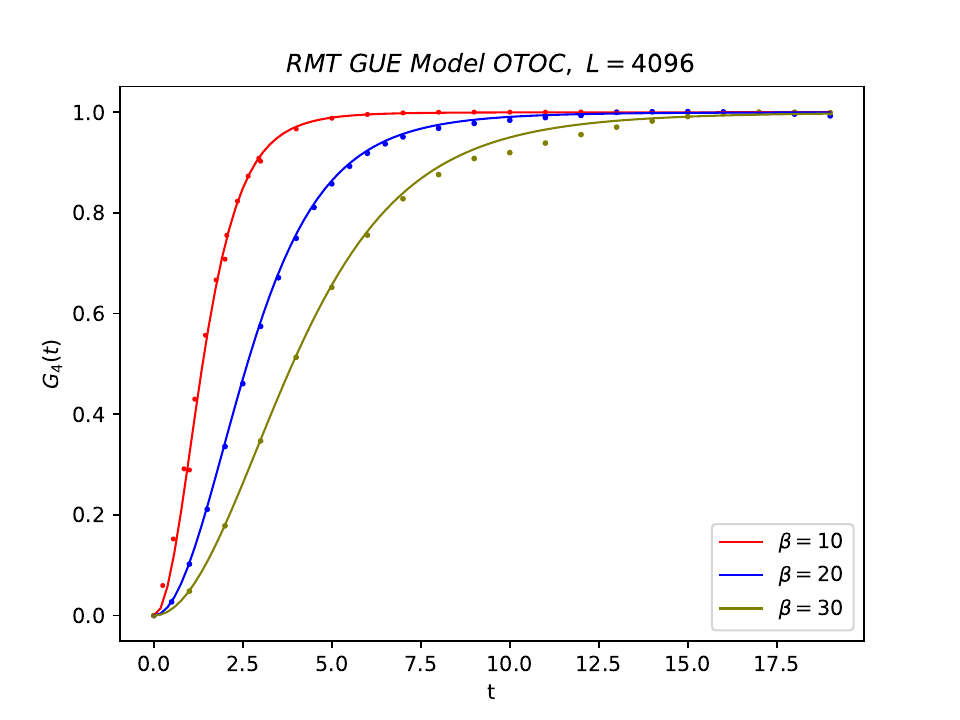}}
\caption{Plot of $G_{4}= 1 - {\tilde F}$  vs $t$. Solid  curves are the theoretical result, eq.(\ref{g4rmt}), in the $L\rightarrow \infty$ limit. Points are numerical data.  }
\label{otocg}
\end{figure}

The OTOC for RMT is plotted in Fig \ref{otocg}.
In appendix \ref{rmt4pt}, eq.~\eqref{4ptlarlfb},  we show that in the $L\rightarrow \infty$ limit, $G_4$ is given by 
\be
\label{g4rmt}
G_4=1- {F(t) \over F(0)}
\ee
where 
\be
\label{ftdef}
F(t)=\left |{{J_1\left(2 \left (t-i{\beta\over 4} \right)\right) \over t - i {\beta\over 4}}}\right|^4
\ee
where $J_\alpha(x)$ is the Bessel function. Comparing eq.~\eqref{g4rmt} with eq.~\eqref{gtilfdef} we have
\be
\label{deftildef}
{\tilde F}={F(t)\over F(0)}.
\ee
In Fig. \ref{otocg}, we show this function as a solid curve for three different values of $\beta$ and also the numerically obtained data for $L=4096$, which is in 
reasonable agreement. 

The important point to note is that $G_4$ reaches  its asymptotic value $1$ in a time scale of order unity, not scaling with $L$. This shows that the GUE exhibits 
``hyperfast" scrambling. As was noted in \cite{Cotler:2017jue} the thermalisation time scale is also of order unity and therefore comparable to this scrambling time. 
This  in contrast to what we saw above for the non-local SYK model where the scrambling time  scales like $\beta \ln N$. 

The underlying reason for this hyperfast scrambling is the $U(L)$ symmetry which is preserved by the GUE, see appendix \ref{rmt4pt} for further discussion and also in \cite{Cotler:2017jue}. There is no notion of ``simple operators" which is invariant under this symmetry. The $\psi_i$ operators under a general $U(L)$ transformation can be turned into arbitrarily complicated operators involving up to $N$ factors of the $\psi$'s. States prepared by acting with such complicated operators are already scrambled at the 
get-go, resulting in the hyperfast scrambling in the GUE. 

%
%
	\section{Sparse Random Matrices}
	\label{sprmt}


In this section, we shall explore the consequences of considering random matrices which are not fully random in the sense that not all elements of the matrix are randomly drawn numbers. We refer to this more general class of matrices as sparse random matrices. 

The reduction of randomness is achieved as follows. 

	We set the diagonal matrix elements $M_{ii}$ to vanish. 
	We then pick randomly and uniformly, $n$ off-diagonal matrix elements, $M_{ij}$, with $i< j$. The value of each of these matrix elements is chosen independently with its real and imaginary parts being drawn from a Gaussian distribution with variance { ${1\over 2L}$}.  All other  off-diagonal matrix elements of $M$ are also set to zero. Finally the matrix is   made Hermitian by taking it to be $\frac{1}{2}(M+M^\dagger)$. 
	 
	 The behaviour of the resulting random ensemble we obtain in this way, as a function of the number of random elements $n$, is then studied below numerically. 

	We now present results of numerical analysis for various quantities such as density of states, specific heat, level spacing and spectral form factor for the resulting ensemble.  
\subsection{The density of states}
\label{srmtdos}
	In the plots below, the density of states (DoS) is normalized to unity unless specified otherwise, {\it i.e.},
	\begin{align}
	\label{normden}
		\int d\lambda \rho(\lambda)=1 \,.
	\end{align}
		The plot for  a matrix of size $L=4096$,  as $ n \,(\le 1.6 \times 10^7)$ is varied,  is shown below. 
		The fully random case corresponds to the GUE with $L=4096$. 
		\begin{figure}[H]
\subfigure[]{		\hspace{-10mm}\includegraphics[width=7.5cm,height=6.5cm]{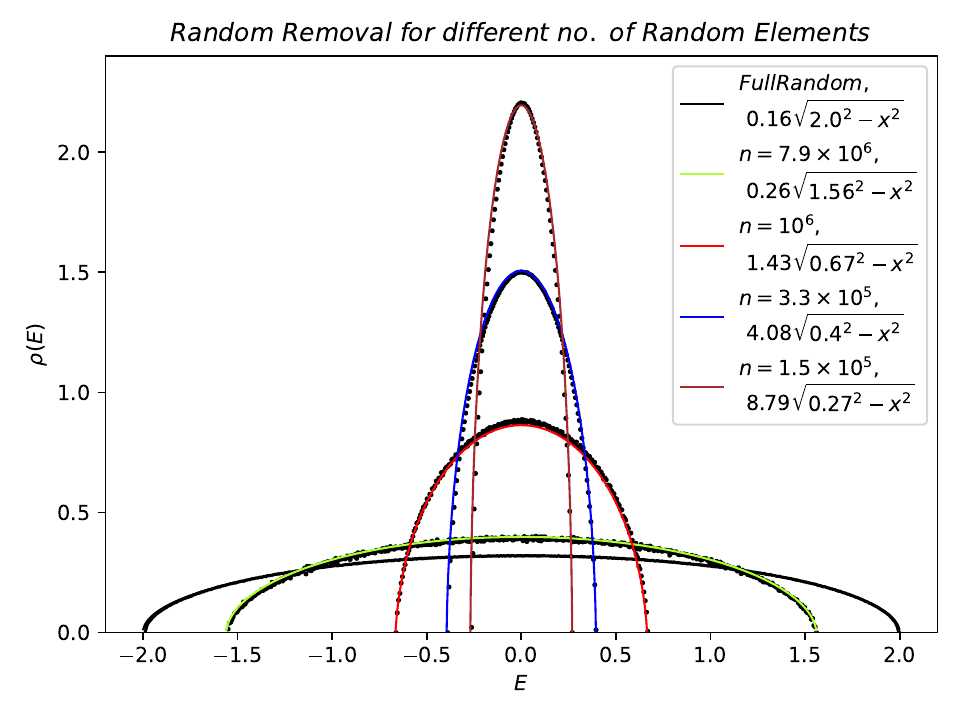}
		}
\subfigure[]{			\includegraphics[width=7.5cm,height=6.5cm]{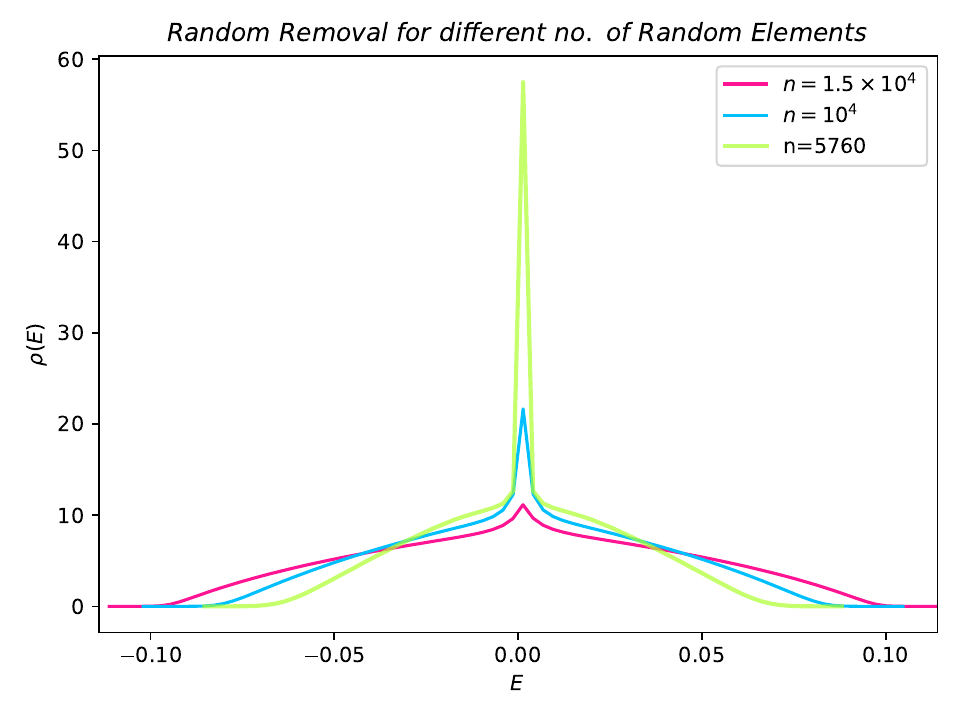}
		}		

				\caption{The density of states for the random sparseness with $L=4096$ and varying amount of randomness,  $n$. Lower values of $n$ implies a more sparse random matrix. For  $n\ge 1.5\times 10^5$ the spectrum is fitted by the Wigner semi-circle, after rescaling $E$, Fig(a) on left. Thereafter, for $n\le 1.5 \times 10^4$ the behaviour changes and this is no longer true, Fig(b) on right. 
			}
			\label{finiteeffectforp}
			\end{figure}
		
			
			
			We see that there is a general tendency for    the eigenvalues to get more concentrated  towards the centre, as $n$ decreases, {\it i.e.}, the sparseness is increased, indicating that the repulsion between eigenvalues decreases. 
			
			However starting from the GUE there is an interesting change in behaviour  as $n$ continues to decrease. Initially, when $n$ is not too small, corresponding to the values, $n\ge 1.5 \times 10^5$,  {\it i.e.}, , ${n\over L^2} \ge 9 \times 10^{-3}$,  in 
			Fig. \ref{finiteeffectforp} (a), the density of states can be fitted to  a Wigner semicircle, after rescaling the energy. We find,  
			\be
			\label{denrara}
			\rho(E)={2\over B} \rho_W\left({2 E \over B}\right),
			\ee
			where $\rho_W$ is the Wigner semicircle, eq.~\eqref{Wignersemicircle}. The parameter $B$ scales with $n$ in a power law fashion,
			{$B\sim \left({n \over L^2}\right)^{0.5}$},   as can be seen in Fig. \ref{sloperspa}. 
			\begin{figure}[H]
				\centering
						\begin{minipage}[b]{0.49\linewidth}
				\includegraphics[width=7cm,height=6.2cm]{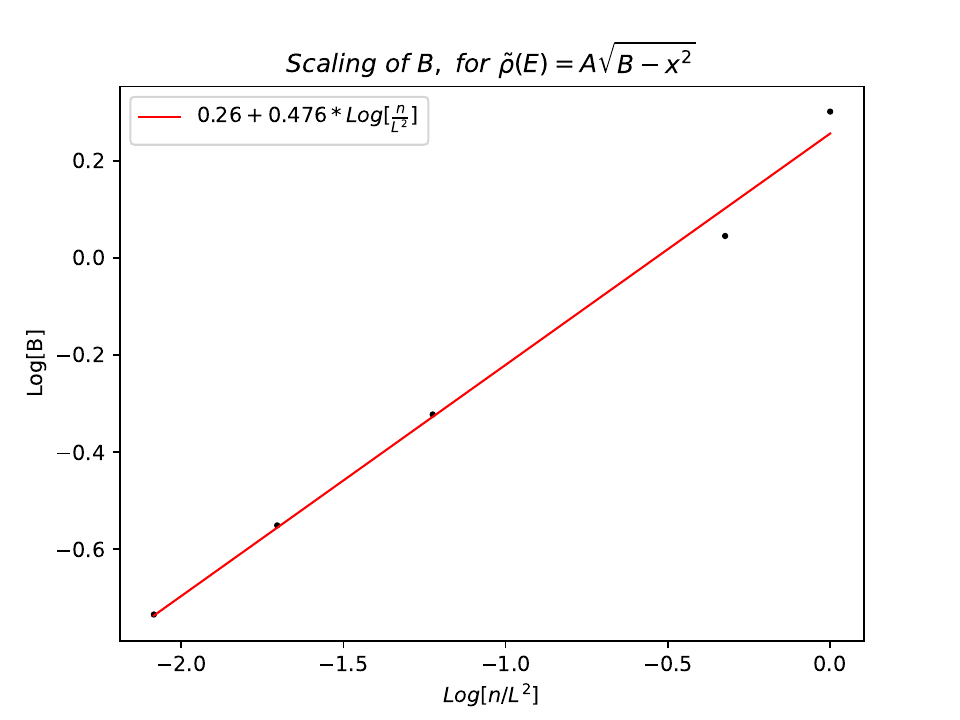}
			\end{minipage}     
			\caption{Dependence of rescaling parameters $B$ on ${n\over L^2}$. }
			\label{sloperspa}
			\end{figure}
			
			However once the randomness becomes very small, $n\le 1.5 \times 10^4$, {\it i.e.},  ${n\over L^2}\le9 \times 10^{-4}$, in Fig. \ref{finiteeffectforp} (b), there is an important  change. The effect of the repulsion among eigenvalues decreases to a point where the eigenvalues now like to clump much more at the center, near $E=0$, resulting in a sort of cusp at $E=0$. This is reminiscent of a condensate forming in bosonic systems.  As $n$ decreases further, the clumping and resulting cusp at $E=0$ get more pronounced. 
			The transition to the regime where the   Wigner distribution stops   describing $\rho$ occurs over a fairly narrow range in $n$. From Fig \ref{finiteeffectforp} we see that the value $n=1.5\times 10^5$  is well fitted with $\rho_W$, but this is no longer true for $n=1.5 \times 10^4$. 
%
			
			It would be very interesting to find out how sudden is this transition, as $L\rightarrow \infty$, but this is not easy to analyse numerically.
			Let us denote by $n_c$ the approximate value of $n$ which separates the  two regimes.  One would like to understand how $n_c$ behaves as $L\rightarrow \infty$; a natural ratio to consider is $n_c/L^2$ which is the fraction of total matrix elements that are non-zero. This ratio is plotted in Fig.\ref{nvsL} as a function of $1/L$. We find that, subject to the limitations of our numerical analysis, an approximate fit is given by 
			\be
			\label{apnc}
{		n_c\propto L^{1.5}, \ \ {n_c\over L^2}\propto L^{-0.5}}
			\ee

					\begin{figure}[H]
						\centering
			\begin{minipage}[b]{0.5\linewidth}
				\includegraphics[width=6.5cm,height=5.5cm]{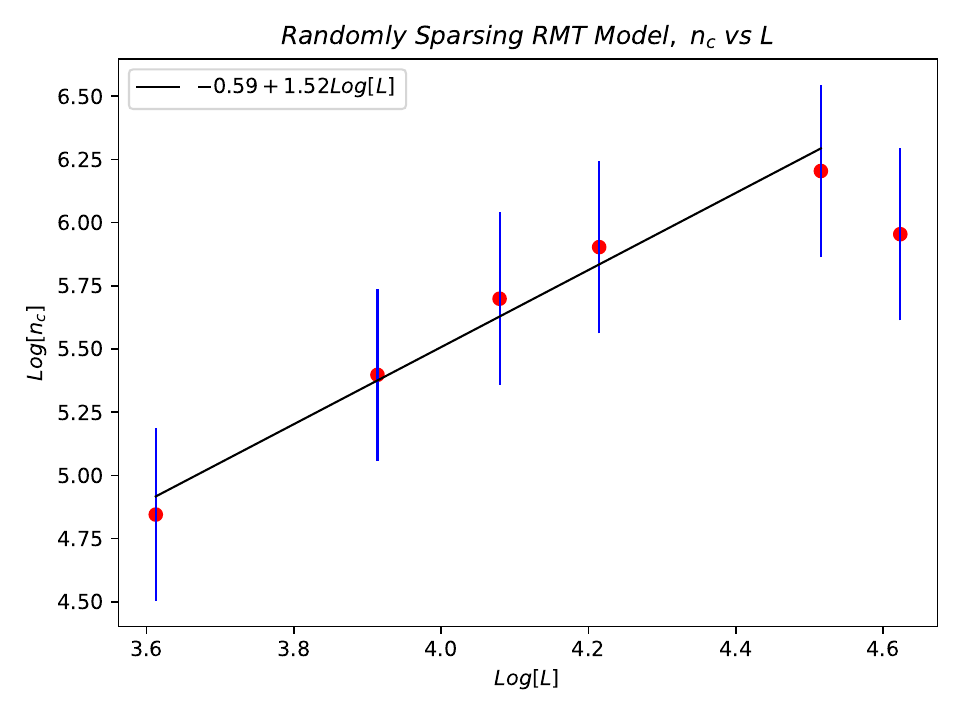}
			\end{minipage}           
				\caption{ Plot of $\log n_{c}$  vs $\log L$. Error bars are at $1 \sigma$. Solid line is the best fit. }
			\label{nvsL}
			\end{figure}

Note that in Fig. \ref{nvsL} the solid line is the best fit obtained after  discarding the last point (at the value of $\log L=4.623$).  { Interestingly, in \cite{kuhn}, a similar cusp-like behaviour in the distribution of eigenvalues for sparse random matrices is analyzed. It would be interesting to investigate the relation between this analysis and ours. We leave this for the future. 

{
\subsubsection{Modeling the weakening of eigenvalue repulsion }
\label{modelwkrep}
Let us end this subsection with one more comment. As mentioned earlier, the results of  fig.\ref{finiteeffectforp} suggests that the eigenvalue repulsion is decreasing as the sparseness in a random matrix is increased. Qualitatively, this can be modeled as a decrease in the repulsion term, the Vandermonde factor, in the matrix integral as we shall now show{\footnote{We thank Shiraz Minwalla for pointing this out to us.}}.  The matrix integral in eq.\eqref{GUE} can be expressed as an integral over the eigenvalues and takes the form 
\begin{align}
	Z_{GUE}&=\int \left(\prod_i d\lambda_i e^{-{\frac{L}{2}\lambda_i^2}}\right)\left(\prod_{i< j}|\lambda_i- \lambda_j|^2\right)\nonumber\\
	&=\int \prod_i d\lambda_i e^{-\sum_i \frac{L}{2}\lambda_i^2+2\sum_{i<j}\ln |\lambda_i -\lambda_j|}\nonumber\\
	&=\int \prod_i d\lambda_i e^{-\frac{L}{2}\int d\lambda \hat{\rho}(\lambda)\lambda^2+\int\int_{\lambda_{i}\neq\lambda_j}d\lambda_i d\lambda_j \hat{\rho}(\lambda_i)\hat{\rho}(\lambda_j)\ln |\lambda_i -\lambda_j|}\label{zgueindos}
\end{align}
where in the last line, we introduced a density of states $\hat{\rho}(\lambda)$ to write the sum as a integral. This density of states is such that its normalization is $L$ corresponding to the fact that there are $L$ eigenvalues for an $L\times L $ matrix. Rewriting it in terms of the unit-normalized density of states, $\rho(\lambda)$ as
\begin{align}
	\hat{\rho}(\lambda)=L \rho(\lambda)\label{hatrhrho}
\end{align}
the partition function for the full random matrix can be written as
\begin{align}
	Z_{GUE}=	&=\int \prod_i d\lambda_i e^{-L^2\left(\half \int  {\rho}(\lambda)\lambda^2+\int\int_{\lambda_{i}\neq\lambda_j} {\rho}(\lambda_i){\rho}(\lambda_j)\ln |\lambda_i -\lambda_j|\right)}\label{fulgzinrl2}
	\end{align}
The derivation of the Wigner distribution proceeds by carrying out a saddle point analysis, justified by the presence of the factor, $L^2$, which is large when $L\rightarrow\infty$. 
The Vandermonde factor $|\lambda_i -\lambda_j|^2$ in the first line of eq.\eqref{zgueindos} is the term that causes the repulsion between the eigenvalues as can be seen by the divergence of the log term in the second line. Thus, taking a cue from this, we can model the decrease in the repulsion by changing the exponent of the Vandermonde factor as follows. Consider the following ansatz for the partition function for the sparse random matrix, 
\begin{align}
		Z_{\text{Sparse-GUE}}&=\int \left(\prod_i d\lambda_i e^{-{\frac{L}{2}\lambda_i^2}}\right)\left(\prod_{i< j}|\lambda_i- \lambda_j|^{2\alpha}\right)\label{zguesprans}
\end{align}
Repeating the same series of steps that led to eq.\eqref{fulgzinrl2}, we now get
\begin{align}
		Z_{\text{Sparse-GUE}}	&=\int \left(\prod_i d\lambda_i\right) e^{-L^2\left(\half \int  {\rho}(\lambda)\lambda^2+\alpha\int\int_{\lambda_{i}\neq\lambda_j} {\rho}(\lambda_i){\rho}(\lambda_j)\ln |\lambda_i -\lambda_j|\right)}\label{fulgzinrlspl2a}
\end{align}
The factor $\alpha$, as can be seen above, appears only in the second term in eq.\eqref{fulgzinrlspl2a}.
At this stage, we can now do a rescaling of the density of states $\rho(\lambda)$ and the integration variables to get to the form in eq.\eqref{fulgzinrl2}. The rescaling is given by 
\begin{align}
	\rho(\lambda)= \frac{1}{\sqrt{\alpha}}\rho_r\left(\frac{\lambda}{\sqrt{\alpha}}\right)\label{rhoresc}
	\end{align}
Since, the range of the integration in eq.\eqref{fulgzinrlspl2a} is $[-\infty,\infty]$, we can rescale the integration variables as $\lambda=\sqrt{\alpha}\hat{\lambda}$ to get
\begin{align}
		Z_{\text{Sparse-GUE}}	&=\int \left(\prod_i d\lambda_i\right) e^{-L^2\alpha\left(\half \int  {\rho_r}(\hat{\lambda})\hat{\lambda}^2+\int\int_{\hat{\lambda}_{i}\neq\hat{\lambda}_j} {\rho_r}(\hat{\lambda}_i){\rho_r}(\hat{\lambda}_j)\ln |\hat{\lambda}_i -\hat{\lambda}_j|\right)}\label{fulgrrrspl2a}
\end{align}
The form of the partition function above for sparse random matrix is similar to the partition function of the full random matrix except for the factor of $\alpha$ as an overall factor. This shows that as long as $L^2\alpha\rightarrow\infty$ when $L\rightarrow\infty$, we can perform the same saddle point analysis as is done for the full random matrices to obtain a Wigner semi-circle distribution for the rescaled density of states $\rho_r$.  Also that the original density of states $\rho(\lambda)$ is simply related to the Wigner distribution by a simple rescaling. 

The above modeling of the sparse random matrices by the weakening of the exponent of the Vandermonde term gives the rescaled Wigner distribution as long as 
\begin{align}
	L^2\alpha\gg 1 \label{alphainl2}
\end{align}
The analytical results presented above show that  the  numerical results for the distribution of eigenvalues, see the left panel of fig.\ref{finiteeffectforp}, can be well modeled by a weakening of the eigenvalue repulsion term. 

By comparing the numerical results with the model discussed above, we can estimate the value fo $\alpha$ as a function of the sparseness parameter $(n/L^2)$, with $n$ being the number of non-zero elements in the random matrix. Comparing eq.\eqref{denrara} and eq.\eqref{rhoresc}, we find 
\begin{align}
	\alpha=\frac{B^2}{4}\label{alphb}
\end{align}
Furthermore, as was discussed below eq.\eqref{denrara} noting that $B\sim \left(\frac{n}{L^2}\right)^{0.5}$ we find that approximately
\begin{align}
\alpha\sim \frac{n}{L^2}	\label{binin}
\end{align}

However, the numerical results suggest that for sufficiently sparse matrices, the distribution deviates away from the Wigner distribution, see right panel of fig.\ref{finiteeffectforp}. This means that the model presented above breaks down at a critical value, $\alpha_c$, obtained when $n=n_c$. Using the result for the scaling of $n_c$ as a function of $L$ mentioned in eq.\eqref{apnc}, we find that the  break down  approximately at 
\begin{align}
	\alpha_c\propto L^{-0.5}\label{alphac}.
\end{align}
which  corresponds to a  value for the number of random variables, $n_c\sim L^{1.5}$. It is interesting to note that this is value  is much bigger than the  breakdown suggested by the model above, which leads from   eq.\eqref{alphainl2} to $n_c\sim O(1)$. This indicates that  while our modeling of the weakening of the repulsion is qualitatively correct it  is not precise enough to capture some key features of the numerical results.  We leave a better understanding of this for the future. 
}

			\subsection{Comments on  Various Properties}
			\label{srmtcv}
			
			Here we discuss some of the properties of ensemble obtained above with  reduced randomness.

			\noindent \underline{\bf Thermodynamics}
			
			The density of states determines the  finite temperature behaviour of the { thermal} partition function. 
			Since for $n>n_c$,  $\rho$ is well fitted by $\rho_W$,  in this case  the thermodynamics will agree with that of the GUE, after an appropriate rescaling of energy or temperature. However once $n<n_c$ this will no longer be true.  The resulting behaviour of the entropy $S$ at low temperatures is shown in  Fig. \ref{sentforrr}. We have not been able to obtain a good analytical fit to this result. Note that in this case $n_c$ lies in between $1.5 \times 10^4$ and $1.5 \times 10^5$. 
			
			It is interesting to contrast the behaviour of entropy, $S$, for  the random sparseness case  with that of the  GUE and the SYK$_4$ theory shown in Fig {\ref{entguenlsyk}}. 
			In particular, for the SYK$_4$ model, in the large $N$ limit,   we saw that there is a range of temperatures, eq.(\ref{lesyk}), where a saddle point approximation is valid resulting in the entropy behaving linearly with temperature, eq.~(\ref{entSYK}).  One would like to study if  something similar happens for the random sparseness case,  for $n<n_c$, $L\rightarrow \infty$.
			This is however difficult to do numerically.

			\begin{figure}[H]
\centering
					\includegraphics[width=7.5cm,height=6cm]{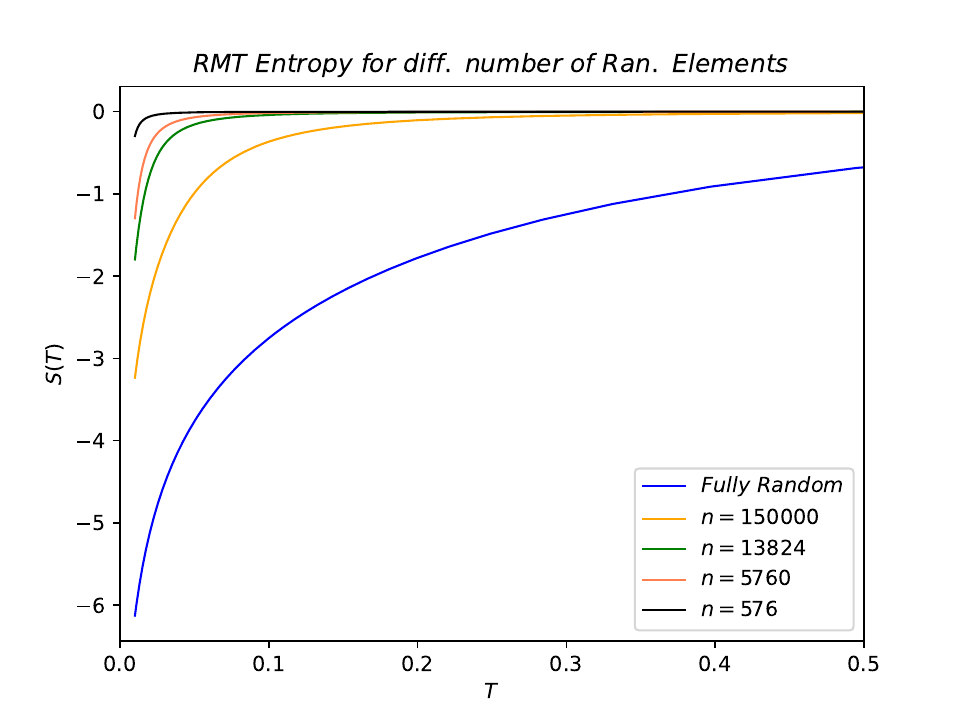}
				
				\caption{  The Entropy at low temperature for varying values of $n$. Here we plotted only the temperature dependent part of the entropy, after subtracting  $\log L $. }
				\label{sentforrr}
								
			\end{figure}

			\noindent \underline{\bf Spectral Form Factor (SFF)}		
			
				We now turn to the spectral form factor, defined in eq.~(\ref{defsff}), with connected and disconnected pieces given in eq.~(\ref{gd}), (\ref{gc}). We shall analyse how the SFF changes with changing randomness, $n$. 
			
			We recall,  eq.~(\ref{cex}), that   the SFF is determined by  the  two-point density correlator and therefore is sensitive to the interactions between the eigenvalues. 

		The numerical results for the random sparseness case are in Fig. \ref{SFFran1}, again for $L=4096$. We take  $\beta=5$ and consider  varying values for $n$.
		 The plots show  much richer behaviour, compared to the GUE, Fig.\ref{SFFRMT}. 
		In  Fig. \ref{SFFran1} besides the fully random case, the SFF is plotted for $5$ other values of $n$. Of these $n=138240$, is roughly comparable 
		to $n_c$, all other values of  $n$ lie  in the range $n<n_c$, where $\rho_W$ is no longer a good fit.  Generally speaking, differences in the SFF compared with  the GUE grow as $n$ decreases.  However, it is interesting to note that the $n=138240$ case already shows differences with the GUE.


			\begin{figure}[H]
			\begin{minipage}[b]{0.3\linewidth}
				\hspace{-19mm}\includegraphics[width=6cm,height=6cm]{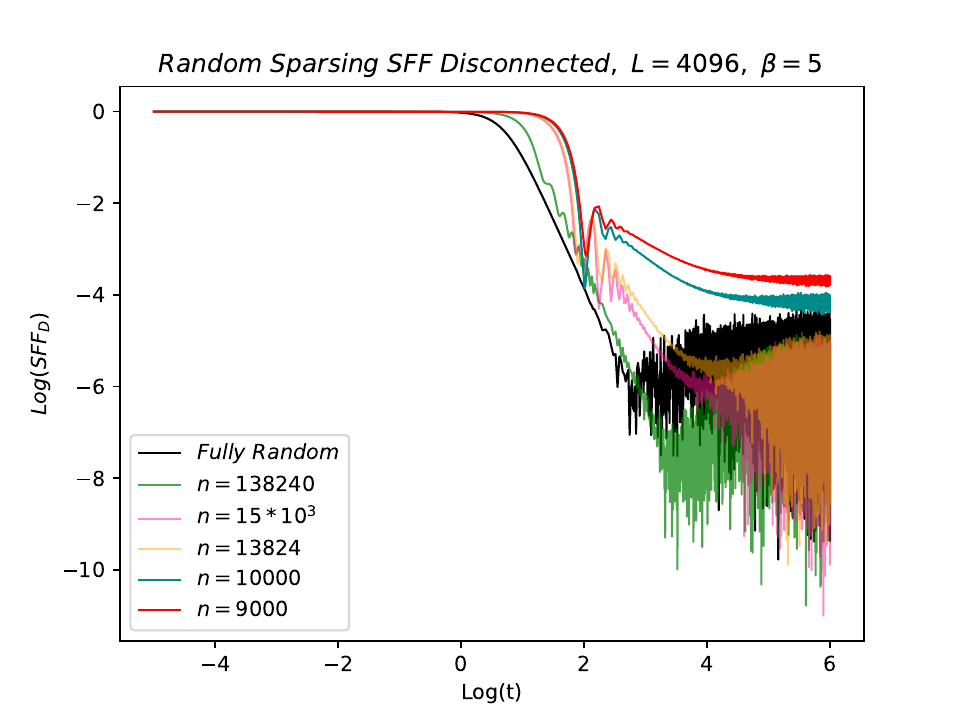}
			\end{minipage}            \hspace{-8mm}
			\begin{minipage}[b]{0.32\linewidth}
				\includegraphics[width=6cm,height=6cm]{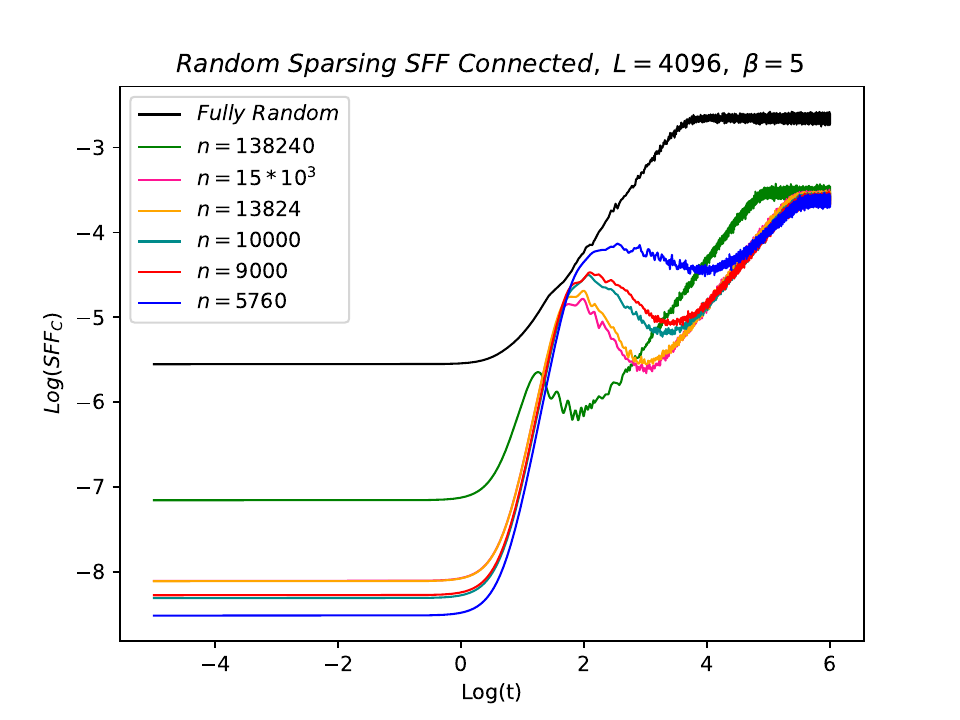}
			\end{minipage}            \hspace{8mm}
			\begin{minipage}[b]{0.32\linewidth}
				\includegraphics[width=6cm,height=6cm]{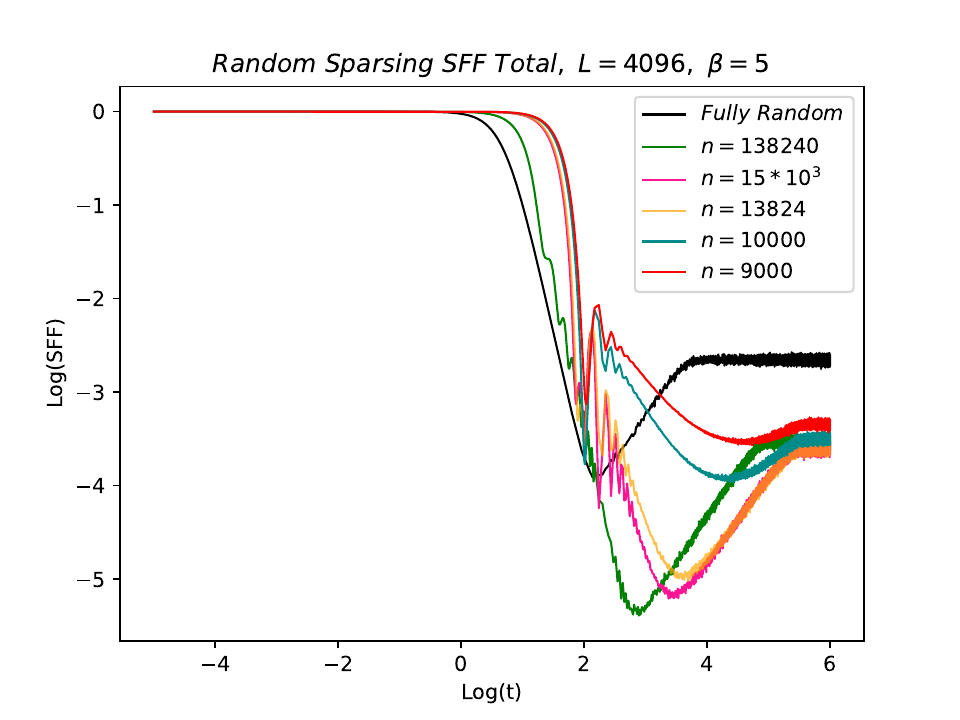}
			\end{minipage} \newline
			\begin{minipage}[b]{0.3\linewidth}
				\hspace{-19mm}\includegraphics[width=6cm,height=6cm]{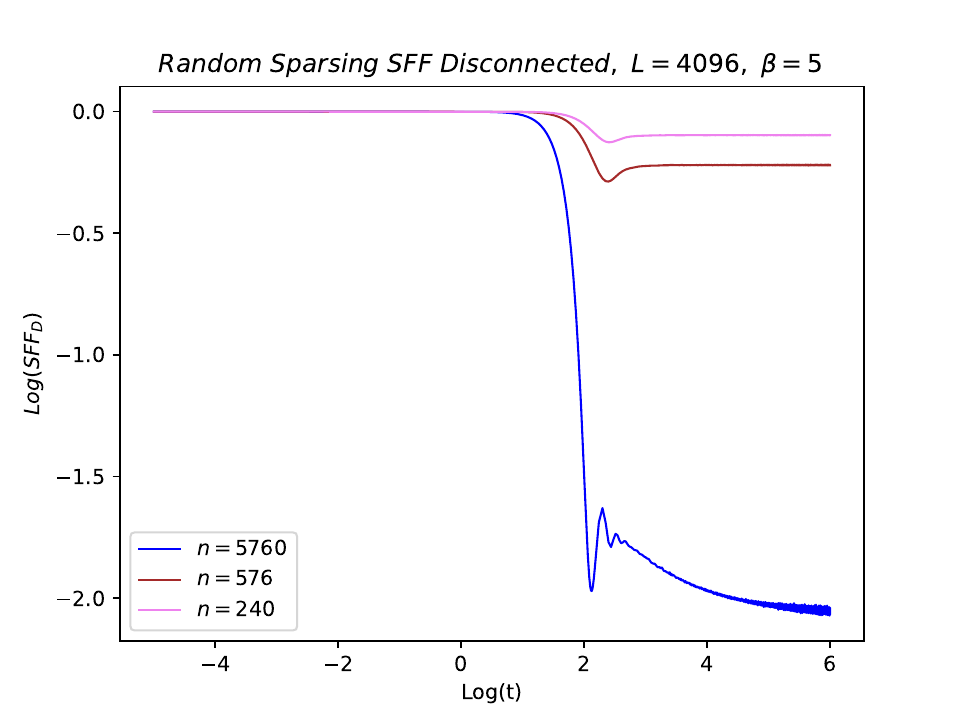}
			\end{minipage}            \hspace{-8mm}
			\begin{minipage}[b]{0.32\linewidth}
				\includegraphics[width=6cm,height=6cm]{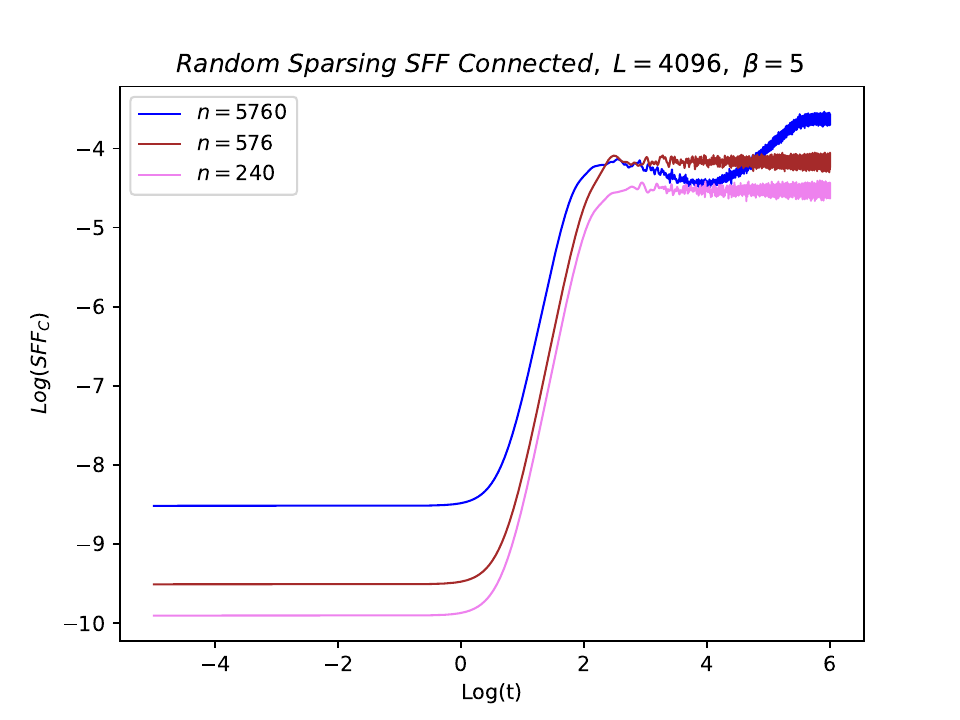}
			\end{minipage}            \hspace{8mm}
			\begin{minipage}[b]{0.32\linewidth}
				\includegraphics[width=6cm,height=6cm]{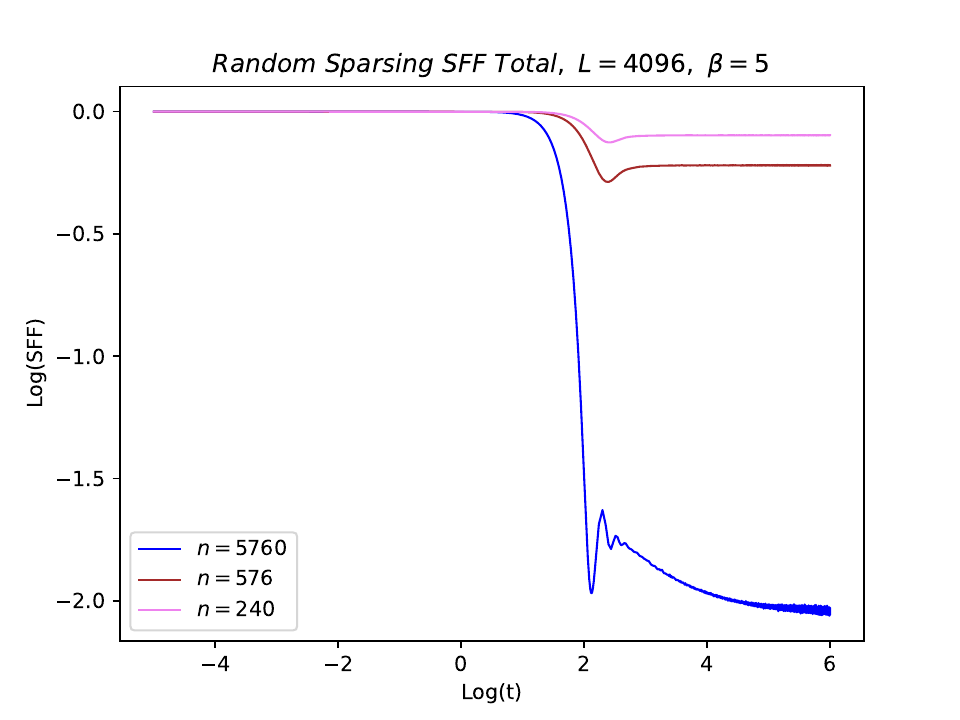}
			\end{minipage}
			
			\caption{Left: disconnected, middle: connected, right: total spectral form factor  with changing number of random elements, parametrised by $n$,  for a matrix of rank $2^{12}=4096$. We have taken an average over 300 ensembles for each cases.  Here $n_c \sim 1.5 \times 10^5$. }
			\label{SFFran1}
		\end{figure}
		While the general features of a dip, ramp and plateau are present, there are interesting differences: 
		\begin{itemize}
		\item
		The top panel has $n$ going down till $9000$. The bottom panel has lower values of $n \le 5760$. 
		In each case we plot the disconnected (left), connected (middle) and total (right) SFF. 
		 We see that in the top panel the ramp region  in the connected SFF has much more structure, and unless $n$ becomes very small and we go to the bottom panel  ($n \le 5760$),  it is not monotonic,  and exhibits an intermediate local maximum, which we call a ``peak",  followed by the local minimum and then grows until the final plateau. 
		 This indicates that while   the nearest neighbour eigenvalues continue to repel, leading to a plateau at late times; at intermediate separation, beyond nearest neighbours,   in fact an attraction develops between eigenvalues. 

		\item
		The height of the plateau decreases with decreasing $n$ suggesting that the repulsion between the nearest neighbours also reduces. This also agrees with what we saw in the density of states, Fig.~\ref{finiteeffectforp}, which get more clumped in the centre as $n$ decreases. 
		
		\end{itemize}

			Some more comments are as follows: 
			\begin{itemize}
			\item The disconnected piece itself has interesting structure. For the GUE and SYK models, the disconnected piece is known to decay and be self averaging, meaning that a single instantiation of the Hamiltonian (for big enough $N$),  reproduces the ensemble average. Here we see that the disconnected piece has an intermediate  `spike' followed by a slower decay, with the intermediate spike being more pronounced for in-between values of $n$.  See Fig.\ref{SFFran1}, left panels,  $n=13824, 10000, 9000$ around $\log(t) = 2$ for spike. 
			 \item The connected piece becomes smaller in magnitude as $n$ decreases, and the full SFF is increasingly dominated, for smaller $n$, by the disconnected term. 
			 \end{itemize}

\noindent \underline{\bf Level Spacing}\\

		Next we turn to the nearest neighbour level spacing. There are a few points to keep in mind. 

		It is important to compute the level spacing after unfolding the spectrum to obtain the Wigner surmise for  the Gaussian ensembles or  SYK$_q$ theory\footnote{We thank Tezuka Masaki for helpful discussion on this point.}. 
		The eigenvalue spacing after unfolding is given by 
		\begin{align}
			s=\Delta\tilde{\lambda}_i=  {\rho(\lambda_i)} \, (\lambda_{i+1}-\lambda_i)\,. \label{dellamt}
			\end{align}
			we see that it is the local average value of the eigenvalue difference.
			The histogram plot of the quantity $\Delta\tilde{\lambda}$ for Gaussian ensembles or the SYK$_q$ theory  then gives the standard Wigner Surmise.
		 
		 Alternatively, another quantity which avoids this procedure 
		 is the ratio of eigenvalue differences, since the dependence of the density of states in eq.~\eqref{dellamt} cancels out in the ratio; 
		 \begin{align}
		 	r_i=\frac{\Delta \lambda_i}{\Delta \lambda_{i+1}}\label{rnde} \,.
		 \end{align}
	 Some more information on how the distribution of $r_i$ behaves for various Gaussian ensembles is given in appendix \ref{Plnrder}. 
	 
		If the eigenvalues are uncorrelated then the statistics will be that of the Poisson distribution  which corresponds to
		\begin{align}
			{ p(r)=\frac{1}{(1+r)^2} }  
			\label{poisrn}
		\end{align}
	  As in \cite{You:2016ldz}, it will actually be convenient to  plot the distribution of the logarithmic ratio $\ln r$, which is,
	  \begin{align}
	  	P(\ln r)=r p(r)	\label{Plnr}
	  \end{align}
  
  Also, we need to be  careful about degeneracies in the spectrum. There will be a degeneracy in the spectrum if the model under consideration possesses symmetries. For example, if the Hamiltonian has fermion parity symmetry, then the eigenvalues will be 2-fold degenerate. Therefore in order to compute the level statistics, one has to first separate the eigenvalues according to the parity sector and plot the differences in eigenvalues corresponding to any one sector. As mentioned in \cite{You:2016ldz}, for $N=0\,\, \text{(mod)} \,\,8$,  which is the case for $N=24$ for which we present the analysis, there are no degeneracies after we do this. 
  
  Let us now turn to the analysis of the plots. 		We have considered matrix of size $L=4096$ and an ensemble size of 1000 in obtaining these plots.

	\begin{figure}[H]
	\hspace{-10mm}
\subfigure[$P(s)$ vs unfolded level spacing]{\includegraphics[width=7.5cm,height=6cm]{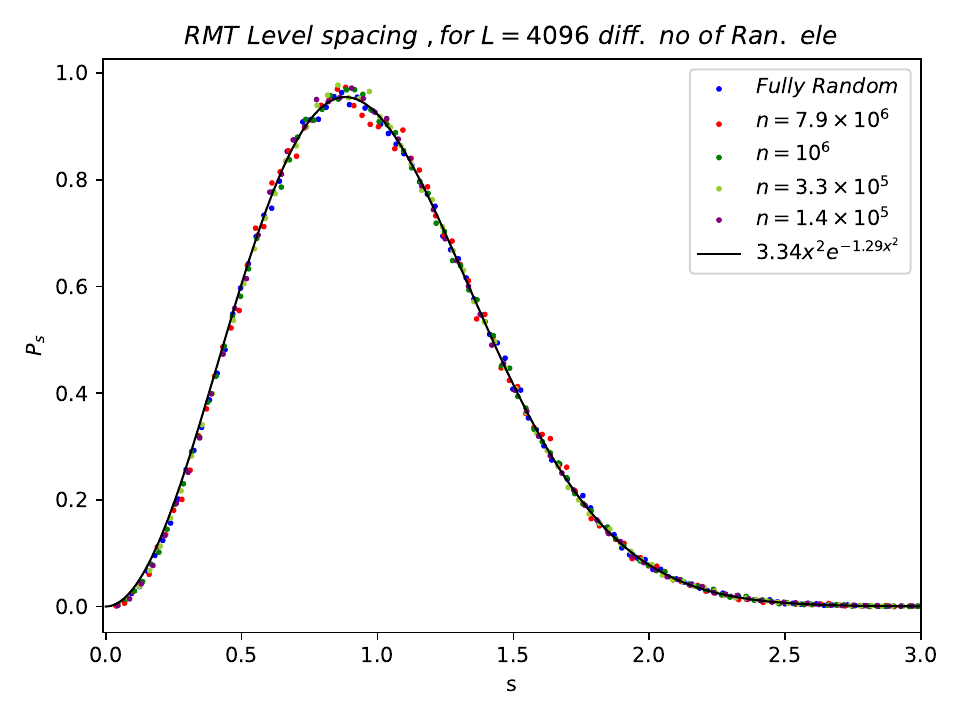}}
\hspace{-2mm}
\subfigure[$P(\ln r)$ vs $\ln(r)$ plot]{\includegraphics[width=7.8cm,height=6.23cm]{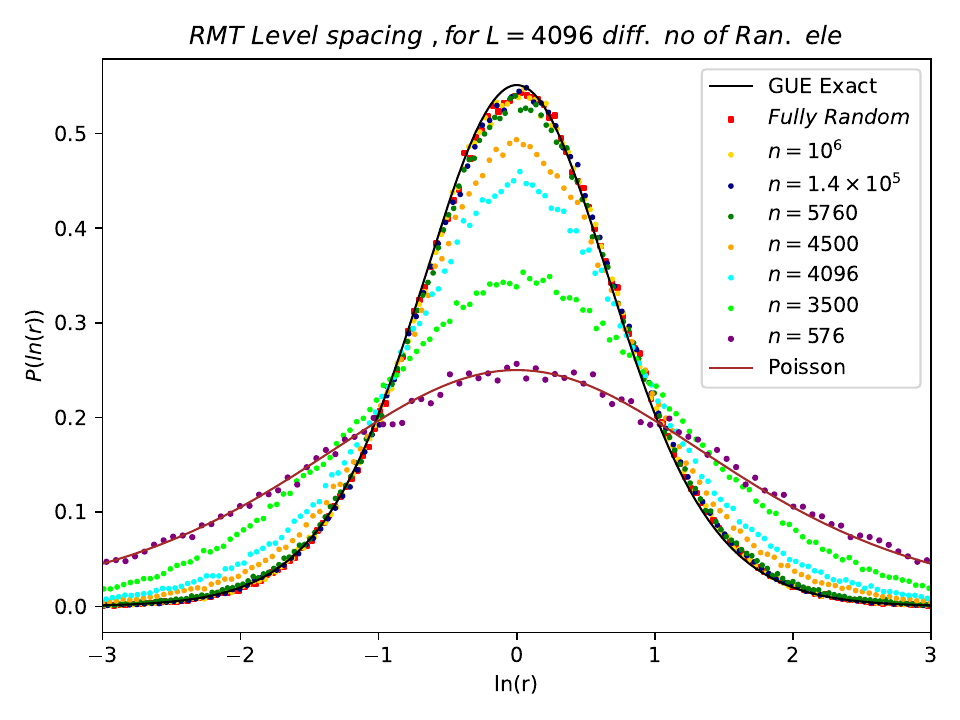}}


	\caption{Level Spacing  with changing randomness ($n$), for a matrix of rank $2^{24/2}=4096$. $n_c \sim 1.5 \times 10^5$.  }
	\label{LEVEL3}
\end{figure}

The left panel is a plot of the probability $P(s)$ versus the unfolded level spacing $s$ given by  eq.~(\ref{dellamt}), for varying values of the randomness parameter { in the range $n \ge n_c$}. 
The Wigner distribution is given by the black curve and corresponds for the GUE to the function,
\be
\label{GUEwig}
P(s)=\frac{32}{\pi^2}s^2e^{-\frac{4s^2}{\pi}}
\ee
 The fully random case is the data for the GUE, and other curves are obtained by reducing $n$   until it reaches a value close to $n_c$.

In the right hand panel we plot $P(\ln(r))$, defined in eq.(\ref{Plnr}), vs $\ln(r)$. Again the black curve is the result for the Wigner surmise, the fully random case is the data for the GUE, and other cases correspond to varying values of $n$, which for $n\le 5760$ are less than $n_c$. 
Note that for the Wigner surmise $p(r)$ is given by 
\begin{align}
	p(r)= N_\beta  \frac{r^{\beta} (1+r)^\beta}{\left( 1 + r + r^2 \right)^{1 + \frac{3 \beta}{2}}} \,,\label{prb}
\end{align}
from which it follows from eq.~\eqref{Plnr} that
\begin{align}
	P(\ln r) = N_\beta  \frac{r^{\beta+1} (1+r)^\beta}{\left( 1 + r + r^2 \right)^{1 + \frac{3 \beta}{2}}} \,,\label{Plnrb}
\end{align}
where 
	\begin{align}
		N_{\beta=1} &= \frac{27}{8} \, \, \quad \quad  \text{for GOE} (\beta =1)  \,, \nonumber\\
		N_{\beta=2} &= \frac{81 \sqrt{3}}{4 \pi} \, \quad \text{for GUE} (\beta =2)\label{Nbetv}.
	\end{align}
For the Poisson distribution it is given by 
\begin{align}
P(\ln r)=\frac{r}{(1+r)^2}   
	\label{Plnrposs}
\end{align}
The result in  eq.~\eqref{prb} is derived in appendix \ref{Plnrder} for the simple case of $3\times 3 $. 

We see from the left panel of fig. (\ref{LEVEL3})  that up till $n=1.5 \times 10^5$, when the density of states fits a rescaled Wigner semi-circle,  the level statistics is also well fitted by the Wigner surmise. However when $n$ decreases further past $n_c$, then differences set it, as we can see from the right panel. These grow, as $n$ decreases  and eventually for $n=576$, when $n/L^2\sim 3\times 10^{-5}$ and the randomness is very sparse,  the statistics becomes Poisson. 

It is not unexpected  that for vastly reduced randomness the statistics becomes close to Poisson. 
Ideally one would have liked to obtain the behaviour with varying $n$ in the limit when $L\rightarrow \infty$, but that is difficult to do numerically and left for the future. 

\noindent \underline{\bf  OTOCs }

Finally, we consider the behaviour of the OTOC, $G_4=1-{\tilde F(t)}$,  eq.~\eqref{gtilfdef},  as a function of time $t$ and analyse it as we vary $n$, the number of random elements\footnote{{ For parrallelized matrix free OTOC computation we have used ``dynamite'' Python frontend, Krylov Subspace methods, PETSc and SLEPc Packages. Detailed discussion for this packages is discussed in \cite{Kobrin:2020xms}}.}. In each case we work with $L=4096$,  and an ensemble size of $2$. Four figures are given below for different values of $n$, $10^6$,$10^5$,$10^4$, $4.5\times10^3$.
In the left panel of each figure the points are the data. { Note that the range of $t$ is different for  $n=10^6, 10^5$ and $n=10^4, 4500$. Comparing the left panels for various $n$, we can see that in general, the growth of the OTOCs slows down as $n$ decreases.} Furthermore, there is a noticeable slowing down once $n$ crosses the value $n_c$, which is somewhat smaller than $10^{5}$.
For example, the OTOCs in Fig.  \ref{otoc111}, with $n=10^6$, exhibit a saturation time scale, at which they approach their asymptotic value, which is  quite similar to that for the GUE, Fig. \ref{otocg}. for example, for $\beta=10$, Fig. \ref{otoc111},   this time scale is   $t\sim 10$, which is comparable to the time scale in fig. \ref{otocg}.
Whereas for $n=10^4$ the saturation  time scale, for $\beta=10$ is $t\sim 50-60$. 

The solid curves in the left panel are obtained by fitting an initial region, $t\le t_{max}$, of the OTOCs to a functional form $A + B e^{\lambda t}$, as in the previous discussion of OTOCs above. The value of $t_{max}$ for each curve has been estimated  by choosing the maximum range in $t$ for which such a functional form is a good fit. The resulting estimate for the Lyapunov exponent is then plotted  in the right panel of each figure. We have done the estimate for $t_{max}$
in a preliminary manner in this analysis and our corresponding estimate for $\lambda$ should also be taken to be preliminary one. 

In particular, we recall that  the GUE exhibits hyperfast scrambling with the OTOCs saturating in a time scale of order unity, Fig \ref{otocg}. One would expect this to continue to be true for sufficiently large values of  $n \sim L^2$. As $n$ beginning to decrease it is possible that exponential growth sets in and a Lyapunov exponent can be defined
in the $L\rightarrow \infty$ limit, once $n$ becomes sufficiently small, perhaps smaller that $n_c$. However we have not been able explore this possibility systematically and leave it for future investigation. 

Let us recall that the underlying reason for the hyperfast growth of the OTOCs for the GUE is $U(L)$ symmetry of the full random matrix, see the discussion above section \ref{sprmt}. On the other hand, for the sparse random matrices we have discussed so far, this $U(L)$ symmetry is broken. This is explained in detail in appendix \ref{symsparuni}. This provides the motivation to study the OTOCs for sparse random matrices. 

\newpage
	\noindent \underline{\bf 1)  Random Elements $n=10^{6}$}\\
\begin{figure}[H]
	\centering
\vspace{-5mm}
\hspace{-10mm}
\subfigure[]{\includegraphics[width=7cm,height=6cm]{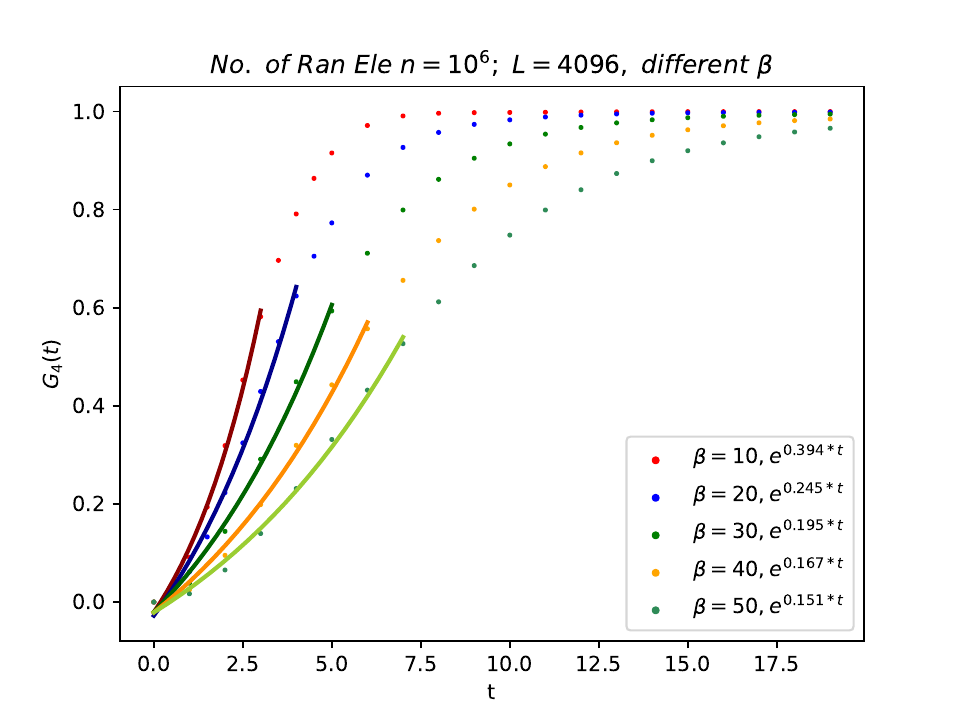}}
\subfigure[]{\includegraphics[width=7cm,height=6cm]{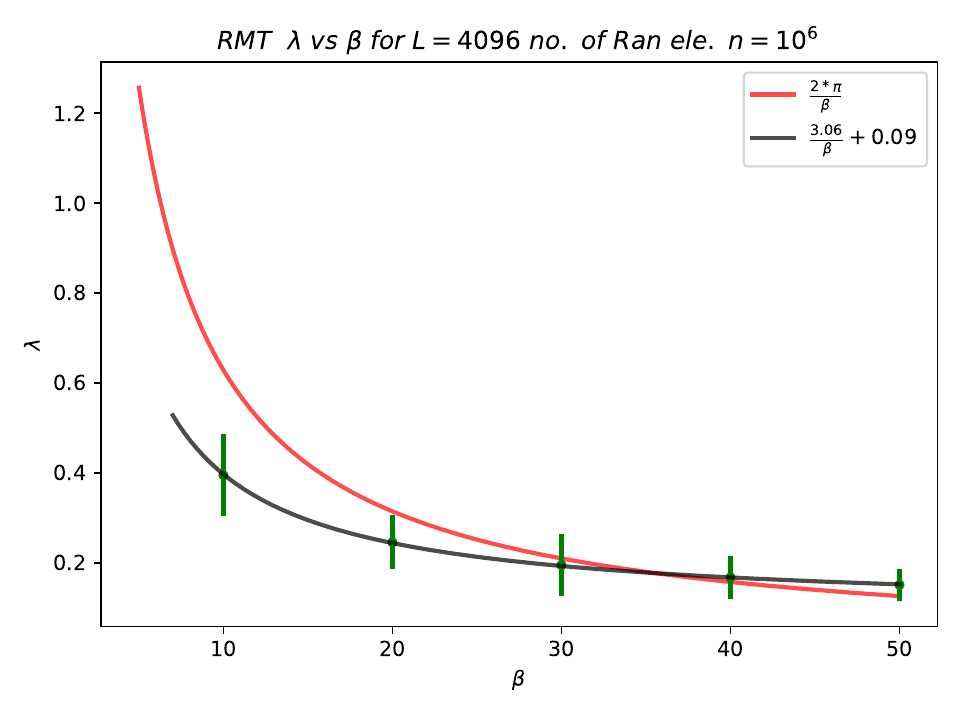}}
\caption{Fig(a) is the OTOC as a function of time $t$ for different $\beta$.  Solid curve is the  best fit at early times. Fig(b)  is the Lyapnov coefficient as a function of $\beta$.}
\label{otoc111}
\end{figure}
\vskip -10pt
\noindent \underline{\bf  2) Random Elements $n=10^{5}$}\\
\vspace{-5mm}
\begin{figure}[H]
	\centering
\hspace{-10mm}
\subfigure[]{\includegraphics[width=7cm,height=6cm]{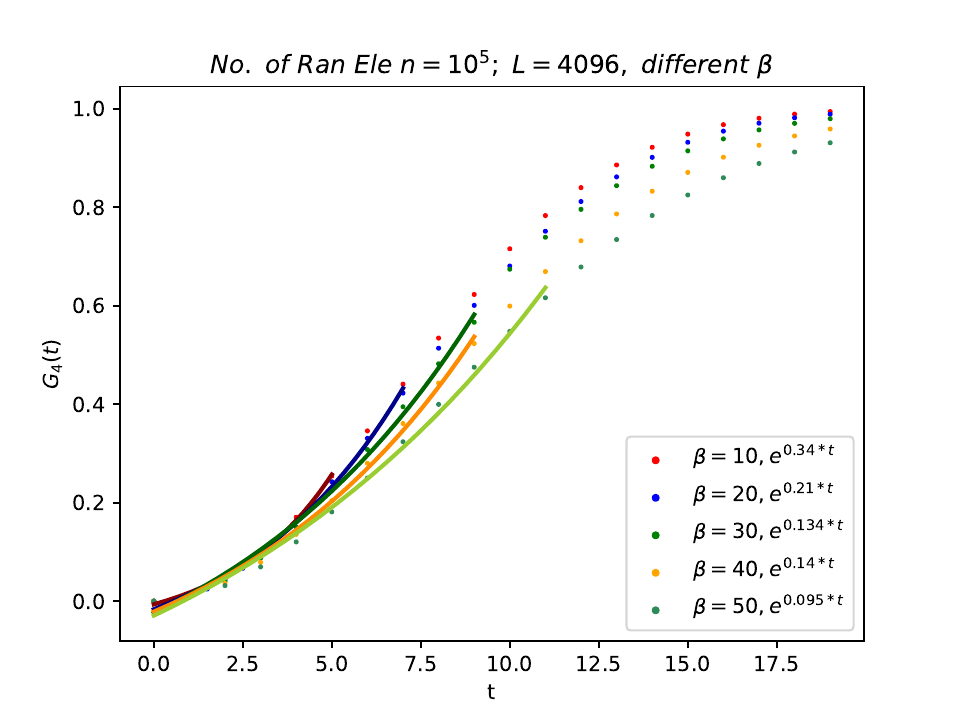}}
\subfigure[]{\includegraphics[width=7cm,height=6cm]{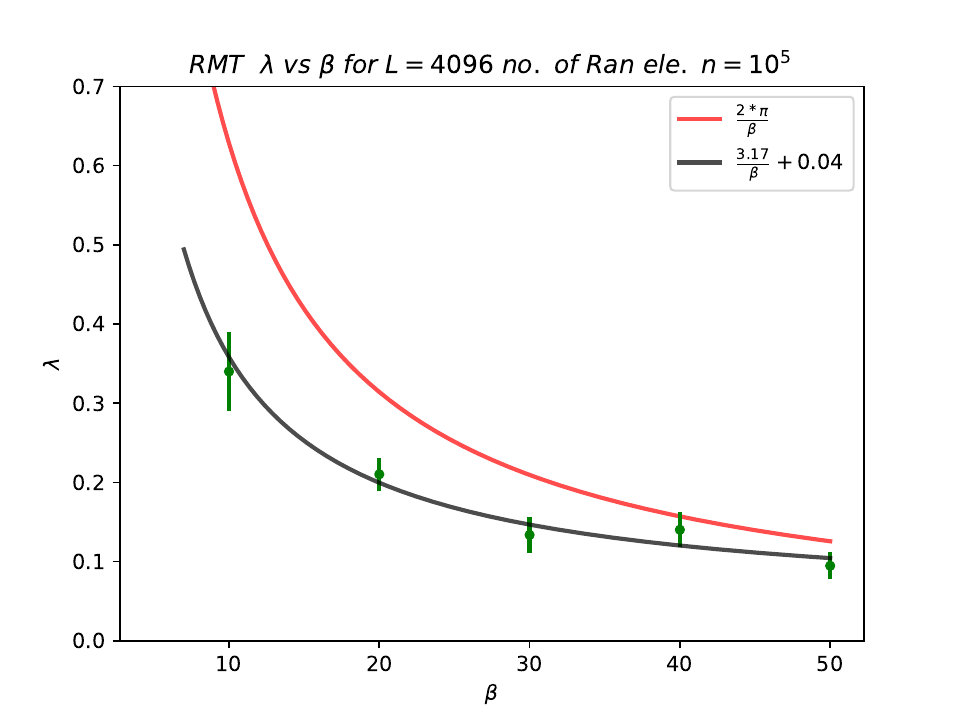}}
\caption{Fig(a) is the OTOC as a function of time $t$ for different $\beta$.  Solid curve is the  best fit at early times. Fig(b)  is the  Lyapnov coefficient as a function of $\beta$.}
\label{otoc112}
\end{figure}
\vspace{10mm}

\noindent \underline{\bf 3)  Random Elements $n=10^{4}$}\\
\begin{figure}[H]
	\centering
\vspace{-5mm}
\hspace{-10mm}
\subfigure[]{\includegraphics[width=7cm,height=6.2cm]{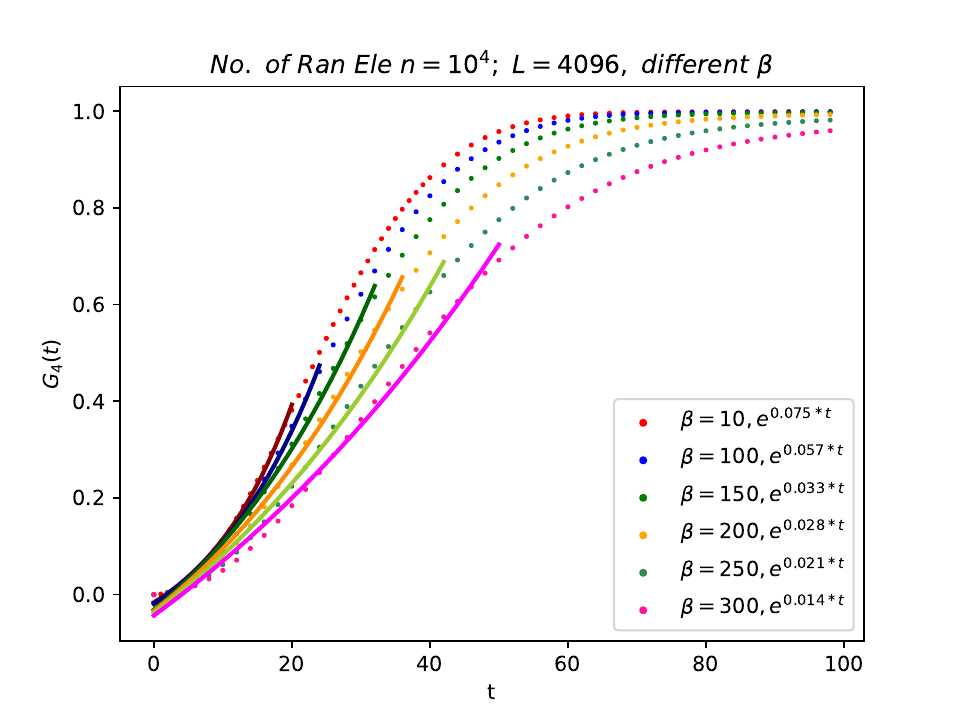}}
\subfigure[]{\includegraphics[width=7cm,height=6cm]{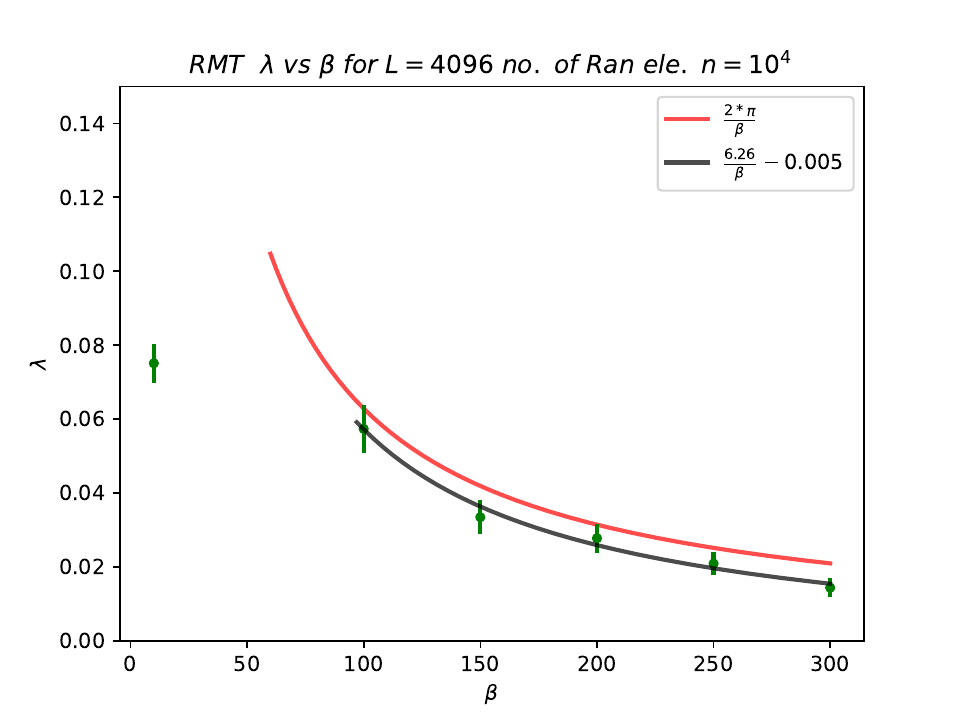}}
\caption{Fig(a) is the OTOC as a function of time $t$ for different $\beta$.  Solid curve is the  best fit at early times. Fig(b)  is the  Lyapnov coefficient as a function of $\beta$.}
\label{otoc113}
\end{figure}
\vspace{-5mm}
\noindent \underline{\bf  4) Random Elements $n=4500$}\\
\vspace{-5mm}
\begin{figure}[H]
	\centering
\hspace{-10mm}
\subfigure[]{\includegraphics[width=7cm,height=6.2cm]{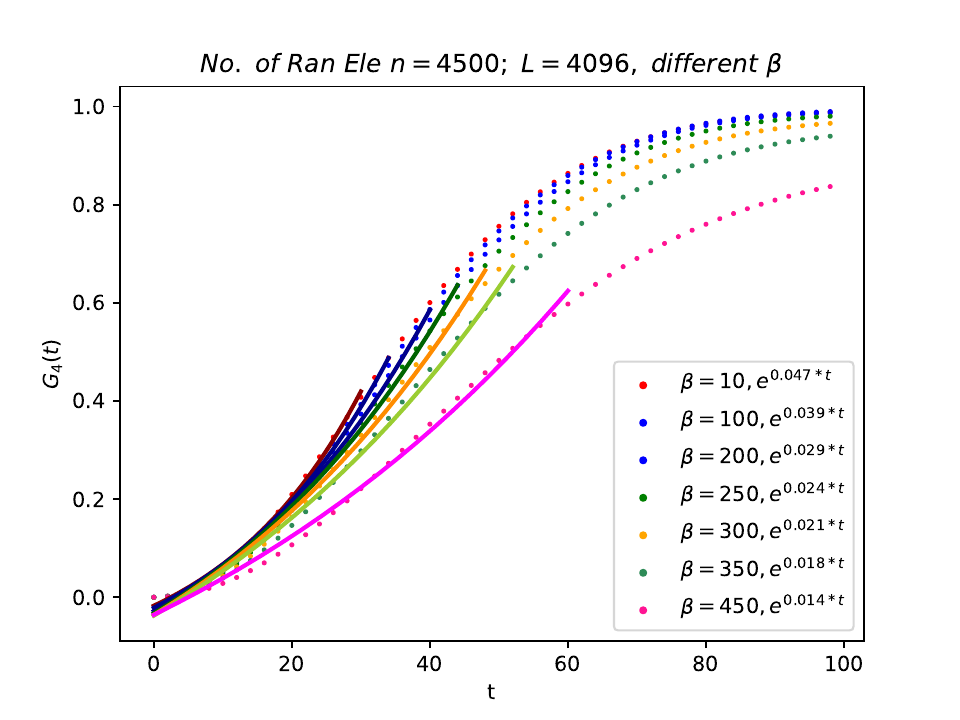}}
\subfigure[]{\includegraphics[width=7cm,height=6cm]{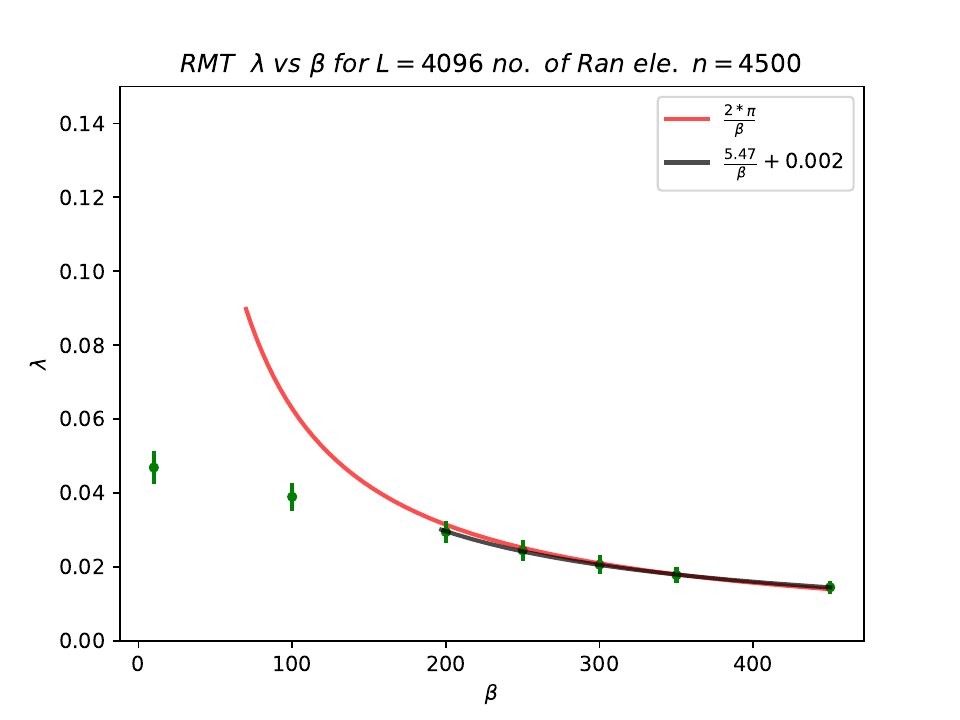}}
\caption{Fig(a) is the OTOC as a function of time $t$ for different $\beta$.  Solid curve is the  best fit at early times. Fig(b)  is the  Lyapnov coefficient as a function of $\beta$.}
\label{otoc114}
\end{figure}

Before closing this section let us make one final comment. Here we have studied the ensemble obtained by randomly choosing some fraction of matrix elements to vanish. In appendix \ref{sysspa} we discuss another   model where the non-vanishing matrix elements are chosen in a more systematic manner. We first choose an integer $q$, 
and then only  take the  matrix  elements, $H_{ij}$, for $i-j=0 \ {\rm mod} \  q$, to be   non-zero. The resulting matrix then can be block diagonalised into $q\times q$ blocks.
Each of these blocks behaves like a GUE.  E.g., the level spacing statistics of each block follows the Wigner surmise, however when we combine all the blocks the resulting level statistics becomes Poisson.

%

	\section{Local SYK model}
	\label{sykmat}
	\subsection{Local SYK model and the reduction of random couplings} 
	\label{lsykdef}
	
	In this section, we consider a variation of the SYK model where the number of random couplings is vastly reduced and only $\order(N)$. 
%
	The fermions in this model can be thought of as lying on a circle and the only couplings present are those between $q$ nearest neighbours. The  Hamiltonian is given by 
%
	\begin{align}
		\label{H_local}
		H_{\rm local-SYK} &= i^{q/2} \sum_{i_1, i_2, \cdots i_q} 
		j_{i_1i_2\cdots i_q} \psi_{i_1} \psi_{i_2}  \cdots \psi_{i_q} \,, \nonumber \\
		j_{i_1i_2\cdots i_q}&=  0  \, \quad (\mbox{except $i_1 = i_{2} + 1 =  i_{3} + 2 = \cdots = i_q + q-1$, }  \\
		& \qquad \quad  \,\,\,\,\,\, \mbox{and \, $i_1 = i_{2} - 1 =  i_{3} - 2 = \cdots = i_q - (q-1)$} ) \nonumber,
	\end{align}
	with the  fermions satisfying  periodic boundary conditions:
	\be
	\label{fermplsyk}
	\psi_{i=N+1} = \psi_{i=1}\,.
	\ee	
	To distinguish between this model and the conventional SYK model, with all to all couplings, we will refer to the model above as the {\it local SYK} model, in contrast to the  conventional SYK model  which we have often referred to  as the  non-local SYK model above.
	The couplings $j_{i_1i_2\dots i_q} $ present are taken to be random with vanishing mean and   variance 
	\begin{align}
		\langle j_{i_1i_2\cdots i_q}j_{j_1j_2\cdots j_q}\rangle =(q-1)!\hat{J}^2\delta_{i_1,j_1}\delta_{i_2,j_2}\cdots\delta_{i_q,j_q}\label{lsykvar}
	\end{align}
	We are particularly interested in the behaviour as $N\rightarrow \infty$, with ${\hat J}$ being held fixed. Here we will also keep $q$ fixed. 
	
	Notice that the variance ${\hat J}^2$ above has a different $N$ dependence  from the non-local SYK model, eq.~(\ref{varsykq}). 
	The choice above is appropriate for a local system and ensures that a good  thermodynamic limit exists with energy, entropy, etc, scaling extensively with the number of sites, $N$, when  $N\rightarrow \infty$, at fixed $T, {\hat J}$. 
	
	 Also notice that 
	 with the choice we made in eq.~(\ref{lsykvar}), we get the dispersion in energy
	\be
	\label{trh3}
	\langle \Tr H^2\rangle =L N  (q-1)!  {\hat J}^2 
	\ee
	which has the same linear $N$ dependence as in  the non-local SYK  model in the large $N$ limit 
	\be
	\label{trh}
	\langle \Tr H^2\rangle= {L N J^2\over q}\,. 
	\ee
	
	One more comment is worth making here. While the four-fermi interaction in eq.~(\ref{H_local}) does preserve a notion of locality, bilinear terms  which would typically be present in a local theory are absent. For Majorana fermions, which we are considering here, the bilinear  terms couple nearest neighbours and result in an additional  ``hopping" contribution to   $H$
	\be
	\label{hop}
	H_{hop}=  \sum_iJ_{2,i}\,\psi^i \psi^{i+1} \,.
	\ee
	More generally one can consider bilinear terms which couple next to nearest neighbours, etc, as well. 
	We will neglect all such terms below. {Study of a related  SYK model with these hopping contributions is done in \cite{Garcia-Garcia:2018pwt}. }

	
	
 We restrict ourselves to the particular case of $q=4$ in this section. We will compare and contrast the behaviour of the  local and non-local SYK model by studying various quantities such as density of states, specific heat, 2-pt Green's functions, spectral form factor and OTOCs.

 Another variant of SYK model that is studied in the literature is the model with a chain of SYKs. In such a model, at each site there is a conventional $q$-SYK model, with the different sites connected by quadratic coupling terms between nearest neighboring sites \cite{Gu:2016oyy}. This is different from the model we study here,  eq.\eqref{H_local}, with the difference being that, at each site we have a single fermion as opposed to a full $q$-SYK and the interactions between different sites being quartic between the nearest neighbour sites as opposed to being quadratic in the SYK chain case.

	Before proceeding let us mention one potentially confusing point while reading the various plots of this section. The numerical results have been obtained for finite values of $N$. { In all the plots, we take $J=1$ for  the non-local model and also take ${\hat J}=1$ for  the local model. } 
	
	\subsection{The density of states (DoS) and specific heat}
	\label{doscv}
	The unit normalised density of states for the non-local and local models are presented in fig.~\ref{spectrum1} below for a few different values of $N$. 
	We see qualitatively that the states are more clumped at the center  (near $E=0$ in the plot) in the local SYK model and the behaviour close to the two ends is  more like the tails  of a normal distribution. 
	The reader will recall  that when we discussed the non-local SYK and GUE models (fig.~\ref{finitenpic}) we saw  that compared to the GUE the spectrum of the non-local model indicates that the repulsion between eigenvalues is weaker, leading to the eigenvalues being more concentrated near the center; we see now that  this tendency is even more accentuated in the local case. 
	We have scaled the axes in fig.~\ref{spectrum1} appropriately with $N$  that ensures the width of the distribution is almost same for all values of $N$. 

%

	\begin{figure}[H]
		\begin{minipage}[b]{0.5\linewidth}
			\hspace{-13mm}
			\includegraphics[width=7.7cm,height=7cm]{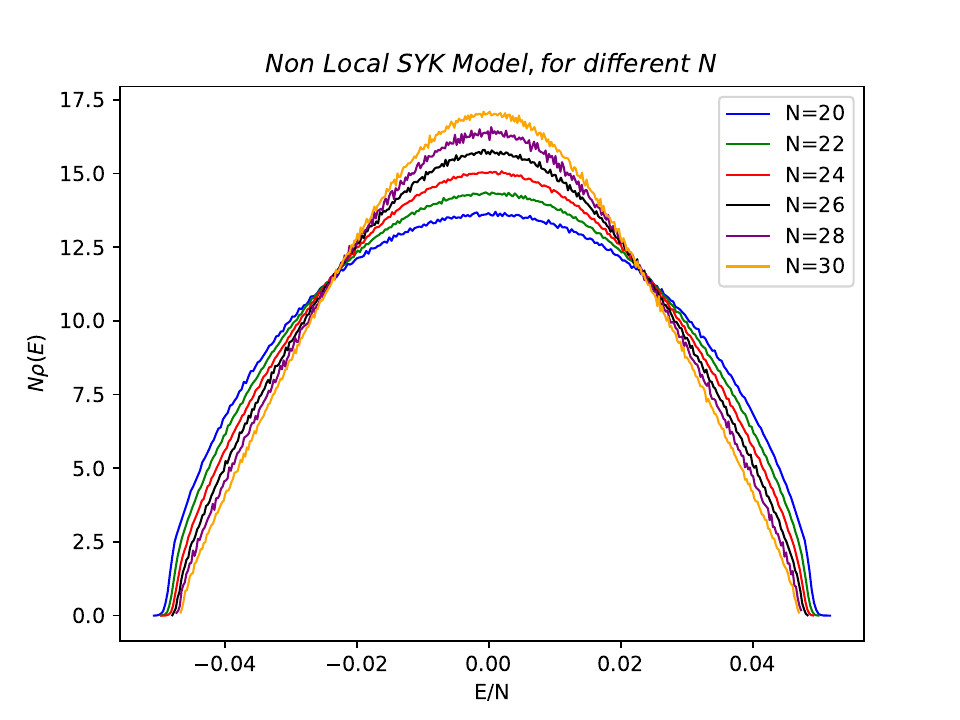}
		\end{minipage}\hspace{4mm}
		\begin{minipage}[b]{0.5\linewidth}
					\hspace{-10mm}
			\includegraphics[width=8cm,height=7cm]{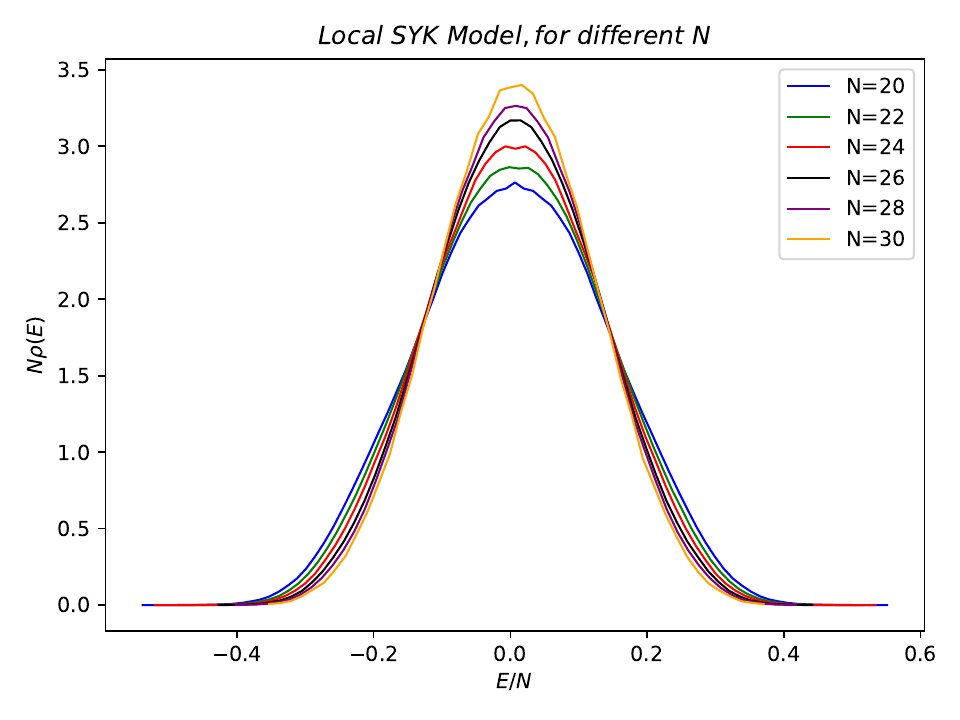}
		\end{minipage}
		\caption{The density of states for non-local SYK model (left) and local SYK model (right) for different $N$. 
			For the non-local SYK model, $J=1$, and for  $N=20, 22, 24, 26, 28$, we take $10^{4}, 4\times10^{3}, 10^{3}, 150, 21$ elements of the ensemble, respectively. 
			  For the  Local SYK model, ${\hat J} = 1$, and for  $N=20, 22, 24, 26, 28, 30$, we take $10^{4}, 4\times10^{3}, 10^{3}, 500, 100, 10$ elements of the ensemble, respectively. 
		}
		\label{spectrum1}
	\end{figure}

	The non-local SYK model has a specific heat which is linear in temperature $T$ in the low energy region, eq.~(\ref{entSYK}), eq.~(\ref{lesyk}). 
	The resulting behaviour of  $\ln(Z)$ in the non-local case also has a constant and $T$ linear term in the temperature regime, eq.~(\ref{lesyk}).
	
	The plot for entropy  $S$ vs $T$ in the non-local SYK and the GUE  was presented in fig.~\ref{entguenlsyk}. The corresponding plot for the local SYK model is in fig.~\ref{pic2}.  
	We see from fig.~\ref{pic2} that there are qualitative similarities between them,  
	 essentially due to the fact that in all cases the entropy vanishes at $T\rightarrow 0$, and increases monotonically achieving its maximum value, $S=N \ln 2$, as $T\rightarrow \infty$. An important feature for the non-local SYK is the low-temperature regime eq.~(\ref{lesyk}), where the entropy has the characteristic behaviour, eq.~(\ref{entSYK}). The presence of the constant $S_0$ and the linear $T$ behaviour are closely tied to the presence of  the  Schwarzian action. 
	 It would be very interesting to check whether the same kind of behaviour arises in the local case. Unfortunately, the numerics which have been carried out at finite $N$ do not allow us to obtain a conclusive answer and we leave this question for the future.

%

	\begin{figure}[H]
%
%
%
			\centering
			\subfigure{\includegraphics[width=7.5cm,height=5.5cm]{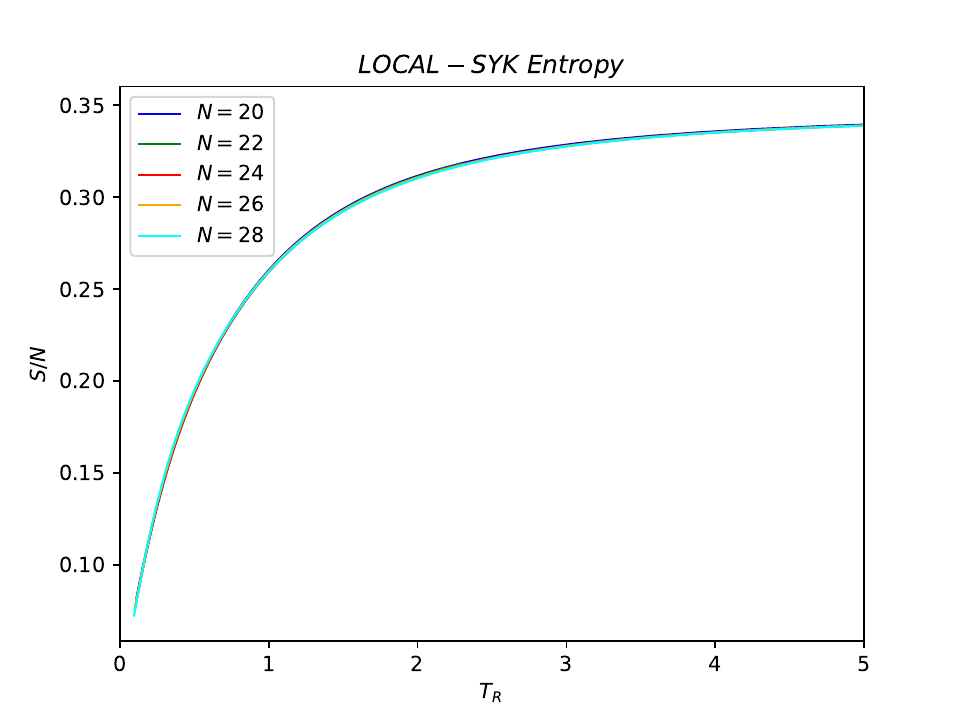}}

		\caption{ $S/N$ vs $T$ plot for Local SYK model. { Here $T_R = T \times N^{3/2}$ is the rescaled temperature.} }
		\label{pic2}
	\end{figure}
	
	Let us make one comment before proceeding. We see from fig.~\ref{pic2} that the rate of change of $S/N$ from its low-temperature value, where it is close to vanishing,  to its asymptotic value as $T\rightarrow \infty$, becomes more rapid as $N$ increases, suggesting that this change should be quite rapid in the $N\rightarrow \infty$ limit.

	\subsection{Green's function}	
	The two point function of fermions is another important quantity  and is given by 
	\begin{align}
		G_{\psi}(t)=\frac{1}{N}\sum_{i=1}^N\psi_i(t)\psi_i(0)\label{grensyk}
	\end{align}
	In the non-local SYK model, the two point function can be computed analytically in the large $N$ limit, in the low-energy regime. Due to the emergent conformal symmetry, it takes the scaling form, 
	\begin{align}
		G_{NL}(\omega)=\frac{g_0}{\sqrt{\omega}}\label{nsygf}
	\end{align}
	where $g_0$ is a constant. 
	
	The corresponding plot for the Green's function in the non-local model  are shown in fig.~\ref{imaginaryGreennonlSYK}. 	The left panel shows $\log(\Im(G(\omega)))$ against 
	$\log(\omega)$. There are three regions which can be fitted to a power law. For $\omega\ge 1$ we have a good fit, shown in green,  to the free field result 
	$\Im G(\omega)={1\over \omega}$. For  $0.22< \omega <0.65$, the red line is a  good fit to the conformal result, eq.(\ref{nsygf}). For sufficiently small $\omega$ we find the fit to eq.(\ref{nsygf}) breaks down, and at very low  $\omega < 0.1$ a new scaling regime arises shown in black, with  $\Im (G) \sim {1\over \omega^{0.16}}$. This break down and the new scaling regime are finite $N$ effects. The middle panel in fig.~\ref{imaginaryGreennonlSYK} is $\Im(G)$ vs $\omega$ showing the region $\omega>1$ in green, and the right panel shows  that the region $0.22<\omega<0.65$  is  well fitted by eq.(\ref{nsygf}).


%
	\begin{figure}[H]
		\hspace{-12mm}
\subfigure[]{\includegraphics[width=5cm,height=4.5cm]{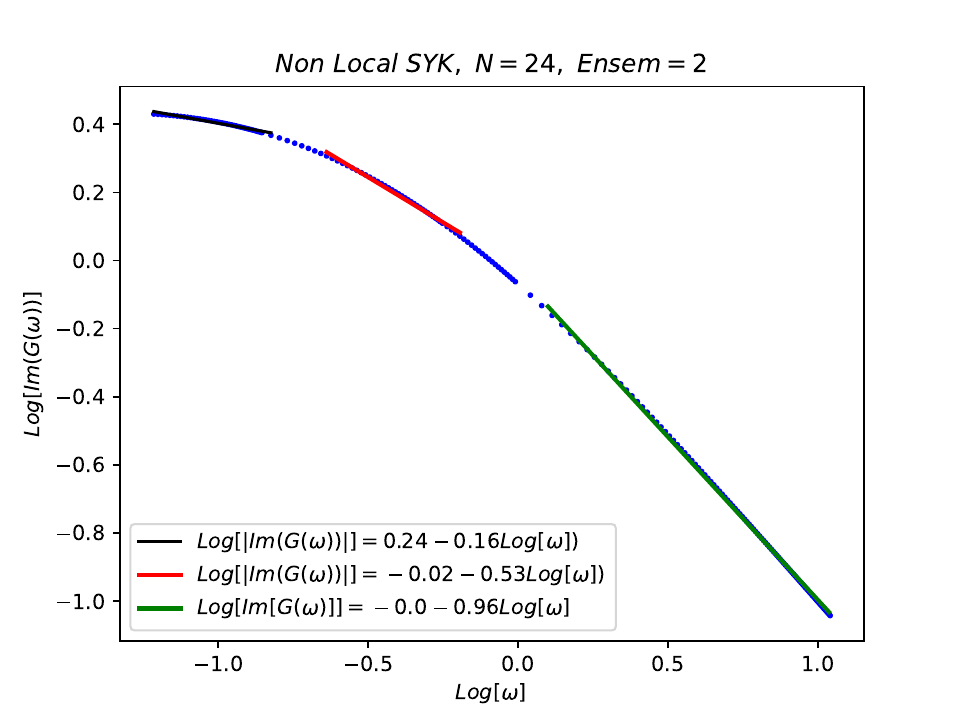}}	
\subfigure[]{\includegraphics[width=5cm,height=4.5cm]{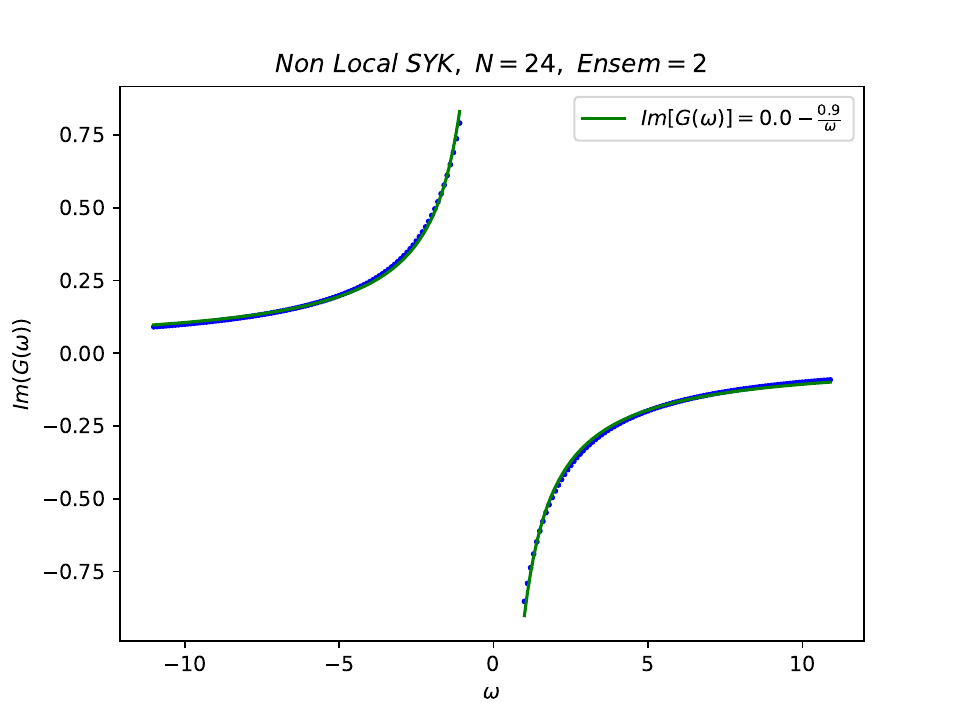}}		\hspace{-5mm}
\subfigure[]{\includegraphics[width=5cm,height=4.5cm]{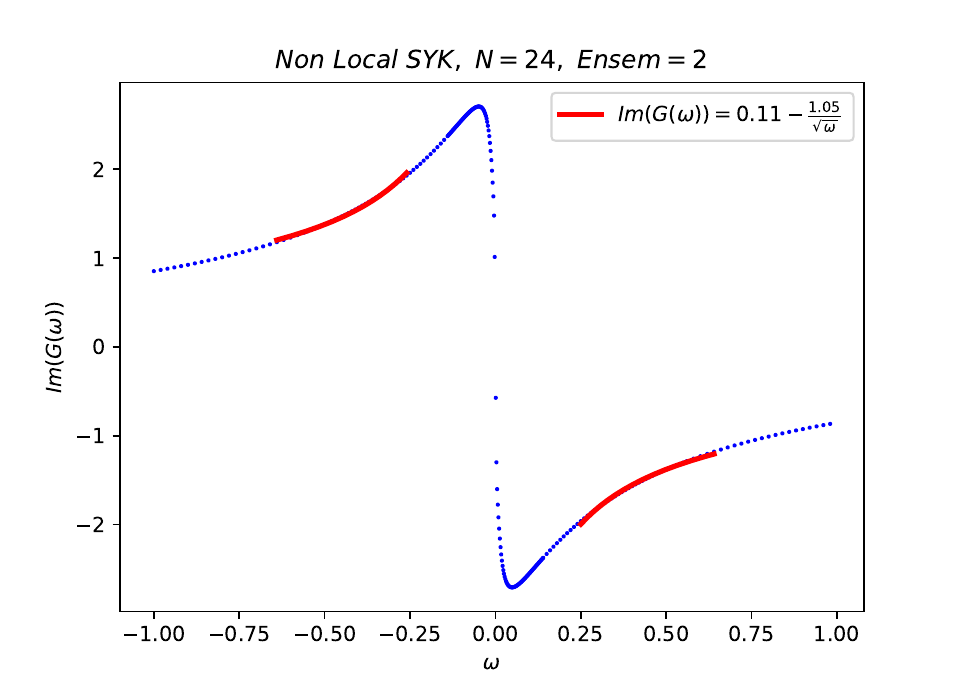}}
\caption{(a) $\log (\Im G(\omega))$ vs $\log \omega$  for $N=24$, in non-local SYK Model. (b)  $\Im(G(\omega))$ vs $\omega$ showing that  the $\omega>1$ region is well fitted by the free field result. (c)  $\Im(G(\omega))$ vs $\omega$  showing that the region $0.22< \omega< 0.65$ is well fitted by the conformal result, eq.~(\ref{nsygf}).} .
		\label{imaginaryGreennonlSYK}
	\end{figure}
                     
             The corresponding plots for the Green's function in the local model are shown in  fig.~\ref{imaginaryGreenlSYK}.	Once again there are three panels. The left panel is a plot of $\log(\Im G(\omega)) $ vs $ \log (\omega)$  showing that for $\omega>0.1$ there is a reasonably good fit with the  free field result, 
             $\Im G(\omega)={1\over \omega}$.  
             The middle and right panel show that the regions $0.1<\omega<1$ and $\omega>1$ respectively, are well fitted by the free field form. There are deviations for $\omega<0.1$. We have not been able to reliably estimate the behaviour in this region.


             \begin{figure}[H]
             		\hspace{-15mm}
\subfigure[]{\includegraphics[width=5cm,height=4.5cm]{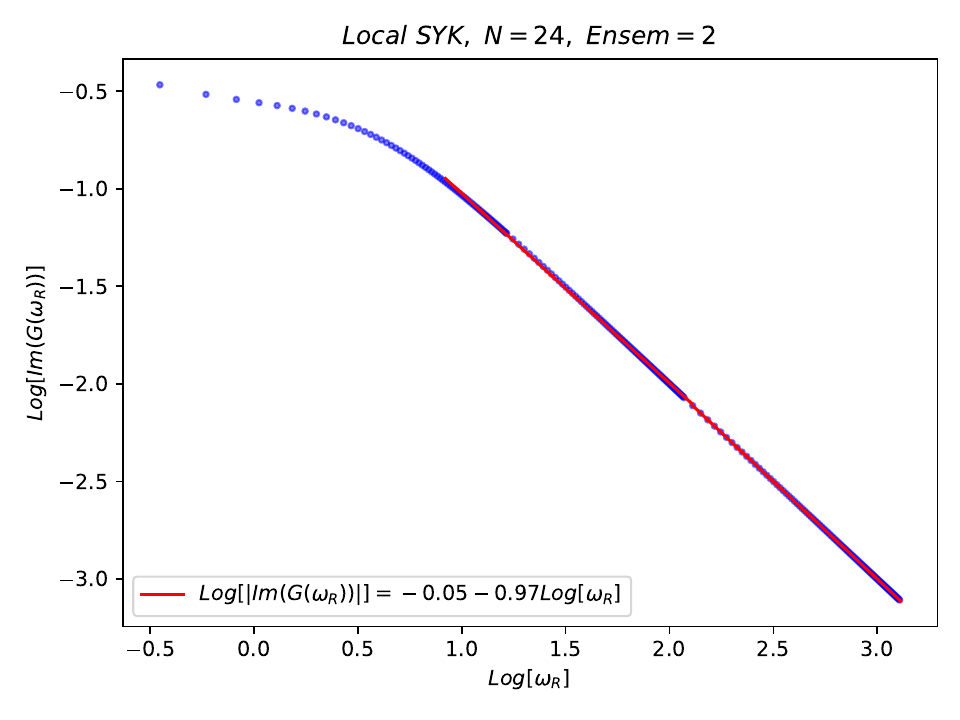}}
		\hspace{-2mm}
\subfigure[]{\includegraphics[width=5cm,height=4.5cm]{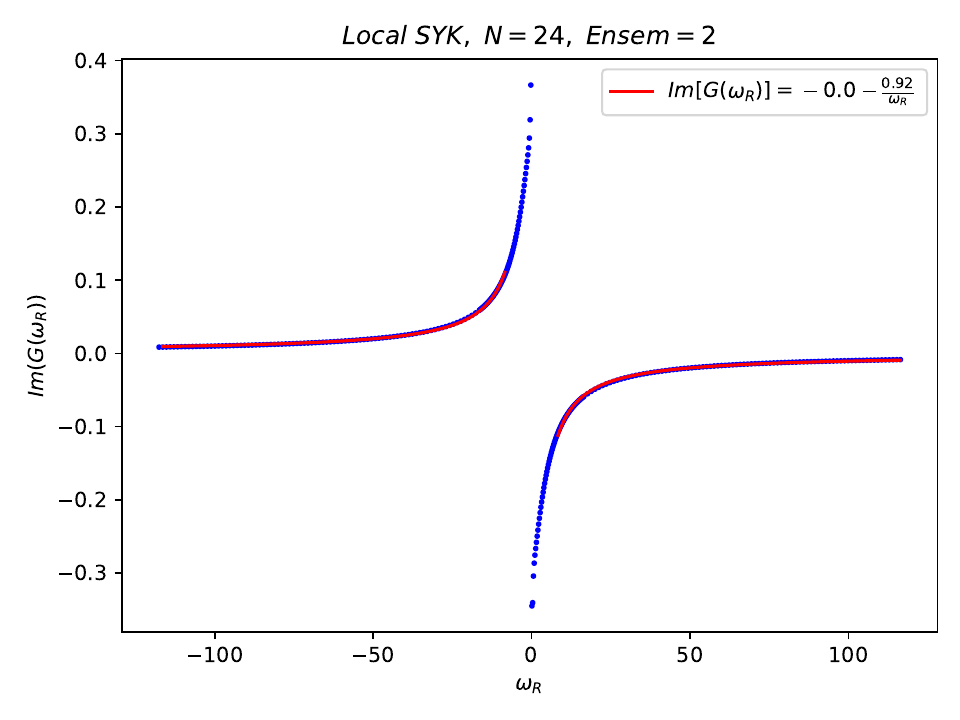}}
		\hspace{-2mm}
\subfigure[]{\includegraphics[width=5cm,height=4.5cm]{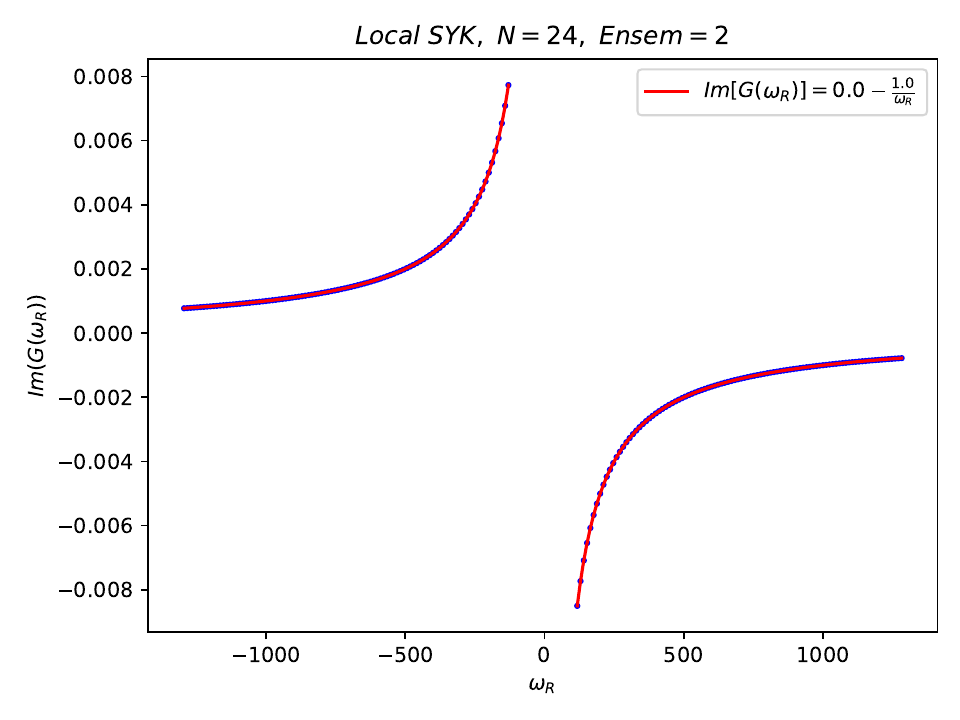}}
\caption{{
(a) $\log(\Im G(\omega_R))$ vs $\log \omega_R$  for $N=24$, in local SYK Model. 
Here we rescale $\omega$ as $\omega_R = \omega \times N^{3/2}$. 
(b) $\Im(G(\omega_R))$ vs $\omega_R$ for the region $0 <\omega_R<100$. (c) $\Im(G(\omega_R))$ vs $\omega_R$ for the region $\omega_R>100$ . 
}
}
		\label{imaginaryGreenlSYK}
	\end{figure}

	\subsection{Spectral form factor (SFF)}
	\label{slyksff}
	The SFF was discussed for the non-local SYK and GUE in subsection \ref{subsffrmtsyk}, fig. \ref{SFFRMT} and for the random sparseness model in section \ref{srmtcv}, fig. \ref{SFFran1}. 
	In the figure below we plot the disconnected, connected and total SFF for the non-local and local models. 
	
\begin{figure}[H]
\centering
\subfigure[]{\includegraphics[width=7.6cm,height=6.5cm]{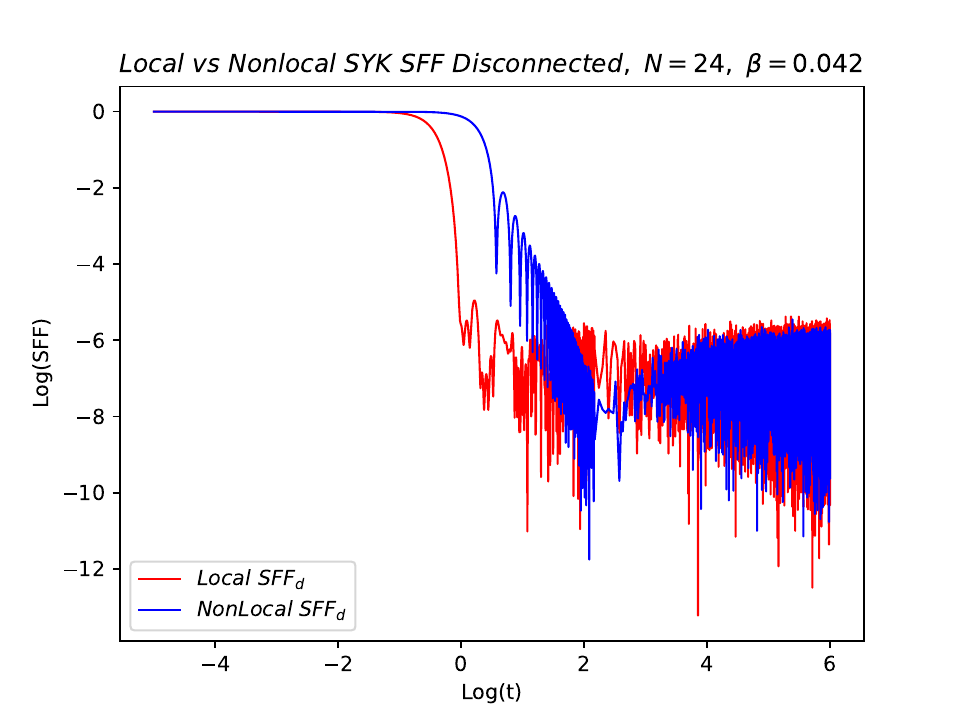}}
\subfigure[]{\includegraphics[width=7.6cm,height=6.5cm]{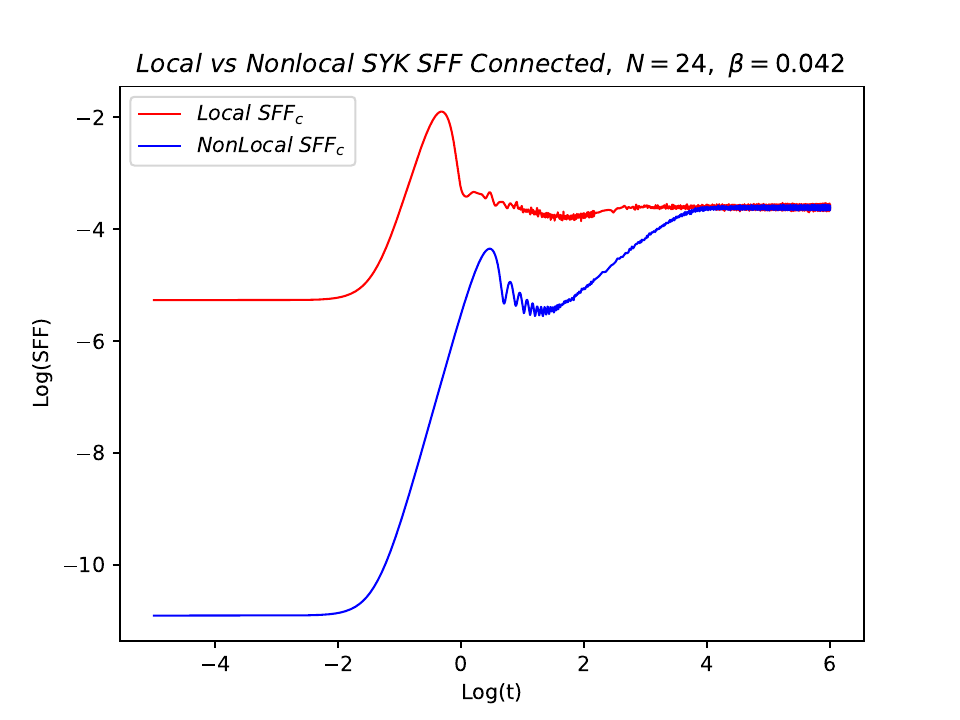}}

\subfigure[]{\includegraphics[width=7.6cm,height=6.5cm]{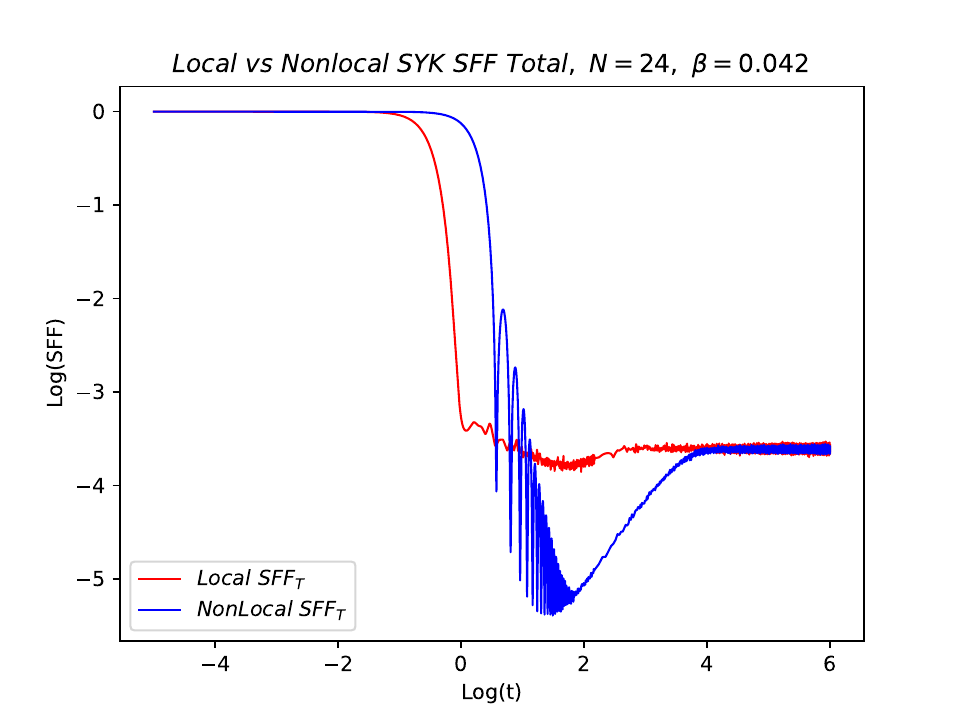}}
\caption{(a) disconnected, (b) connected, (c) total spectral form factor respectively for the local (red) and  non-local (blue) SYK models, at $N=24$, {$\beta =0.042$}, and 1000 ensembles. } 
\label{sffloc2}
\end{figure}

 The most immediate difference that catches the eye  is in the total SFF ((c) panel) where we see that the  ramp in the non-local model is almost flattened in the local SYK. 

For the case with random sparseness the reader will recall, see fig.~\ref{SFFran1} that we had seen a similar behaviour for the SFF, namely, the flattening of the ramp as we increase sparseness, in particular for $1\ll n\ll n_c$. It therefore seems that the  reduction of eigenvalue repulsion which occurs  in the random sparseness case,  is also present   here with $ \order(N)$ randomness. We had noticed in the case of sparse random matrix analysis that as the sparseness is increased, when $1\ll n\ll n_c$, the ramp in the SFF starts to disappear along with the level spacing becoming more Poisson-like, see fig.\ref{LEVEL3}. 

{
Recall that the SFF of the non-local SYK model is quite similar to that of the GUE where the ramp can be understood as a consequence of the sine kernel for the two point density correlator.
The absence of a ramp in the local SYK model indicates that the two-point density correlator is quite different for this system. 
}
One more comment worth making about fig.~\ref{sffloc2} pertains to the disconnected part ((a) panel). 
For the non-local model,  the disconnected piece falls like $1/t^3$ (at $t\gg\beta$). We see from the figure that  in the local model, the  disconnected term follows a similar fall-off. 

From fig.~\ref{sffnloc}  we also see that the local maximum in the connected SFF for the local case occurs  somewhat early, at a time when the disconnected part has still not decayed away and is comparable in magnitude to the connected one. 
As a result such a local maximum does not appear in the full SFF.


%
%

	\begin{figure}[H]
		\centering
\subfigure{\includegraphics[width=8.4cm,height=7.5cm]{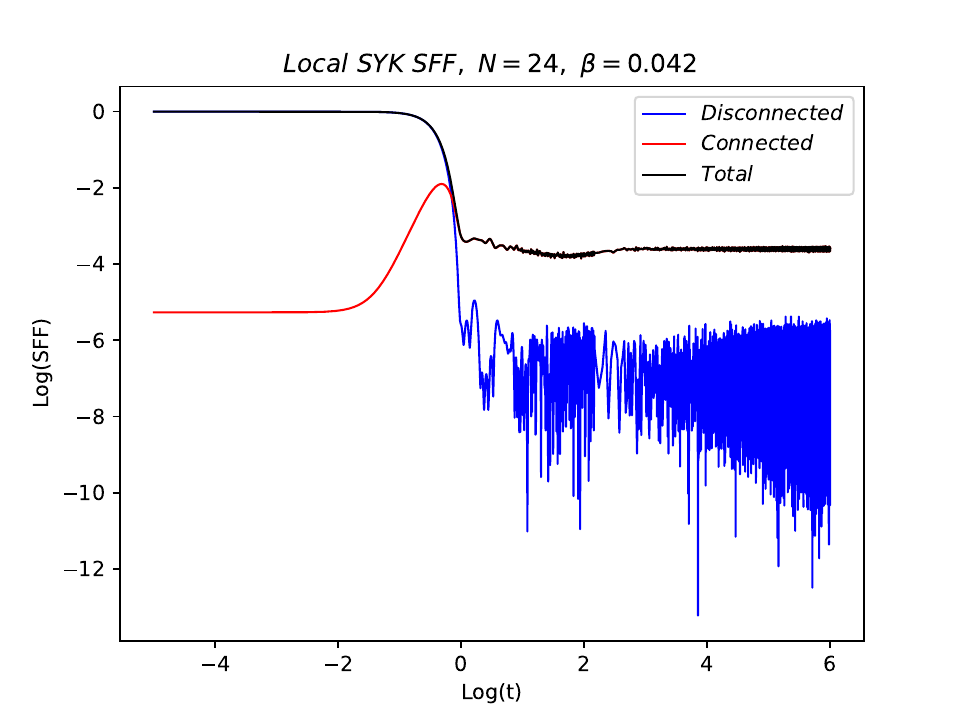}}
		\caption{Log-log plot of total (black), disconnected (blue), connected (red), SFF  for  the  local SYK model.
		 }
		\label{sffnloc}
	\end{figure}

Finally, we have also analyzed the dependence of the SFF for varying values of the  inverse temperature $\beta$, as shown in fig.~\ref{sffdiffbeta}. From the (b) panel (connected component) we see that the ramp and intermediate maximum get less and less pronounced in the local model as $\beta$ increases, and finally disappear around $\beta\simeq 70$. This is  connected to the fact that the starting value, at early times, for the connected component increases with increasing $\beta$ while, in contrast, the height of the intermediate maximum roughly stays constant. The final value of the plateau also increases with increasing $\beta$, as can also be see in the right panel.

%

\begin{figure}[H]
\subfigure[]{\includegraphics[width=7.6cm,height=6.5cm]{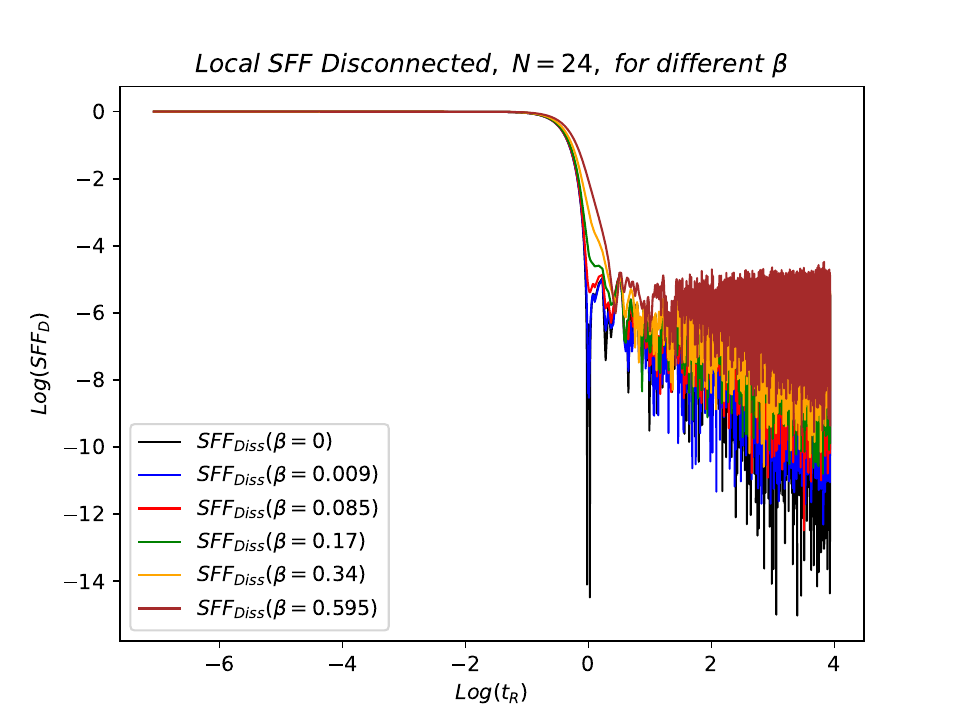}}
\subfigure[]{\includegraphics[width=7.6cm,height=6.5cm]{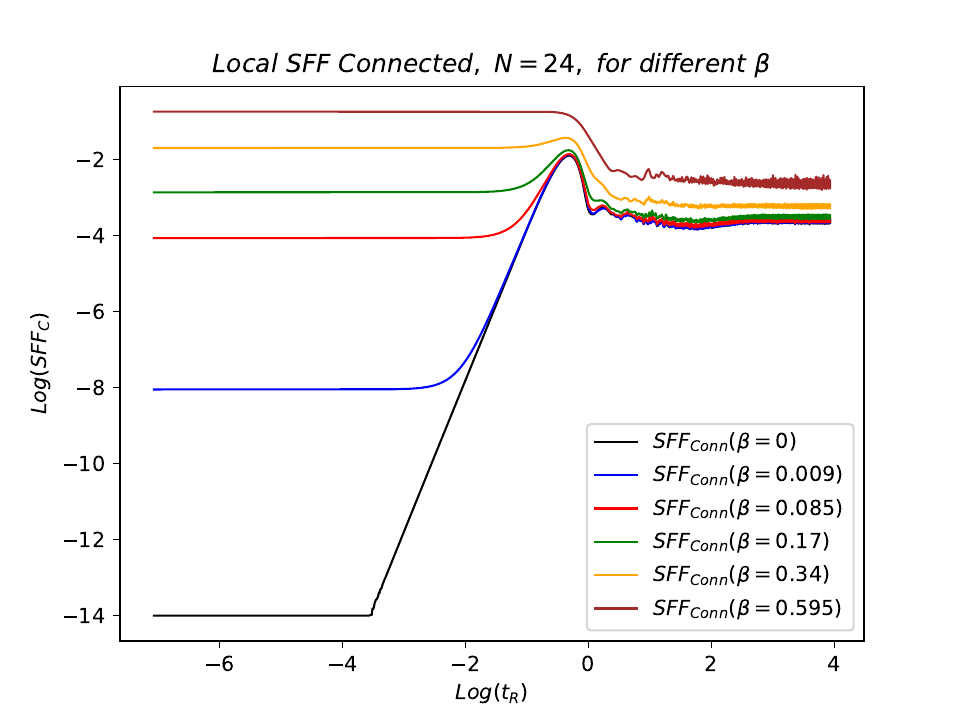}}

\centering{\subfigure[]{\includegraphics[width=7.6cm,height=6.5cm]{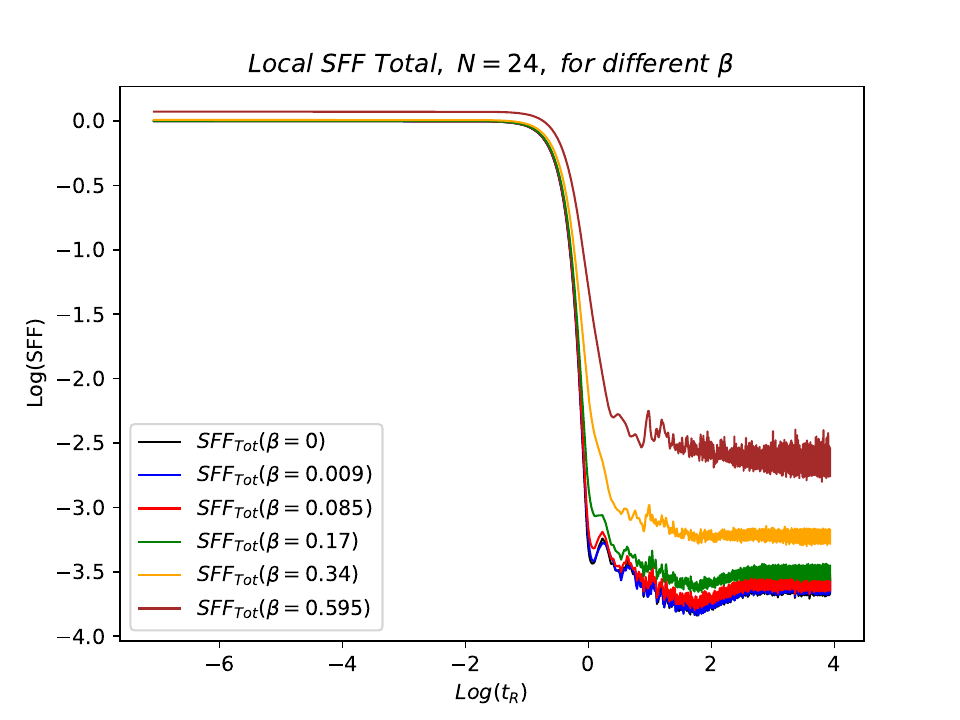}}}
\caption{(a) disconnected, (b) connected, (c) total SFF, respectively for local SYK model, at $N=24$, 600 ensembles, and varying  $\beta$. 
This $\beta$ is in units of $\hat{J}$. And we rescale the time as $t_R = t \times N^{-3/2}$. 
} 
\label{sffdiffbeta}
\end{figure}

\subsection{Level Spacing}
The nearest neighbour level spacings statistics is also a useful diagnostic of chaos. 
After unfolding the spectrum, as described in subsection \ref{srmtcv}, and taking care to separate the eigenvalues for eigenvectors of  different parity, the resulting statistics is plotted in fig.~\ref{oprtar1}. 

The left panel is the probability $P(s)$ vs the spacing $s$. The  green, blue and brown points   are the data for the GOE, non-local SYK and local SYK models respectively (the GOE is  the relevant ensemble since $N=0\,\text{mod} \,8$,\cite{You:2016ldz}). The solid curves are the best fits to this data, with the green, blue and brown  curves being the fit to the green, blue and brown  points, respectively. We see that the non-local SYK model is in good agreement with 
 the GOE RMT result, eq.~(\ref{pws}), see subsection \ref{nlrmtsyk} and \cite{Cotler:2016fpe, Garcia-Garcia:2016mno}, while the local SYK model is in good agreement with Poisson statistics,
\be
\label{psta}
P(s)\propto e^{-s}.
\ee
In the right hand panel the same nearest level statistics is presented now as a plot of the variable $r$, eq.~(\ref{rnde}), along $x$-axis,  vs $P(\ln(r))$, eq.(\ref{Plnr}). 
The GOE, non-local SYK and local SYK data is in  green, blue and orange respectively, and the corresponding curves for the GOE RMT, eq.~(\ref{pws}) and Poisson statistics, eq.~(\ref{psta}), are in green and red respectively. 
We see that again the non-local and  local SYK model are  in good agreement with  GOE and Poisson statistics respectively. 

The bottom line is that we find for the local SYK model that the nearest neighbour level statistics agrees with the Poisson distribution to very good approximation, showing, at least as far as this diagnostic is concerned,  that the model is {\it not} chaotic.  { This is also in agreement with what we have observed in the analysis of SFF  earlier in subsection \ref{slyksff}. }

\begin{figure}[H]
\begin{minipage}[b]{0.45\linewidth}
\hspace{-8mm}
\includegraphics[width=7.5cm,height=6cm]{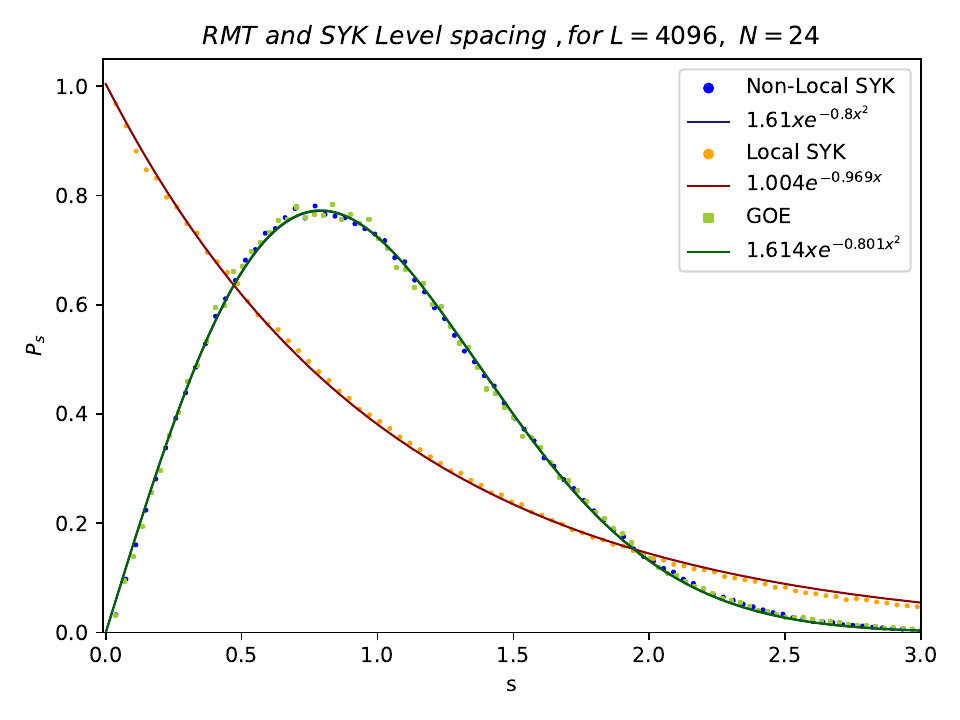}
\end{minipage}       \hspace{2mm}\vspace{-3mm}
\begin{minipage}[b]{0.49\linewidth}
\centering
\includegraphics[width=8cm,height=6.2cm]{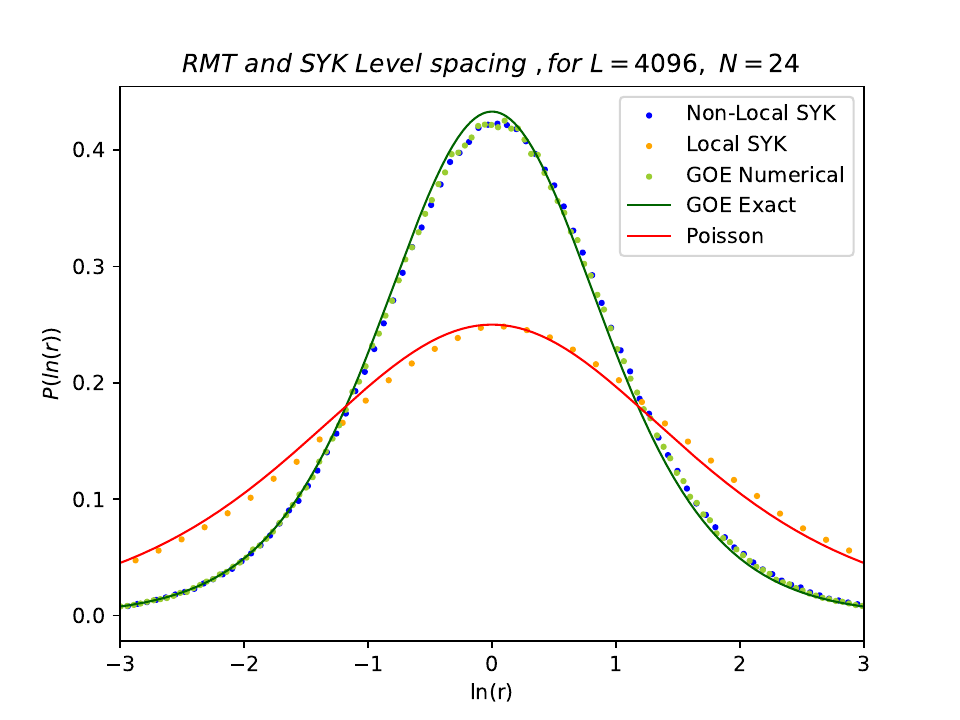}
\end{minipage}      
\caption{Nearest neighbour level spacing for the GOE, Non-local SYK Model and  Local SYK Model.}
\label{oprtar1}
\end{figure}

\subsection{Out-of-time-ordered correlators}
\label{OTOCLOCAL}
Next we turn to OTOCs.

{ As was discussed in \cite{Maldacena:2015waa,Khemani:2018sdn}} , one needs a small parameter to be  present, analogous to ${1\over N}$ in a large $N$ vector model or gauge theory, or to $\hbar$, for a  well defined notion of  scrambling time and the associated Lyapunov exponent  in OTOCs, to exist. The small parameter suppresses the connected component of the OTOC compared to the disconnected one. 
We do not have such a parameter in the local model and in its absence it is unclear  if the  OTOCs  can serve as a useful diagnostic of chaos. 
The parameter $N$ present in the local model, as was discussed at the beginning of this section,  should be thought of as labelling a site with one fermion being present per site. The $N\rightarrow \infty$ limit in the local model is therefore an infinite volume limit which  is different from the vector model or gauge theory cases where in this limit the  degrees  of freedom per site diverge.

Despite this  observation it is worth examining the behaviour of the OTOCs in the local model and comparing it to the non-local case. An important limitation of our analysis below is that we will be working numerically at finite $N$  and our conclusions  will mostly be of  a  qualitative nature. 

Let us also note the following. Drawing on the discussion about quantum chaos in higher dimensional local systems \cite{Khemani:2018sdn,Han:2018bqy}, even in systems without a small parameter of the type mentioned above,  one can often   define a velocity dependent Lyapunov exponent $\lambda(v)$ which vanishes at the butterfly velocity $v_B$. 
For $v=|x|/t$ where $|x| $ is the spatial distance, with $v > v_B$, 
the OTOC $C(x,t)$, defined for an local operator $O(t)$ at $x=0$ and another local operator $W_x$ at $x$, takes the form 
\begin{align}
C(x,t) &  = \frac{1}{2} \braket{ \left[ O_0(t)  \,, W_x \right]^\dagger \left[ O_0(t)  \,, W_x \right]   }     \sim   e^{\lambda(v)  t }  \\
\lambda(v) & =  -c(v-v_B)^\alpha 
\label{deflone}
\end{align}
with $\lambda(v)$ being the velocity dependent Lyapunov exponent. 
When there is a large $N$ (or small $\hbar$) parameter which suppresses the connected component, $\alpha$ is unity and the behaviour of $C(x,t)$ also continues to be of the same form when $v<v_B$, resulting in exponential growth of the OTOC. 
However, in the absence of  such a parameter suppressing the connected component, often  $\alpha$ is  bigger than unity  and the growth of $C(x,t)$ does not continue for $v<v_B$. 
 
The local SYK model we are studying is not a conventional local system since there is no quadratic hopping term in the Hamiltonian, eq.~(\ref{hop}). 
 Even so, it would be interesting to ask whether a butterfly velocity exists in the system and the  behaviour of the OTOC close to it. We will not be able to address this question in a definitive manner here and leave it for the future. 
 

For the non-local model the OTOCs were discussed in section  \ref{rmtotoc}.
By averaging over flavours we defined the function $G_4$, eq.~(\ref{gtilfdef}) whose behaviour is plotted in figure \ref{nlsykotoc}, obtaining, at $N=24$, good agreement for the Lyapunov exponent with the $N\rightarrow \infty$ result. 

The plots for local SYK are shown in fig.~\ref{lotoc1}, \ref{lotocp}. 
In obtaining the plots for the OTOC, we have considered 
\be
\label{defgija}
G_{ij}=1-{\tilde F}_{ij}
\ee
where ${\tilde F}_{ij}$ is given in eq.~(\ref{defgij}),
then averaged the value of $G_{ij}$ for all possible pairs $(i,j)$ keeping the distance $p= |j-i|$ fixed.   This averaged correlation is denoted as ${\bar G}_{i,i+p}$ below.
  Fig.~\ref{lotoc1} (a) is for $p=1$, and varying $\beta$, fig.~\ref{lotoc1} (b) is for $p=2$, and varying $\beta$ and fig.~\ref{lotoc1} (b) is for various $p$ and two different values of $\beta$. 

We see that the OTOCs  in general grow rapidly before  saturating to a maximum value, in a manner, qualitatively speaking,  which is  similar to the non-local SYK model and also the GUE RMT, section \ref{rmtotoc}.
The rate of growth is fastest for ${\bar G}_{i,i+1}$, {\it i.e.},  the nearest neighbour sites and then slows down as $p$ increases. It is worth noting in this context  that for $p=4,5$ there are no terms in the Hamiltonian which directly connect the two sites, and this might account for some of the slow down. 
Also for $p=1,2$ the rate of growth slows down as $\beta$ increases, this is similar to what happens for the non-local SYK model too. 

In conclusion, qualitatively the OTOCs in the local case are similar to the non-local and GUE cases.  A more  systematic analysis of the OTOCs in the local model, especially in the large $N$ limit,  is left for the future. 
It will be especially interesting to study whether  a butterfly velocity can be defined in the system, and whether there is evidence for exponential growth of the OTOCs inside the light cone, $v<v_B$. Some further discussion on the OTOCs in this model can be found in appendix \ref{ExtraLyapunov}.

\begin{figure}[H]
\hspace{-10mm}
\subfigure[Local SYK, $\bar{G}_{i,i+1}$ vs t]{\includegraphics[width=8.4cm,height=6.5cm]{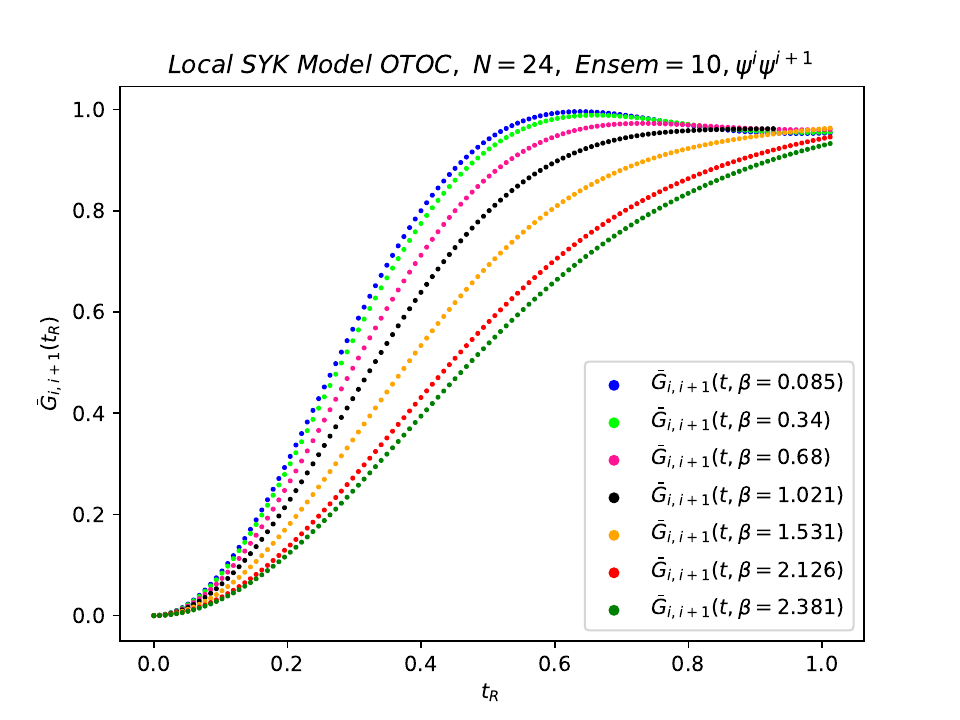}}
\hspace{-9mm}
\subfigure[Local SYK, $\bar{G}_{i,i+2}$ vs t]{\includegraphics[width=8.4cm,height=6.5cm]{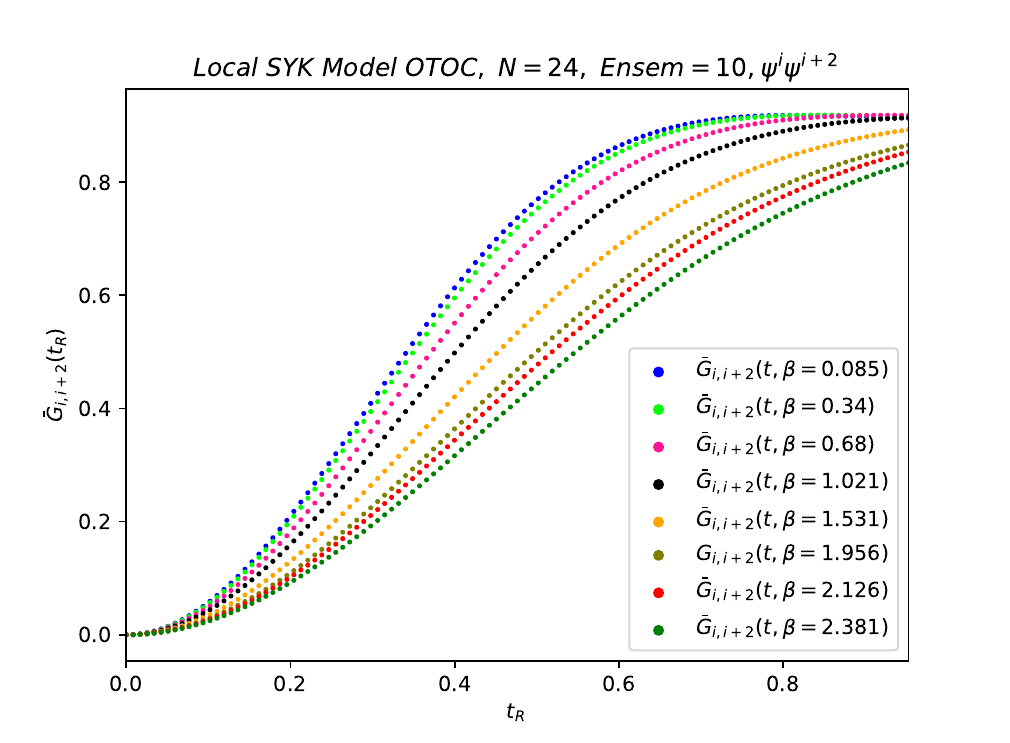}}
\caption{Local SYK model ${\bar G}_{i, i+1},~\&~\bar{G}_{i,i+2}$ as a function of time, for varying values of $\beta$.  $\beta$ is in $\hat{J} = 1$ unit, and we rescale $t$ as $t_R = t \times N^{-3/2}$.}
\label{lotoc1}
\end{figure}

\begin{figure}[H]
\begin{minipage}[b]{0.45\linewidth}
\hspace{-12mm}
\includegraphics[width=8.0cm,height=6.5cm]{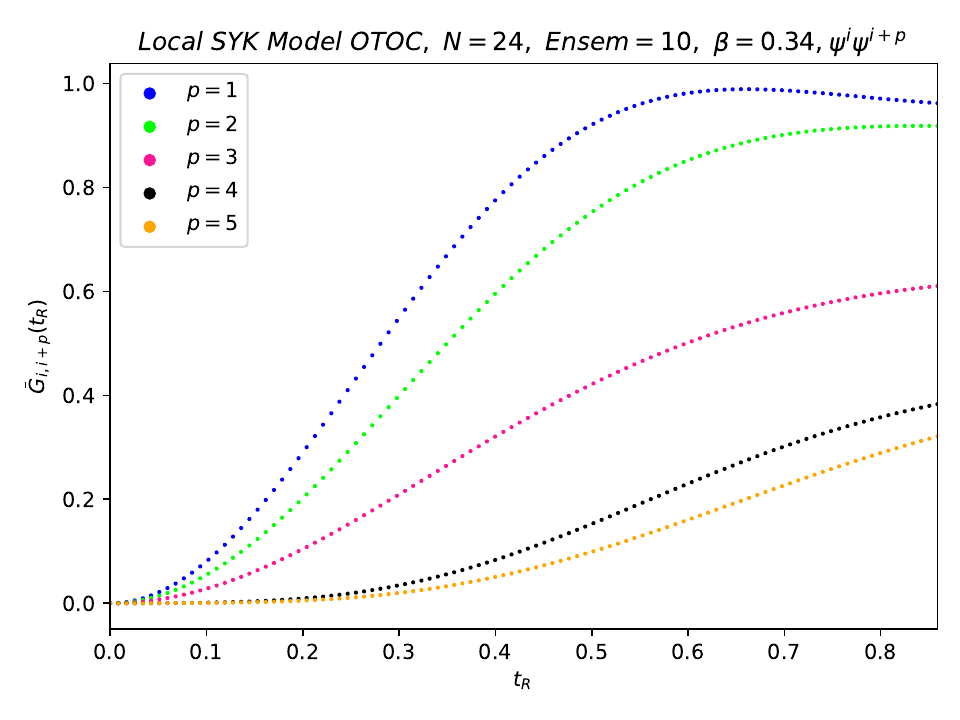}
\end{minipage}            \hspace{3mm}
\begin{minipage}[b]{0.45\linewidth}
\includegraphics[width=8.0cm,height=6.5cm]{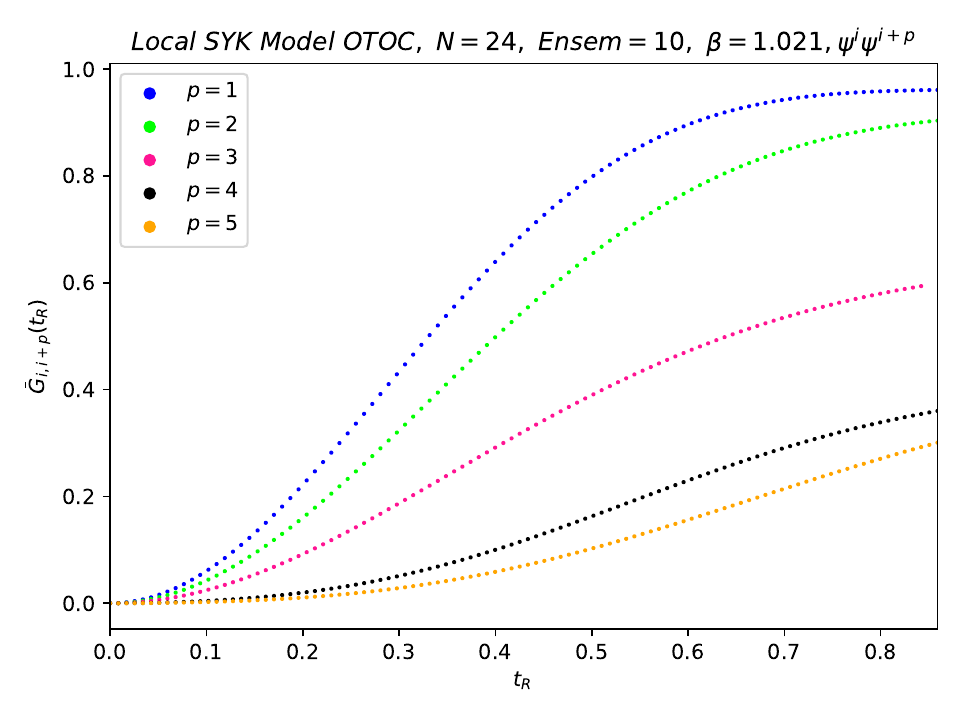}
\end{minipage}            \hspace{2mm}
\caption{Local SYK  ${\bar G}_{i, i+p}$ as a function of time for different values of $p$. Left panel  $\beta=0.34$, Right panel $\beta=1.021$. 
 $\beta$ is in $\hat{J} = 1$ unit, and we rescale $t$ as $t_R = t \times N^{-3/2}$.}
\label{lotocp}
\end{figure}

\subsection{Operator Growth in SYK Models}

The growth of operators from simple to complex ones is also a way to characterise the scrambling properties of a system, \cite{Roberts:2018mnp,Susskind:2020gnl}.

In the non-local SYK model it was suggested that the number of fermionic terms which appear in an operator can be a measure of its complexity.
Thus a single fermion $\psi_i$ is a simple operator whereas another having a product of many fermions would be complex. Staring with the single fermion $\psi_i$ at $t=0$ its time evolution is given by 
\be
\label{growo}
\psi_{i}(t)=e^{i{H}t}\psi_{i}(0)e^{-i{H}t}
\ee
Products of fermions of the type $\psi_{a_1}, \psi_{a_2} \cdots \psi_{a_s}$ for a complete basis in the space of operators and one can therefore expand 
$\psi_i(t)$ in this basis, 
\be
\psi_{i}(t)=\sum_{s,a_{1}<\cdots<a_{s}}2^{\frac{s-1}{2}}c^{\{i\}}_{a_{1}\cdots a_{s}}(t) \psi_{a_{1}}\psi_{a_{2}}\cdots \psi_{a_{s}}
\ee
The probability of having an operator of size $s$ for a particular combination can then be computed by finding all the $c^{\{i\}}_{a_{1}\cdots a_{s}}(t)$.
\newline
In the Hilbert space of operators we can define an inner product 
\be
\langle A, B\rangle=\frac{1}{2^{N/2}}\Tr(A^{\dagger}B)
\ee
and this leads to
\be
c^{\{i\}}_{a_{1}\cdots a_{s}}(t)=2^{\frac{1+s}{2}}\langle\psi_{a_{1}}\cdots\psi_{a_{s}},\psi_{i}(t)\rangle
\ee
The resulting  probability distribution of having operators of size s is then
\bea
P_{s}(t)=\sum_{a_{1}<\cdots<a_{s}}|c^{\{i\}}_{a_{1}\cdots a_{s}}(t)|^{2} \\
P_{s}(t)=\sum_{\mathcal{O}}\frac{|\langle\mathcal{O},\psi_{1}(t)\rangle|^{2}}{\langle \mathcal{O},\mathcal{O}\rangle \langle \psi_{1},\psi_{1}\rangle} 
\eea
In the equation above the  operator $\mathcal{O}$ denotes all the $N \choose s$ operators of size s; $\psi_{a_{1}}\psi_{a_{2}}\cdots \psi_{a_{s}}$, with condition $a_{1}<a_{2}<\cdots <a_{s}$.

Starting with the operator $\psi_1$ at $t=0$ (setting $i=1$ entails no loss of generality), we can find $P_s(t)$ numerically. 
Fig. \ref{oprt1} and \ref{oprt2} are plots obtained for the non-local and local SYK models for $s=1,3,5$.  
We see that in both cases the operator growth is qualitatively similar. $P_1(t)$ decays with time. In the non-local case,   $P_s(t)$ increases, reach a maximum and then decreases, with $P_3(t)$ peaking before $P_5(t)$. In the local case,  the behaviour is similar, but there is a longer tail characterising the slow decrease of $P_5(t)$.
\begin{figure}[H]
\begin{minipage}[b]{0.9\linewidth}
\centering
\includegraphics[width=10cm,height=7cm]{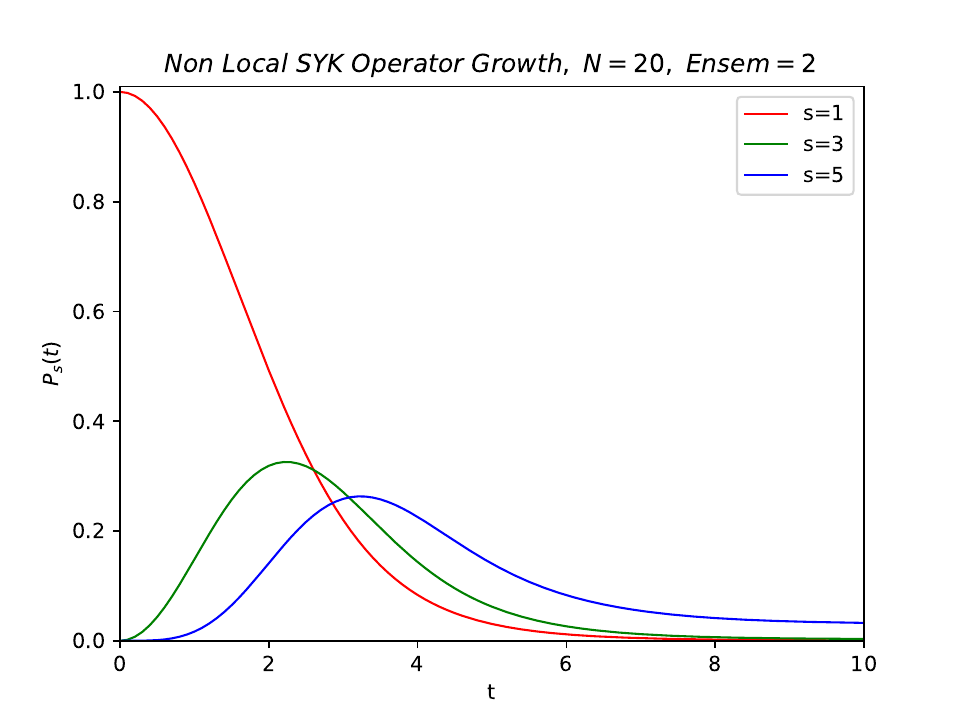}
\end{minipage}      
\caption{Non-local SYK Model:  operator growth as a function of time for different size of operators, $s$.}
\label{oprt1}
\end{figure}
%
%



%
\begin{figure}[H]
\begin{minipage}[b]{0.9\linewidth}
\centering
\includegraphics[width=10cm,height=7cm]{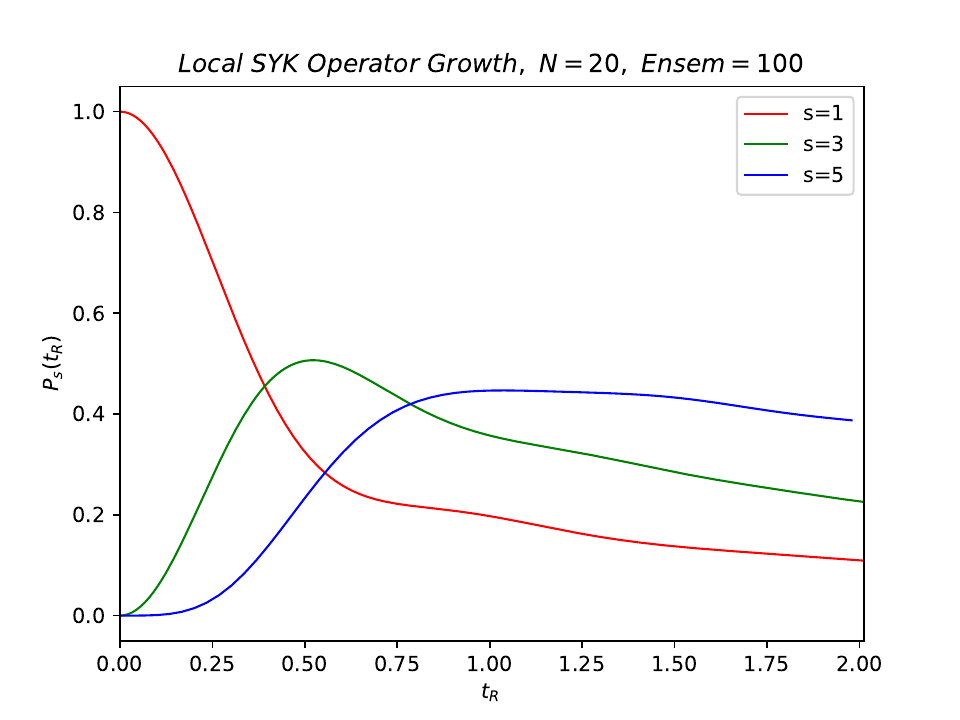}
\end{minipage}      
\caption{Local SYK Model: operator growth as a function of time for different  sizes of operators, $s$.  Here we rescale $t$ as $t_R = t \times N^{-3/2}$.}
\label{oprt2}
\end{figure}

A quantitative understanding of this growth of operators, in the local case for $N\rightarrow \infty$, is left for the future. 


%
	 
	\section{Gaussian Ensembles with Varying Randomness}
	\label{rmtassyk}
	As was discussed in the introduction, the SYK model, like the GUE,  can be thought of as a random Hamiltonian with  matrix elements being drawn from a suitable Gaussian distribution. We explored some other random systems above where the random Hamiltonians were obtained by choosing other Gaussian distributions, with interesting differences in their behaviour. Here, we will make some additional comment 
	on  Gaussian ensembles. 
	
	Before proceeding let us note that 
	the low-energy dynamics of the non-local SYK model,  which is JT gravity, is known to also arise in Random Matrix theory by choosing a non-Gaussian potential $V(H)$
	and considering a suitable double scaling limit \cite{Saad:2019lba}. We  restrict ourselves to commenting on Gaussian ensembles below, since this class of random systems is already quite rich.

	\subsection{General Observations}
	
	
	An Hermitian matrix $H$  can be thought of as  being  a vector in an $L^2$  dimensional Hilbert space  ${\cal H}$ of $L\times L$ Hermitian matrices.   $H$ can be expanded in any basis of ${\cal H}$, 
	\be
	\label{basis}
	H=\sum_{a=1}^{L^2} c_aT^a,
	\ee
	where $T^a$ are Hermitian matrices which are the basis elements,  and  $c_a$ are real coefficients. 
An  inner product in this space is given by the trace, 
	\be
	\label{inprod}
	\langle T^a|T^b\rangle=\Tr(T^a T^b).
	\ee
	
	By drawing the coefficients $c_a$ from various Gaussian distributions one can get different types of Gaussian ensembles. 
	Under a change of basis the  $T^a$'s, and  $H$, transform by conjugation,
\be
	\label{ultrans}
	T^a\rightarrow U T^a U^\dagger,  \quad 
	H \rightarrow U H U^\dagger,
	\ee
	where $U$ is the 	
	 $L\times L$ unitary  matrix  specifying the change of basis. 
	
A few different bases for ${\cal H}$ -  are :
	\begin{itemize}
	
	\item A standard basis corresponding to the root vectors of $U(L)$   \cite{Georgi:1999wka}. 
	
	\item For $L=2^N$ :  a basis made out of tensor products of the Pauli matrices. More precisely, the
	 basis of matrices 	
	$\sigma^{\alpha_1}\otimes \sigma^{\alpha_2} \cdots \otimes \sigma^{\alpha_N}$
	where the indices $\alpha_1, \alpha_2\cdots \alpha_N$ each take $4$ values, and $\sigma^\alpha, \alpha=1,2,3,4,$ are four $2\times 2$ matrices, given by the Pauli matrices,   $\sigma^i, i=1,2,3$, with  $\sigma^4=I $, where  $I$  is the identity. 
	
	\item
	When $N$ is a multiple of $2$ and $L=2^{N\over 2}$  : a basis made out of products of $N$ Majorana fermions, $\psi^i, i=1, \cdots N$ described in eq.~(\ref{psiaspaul}), eq.~(\ref{sumh}) and  eq.~(\ref{defalphaq}). 
	
	\end{itemize}
	Each of these  bases is useful in their own way, since they  allow us to use      different types of physical intuition in the study of randomness. The first basis
	connects to Random matrix theory, the second to  the study of spin systems, and the third to  the study of the SYK model. 
	
	 An important property of the GUE is that the resulting probability distribution is $U(L)$ symmetric. 
	Many  key properties of the GUE, including the fact that OTOCs and operators grow in an hyper-fast manner, as discussed in section \ref{revsec}, are tied to this $U(L)$ invariance. For other random ensembles  the $U(L)$ symmetry will be broken in general.


%
%
%
%
	
As mentioned in the discussion of OTOCs in  \ref{srmtcv} and also in  appendix \ref{symsparuni}, this symmetry is broken for the sparse random matrices studied in section \ref{sprmt}. 
This is true for  the  non-local SYK model, and the breaking of the $U(L)$ symmetry is in part  responsible for interesting differences in behaviour with the GUE.
	It is also true for the local SYK model studied in section \ref{sykmat}, where we keep only $\order(N)$ random variables.

	We  can also consider other ways of breaking the $U(L)$ symmetry, these could  also yield  interesting properties from the point of view of  thermodynamics or chaos. For example we can take $H$ to be block diagonal with two blocks of size $M\times M$ and $(L-M)\times (L-M)$ each, breaking $U(L)\rightarrow U(M)\times U(L-M)$. Or more generally consider replacing the exponential term, eq.(\ref{GUE}),  by 
	\begin{align}
	{\Tr H^2 \over 2 \sigma^2}\rightarrow \sum_{I,J=1}^M {|H_{IJ}|^2 \over 2 \sigma_M^2} + 
	\sum_{\alpha, \beta=M+1}^L {|H_{\alpha \beta} |^2\over 2 \sigma_L^2 }+
	\sum_{I=1}^M \sum_{\alpha=M+1}^L { | H_{I\alpha}|^2\over  \sigma_{ML}^2} 
	\end{align}
	If $\sigma_{ML}=0$, the off diagonal terms in the matrix integral  go to zero and the distributions breaks up into two GUEs of rank $M, L-M$ respectively. But when  $\sigma_{ML} \ll \sigma_M, \sigma_L$, the system behaves like two GUEs with a weak coupling between them. E.g., the density of  states is  only perturbed slightly by the off-diagonal terms, which are ``very massive", { where mass $m \sim \order\left( {1/\sigma_{ML} } \right)$}.  This perturbation can have interest effects for chaos  related properties though. The off-diagonal terms could alter the nearest-neighbour spacings which are of $\order(1/L)$. And for operator growth, one expects that a   simple operator   acting in the $M\times M$ block will at first scramble  in an hyperfast manner into a matrix filling out the $M\times M$ block,  but its subsequent growth   into  a full  $L\times L$ matrix  would be much slower and governed by the off-diagonal terms\footnote{ 
	To be more precise, in making the above  statements we have in mind   the limit where $L\rightarrow \infty$, with $M/L$ being held fixed.}.  By changing the strength of the  off-diagonal term compared to the diagonal ones, once can in this way explore  a range of behaviours. 
	
	Other random ensembles are also worth studying. The GUE eq.~(\ref{GUE}), eq.~(\ref{sigrmt}), corresponds to  choosing the coefficients $c_a$ (for unit normalised $T^a$'s) to be independent  Gaussian random variables with the probability distribution
	\be
	\label{pdfa}
	P(c_a)= \sqrt{L \over 2 \pi} e^{-{Lc_a^2 \over 2}},
	\ee
	{\it i.e.}, with zero mean and variance, $\sigma={1\over \sqrt{L}}$, independent of $a$.
	More generally we can consider distributions where there is a non-zero mean and both the mean and the variance
	vary for different generators. In this case eq.(\ref{pdfa}) is replaced by 
	\be
	\label{pdfb}
	P(c_a)= \sqrt{1 \over 2 \pi \sigma_a^2} e^{(c_a-{\bar c}_a)^2 \over 2 \sigma_a^2}
	\ee
	resulting in   a two-point correlator
	\be
	\label{tpf}
	\langle  c_ac_b\rangle = (\sigma_a^2 + {\bar c}_a^2)\delta_{ab}
	\ee
	Taking $\sigma_a \rightarrow 0$, we will  have  a  highly ordered matrix, where the dispersion about the mean ${\bar c}_a$ is small.  As ${\sigma_a\over {\bar c}_a}$  increases the amount of disorder increases, and in  this manner one can study the change in behaviour from the ordered to disordered situations.

	We will report on the behaviour of some of these systems in a subsequent study \cite{Arka2023}.
	
\subsection{More on SYK Models}
	Let us now turn to the SYK model in the context of the wider discussion above. 
	

	
	As was noted in the introduction,  eq.~(\ref{sumh}), eq.~(\ref{dall}), the GUE can be obtained by summing over all  $\psi^q$ terms with equal variance. 
			 The non-local SYK model  corresponds to only keeping  the terms of the  $\psi^q$ kind, {\it i.e.},  involving products of $q$ number of $\psi_i$ fields. The large $N$ saddle point arises when the coefficients of these terms  are drawn from a Gaussian distribution with variance given in eq.~(\ref{varsykq}). The fixed $q$ model only preserves an $SO(N)$ subgroup of   the $U(L)$ symmetry mentioned above.
			 
			 In \cite{Xu:2020shn}    a version of the  fixed $q$ model was studied where the number of couplings  was reduced  in a random manner. This model is reviewed in appendix \ref{quadbosnmodel}. In the large $N$ limit the fixed $q$ model, to begin, has $\order(N^q)$ couplings. 
			 As long as the number of random couplings after reduction  scale  like $N^\alpha, \alpha>1$, a saddle point analogous to the fixed $q$ case exists in the large $N$ limit, for a suitably chosen variance for the non-zero couplings. In fact, this is also true when $q$ does not take a fixed value, as long as it grows with $N$ as a power 	 $q\sim N^\beta$,  with $\beta<{1\over 2}$. 
			 
			 This  sparse SYK model, \cite{Xu:2020shn}, is analogous to the sparse random matrix model studied in section \ref{sprmt}. 
			 If we think of the GUE, as was mentioned above, by summing over all   $\psi^q$ terms, then at large $N$ the maximum number of terms  are contributed by operators with $q=N/2$. 
			 { In fact, the number of random variables for $q=N/2$ is 
			 \begin{align}
			 _NC_{N/2} \sim \sqrt{\frac{2}{\pi N}} \, \, 2^N \,, 
			 \end{align}
			 which is exponential to $N$. 
			 Thus, instead of starting with the GUE and making the matrix $H$ sparse, we could at large $N$,} only take terms with  $q={N\over 2}$,  and consider the effects of making the randomness sparser by setting some terms to vanish in a random fashion\footnote{We thank Subir Sachdev for emphasising some of these points to us.}. 
			 The difference with the \cite{Xu:2020shn} case is that the value of $q={N\over 2}$ now scales  linearly with $N$. 
			 As we saw in section \ref{sprmt} as the number of terms $n$ is reduced an interesting change in behaviour occurs in this case when $n\sim n_c$, with ${n_c\over L^2}$, scaling with $L$ as given in eq.~(\ref{apnc}). It would be interesting to try and understand this transition using saddle point methods but we have unfortunately not been able to do so\footnote{It could be that a saddle point arises only when the variance of the non-zero couplings also scales with $N$ suitably, we did not consider this possibility in section \ref{sprmt}.}. 
			 
			 The saddle point referred to above, obtained by starting with the fixed $q$ model and keeping $\order(N^\alpha) $ terms is analogous to the one obtained for the  fixed $q$ case (where all terms are kept). In particular, the saddle point continues to break time reparametrisation invariance giving rise to the Schwarzian mode and  the low-energy behaviour is a CFT with the fermionic fields having the anaomalous dimension $1/q$. 
			 
			 In this example we then see that  most of the essential features  found in the non-local  SYK model - tied to the realisation of the time reparametrisation symmetry and the Schwarzian action are quite robust and continue to occur even for much smaller number of random couplings and are independent of the underlying Hamiltonian. Some of the more detailed features though depend on the underlying model. For example, we can have two different models with the same number of couplings $\order(N^\alpha)$, obtained by starting with two different values of $q$, $q_1,q_2$, and reducing  the number of couplings to different extents. The anamolous dimensions of the  $\psi_i$ fields in the low energy CFT will then be different, ${1\over q_1}$ and ${1\over q_2}$ respectively. 
			 
			 Let us also comment on symmetries. 
			 When we keep only even $q$ terms in the sum to obtain $H$ there is a fermion symmetry, $\psi\rightarrow -\psi$ symmetry, which is preserved. This symmetry is broken  once odd $q$ terms are included. Also, among   even $q$ terms if we only keep $q=0 \,(\,\text{mod}\,\, 4)$ terms then $H$ has time reversal invariance - as can be seen from the fact that the couplings are 
			 real. Terms of the same symmetry type can be transformed into one another under $U(L)$ transformations. 
			 Nevertheless and quite interestingly, as mentioned in the previous paragraph, if we take terms for two different powers $q_1$, $q_2$ both of which preserve the same class of symmetries and also take the same number of couplings in both cases, the low energy limit has some differences in the resulting anomalous dimensions of the fermionic fields. 
			 
			 
			 We can now comment on the figures plotted below.
			 Fig.~\ref{RMTvsWigner}~(a) plots the density of states for the GUE, the $\psi^q$ model where all even powers of $q$ are included with the same variance, and the  model where all $\psi^q$ terms with odd powers and the same variance are included. 
			 Fig.~\ref{RMTvsWigner}~(b) included  the three models again, but with a rescaling carried out along the $x$ (energy axis) so that the extent along the $x$ axis of the three curves coincides. The  $y$ axis is also  correspondingly rescaled  to keep the area the same under each curve. We see that after this rescaling, the 
			 density of states for the models obtained by keeping only even or odd terms agree with the GUE.  The density of states for the model obtained by keeping all $q$ terms agrees with the GUE even before rescaling, as  is to be expected.

			 
			 \begin{figure}[H]
			 \hspace{-10mm}
\subfigure[]{\includegraphics[width=7.5cm,height=6cm]{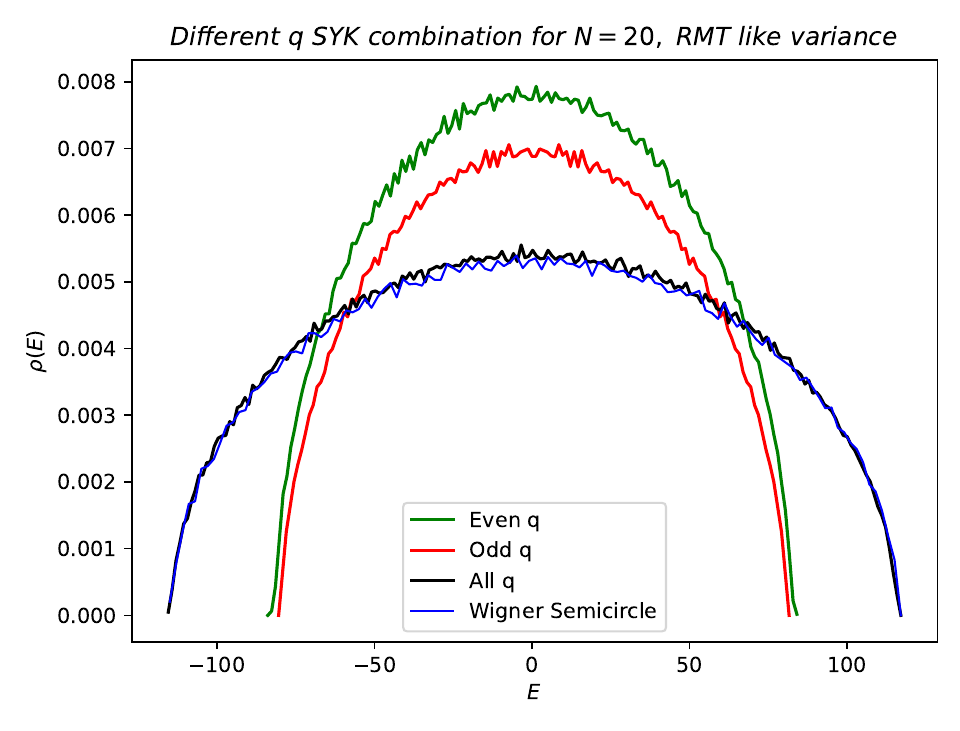}}
			 \hspace{-2mm}
\subfigure[]{\includegraphics[width=7.5cm,height=6cm]{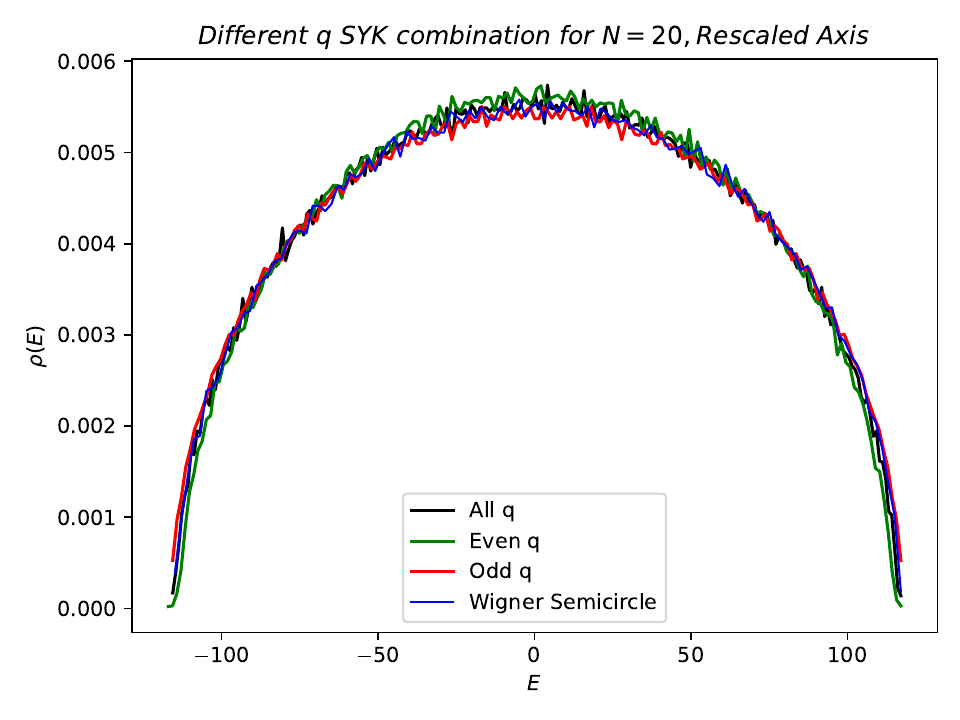}}
\caption{(a) A comparison between models with different $\psi^q$ terms included. (b) After rescaling, all  the different cases matches quite well. }
		\label{RMTvsWigner}
	\end{figure}

	In fig.~\ref{RMTransparsing}  working with the $N=20$ case, we show the density of states  when only some terms of the  $\psi^q$ type, with $q=0$ $({\rm  mod} \,\,4)$, are kept in $H$.
	From our discussion of symmetries above we see that in this case all the terms we retain preserve the fermion and time reversal symmetries.   
		The left panel of the figure shows the density of states for various values  of $n$ - the number of terms which have been kept, with the black curve showing density of states, $\rho$, for the SYK $q=4$ model. The right hand panel is the same data after rescaling the energy  $E$ and $\rho$ axis, so that all the curves coincide in their extent along the $x$ axis,  as  discussed above. In addition, in the right panel we have included 	the RMT  (all $q$)  case and also the case where all $q=0$ $\,({\rm  mod} \,\,4)$ terms are included - these are shown as  yellow and red curves respectively.   We see after the rescaling, $\rho$ for  the GUE and the model where all  $q=0$ $\,({\rm mod}\,\, 4)$ terms are kept,   are in good agreement. As the total number of terms in $H$  is decreased $\rho$ begins to change.
		The black curve in the right panel  is also for the $q=4$ SYK model (where all $q=4$ terms are included) - this model has  a total of  ${}_NC_4= 4845$  number of terms. Qualitatively, the right panel   shows that for intermediate values of  $n$, $\rho$ near the center, around $E=0$,   agrees more with the $q=4$ case.
		Fig.~\ref{RMTransparsingb}  shows a similar plot, with rescaled energy and $\rho$ axes, for $N=24$. The yellow curve is for the GUE and the black curve is  for the $q=4$ SYK.
		We see in this plot more clearly that  at   intermediate values of $n$,  for $E$ close to the edges, $\rho$  differs from both the GUE and  the SYK$_4$
		 theories, while for $E$ close to the center, $\rho$ lies in between the GUE and SYK$_4$ cases. 
		 It could be that some of these differences are due to finite $N$ effects. 
		
		\begin{figure}[H]
\hspace{-10mm} \hspace{-5mm}
\subfigure[]{\includegraphics[width=8.0cm,height=6.5cm]{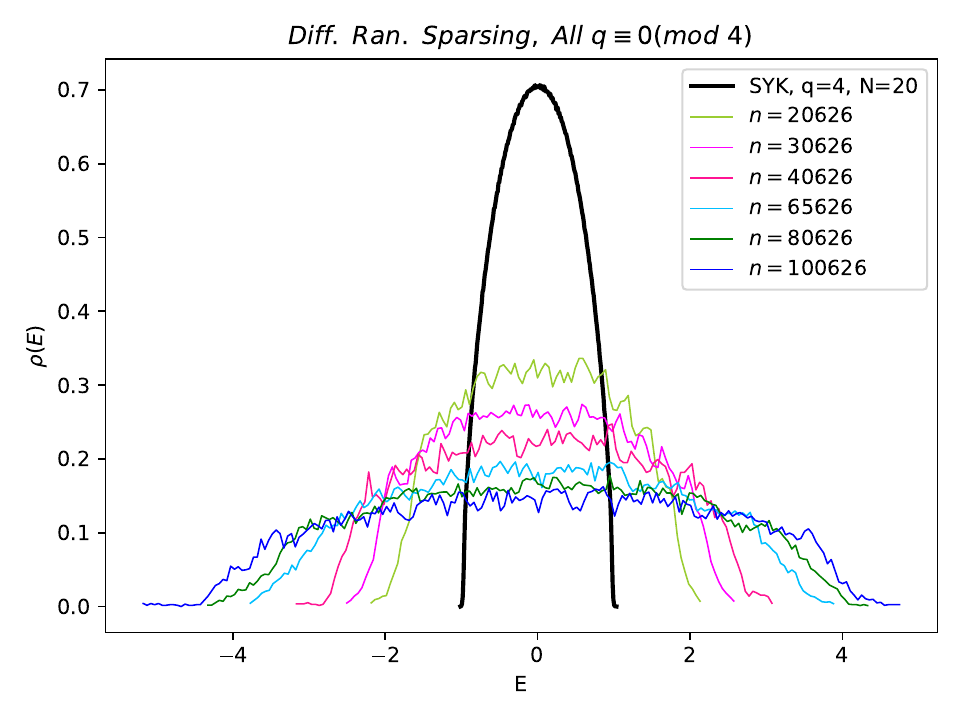}}
\, \hspace{-5mm}
\subfigure[]{\includegraphics[width=8.0cm,height=7cm]{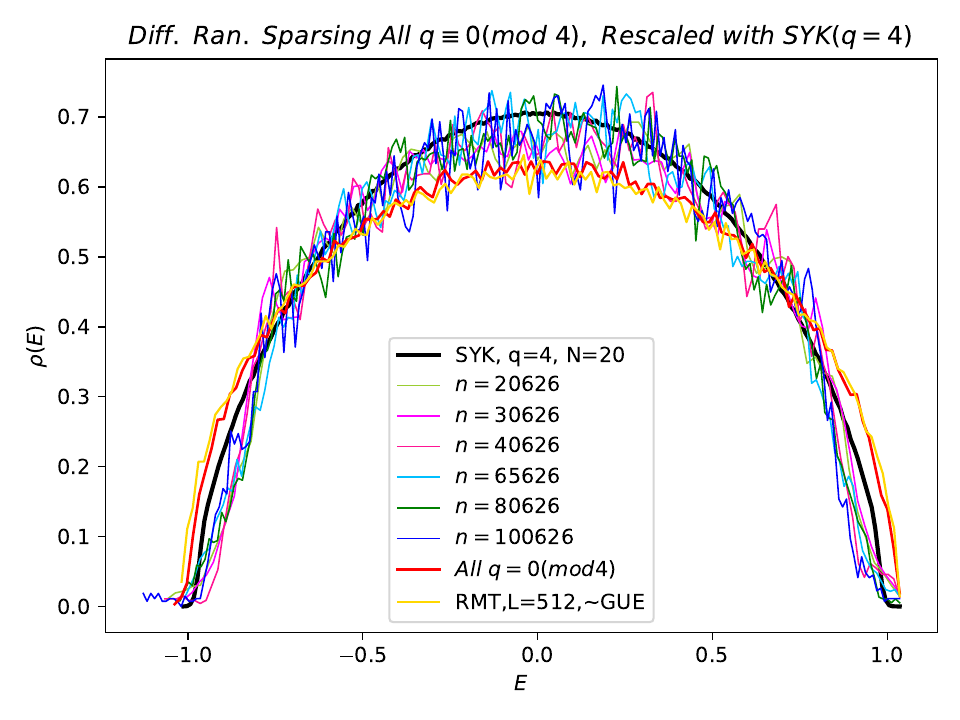}}
\caption{Density of states for models with $N=20$ and varying number of terms, $n$,  in $H$. All terms included are  of $\psi^q$ type with $q=0$\,\,(mod 4).}
		\label{RMTransparsing}
	\end{figure}

\begin{figure}[H]
\centering
\subfigure{	
	\includegraphics[width=8.5cm,height=7cm]{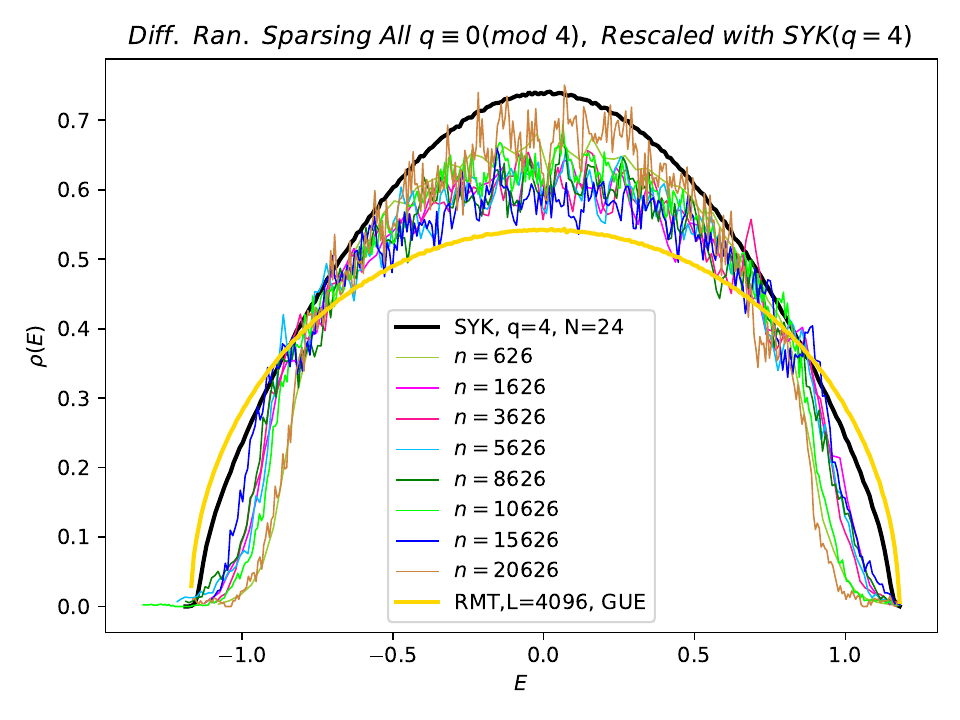}}
	\caption{Density of states for models with $N=24$ and differing number of terms, $n$,  in $H$. All terms included are  of $\psi^q$ type with $q=0$\,\,(mod 4).}
		\label{RMTransparsingb}
	\end{figure}

			In all the figures above the variance for all the non-zero couplings was taken to be  the same. One can also examine what happens if we keep terms whose couplings are normal distributed with different values for the variance.  
			An example of this is shown in Fig.~\ref{SYKq4q6transition} where we have kept all terms of $\psi^4$ and $\psi^6$ type, for $N=28$. 
			The $\psi^4$, $\psi^6$  terms are drawn from  distributions, eq.~(\ref{varsykq}), with $q=4, 6$ respectively, and we  have set $J=1$.
			The full Hamiltonian is taken to be 
			\be
			\label{fH}
			H_{\text{full}}= \lambda H_{\text{SYK}_4}+ (1-\lambda) H_{\text{SYK}_6}
			\ee
			The resulting density of states is plotted in fig.\ref{SYKq4q6transition} for various values of  $\lambda$, { where $\lambda$ is an interpolating parameter.}
			
			Let us also note that   a large $N$ saddle point in terms of the $G$ and $\Sigma$ fields, \cite{Maldacena:2016hyu}, exists in this case. However solving for the 
			values of the $G, \Sigma$ fields analytically, at the saddle point,  is difficult in general. When $\lambda$ is small it is easy to see that the $\psi^4$ coupling is relevant in the conformal theory to which the $\psi^6$ model flows, thus at very low energies the behaviour will be governed by the $\psi^4$ fixed point, while at intermediate energies $E\ll J$ one will have the Schwarzian mode coupled to the $\psi^4$ perturbation. As $\lambda$ is increased one expects that the model will flow from the UV more directly to the CFT governed by the $\psi^4$ coupling. 
			
			The left panel of Fig.~\ref{SYKq4q6transition} is a plot of  the density of states for varying values of $\lambda$ and  the right hand panel is a plot where the density of states is plotted after rescaling the $E$ axis and also $\rho$, as discussed above. We see from both plots that the density of states interpolates smoothly from the SYK$_4$ to the SYK$_6$ case as  $\lambda$ goes from $1$ to $0$. It is also worth noting that in the right hand panel, once the rescaling is carried out, the density of states at the edges  ({\it i.e.},  at  low and high energies), for all values of $\lambda\ne 0$, agrees with the 
			SYK$_4$ model. This connects with our comment above that the low-energy behaviour should be of the SYK$_4$ type.

%


%
%
%
	\begin{figure}[H]
	\hspace{-10mm}
\subfigure[]{\includegraphics[width=8cm,height=6.5cm]{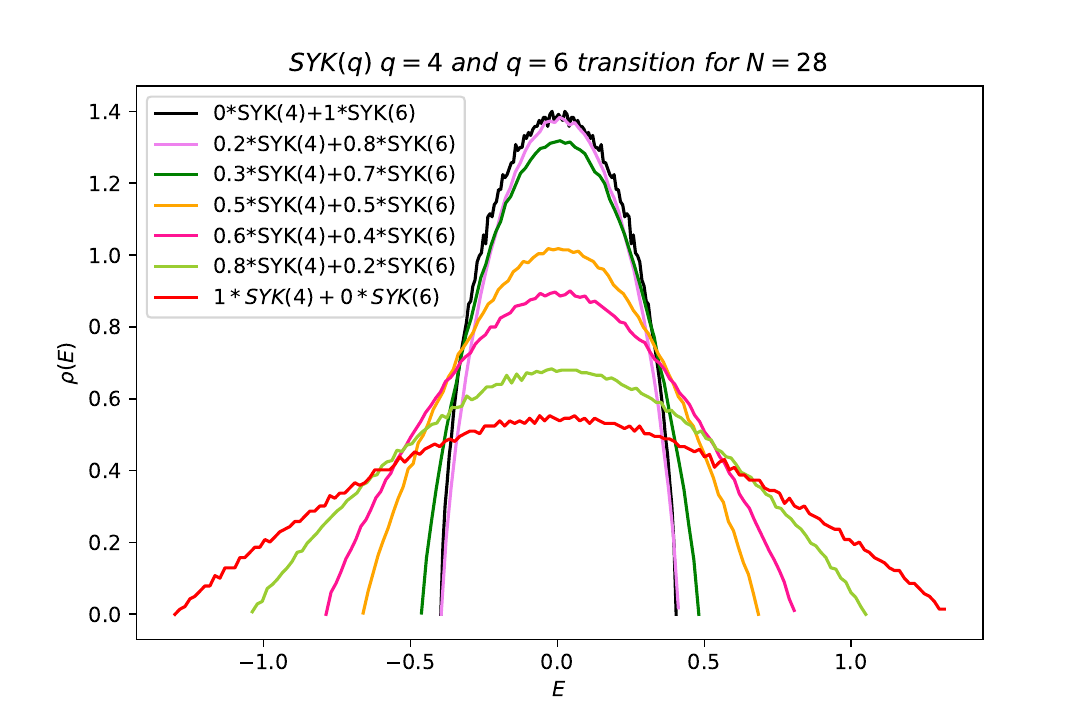}}
\subfigure[]{\includegraphics[width=8cm,height=6.5cm]{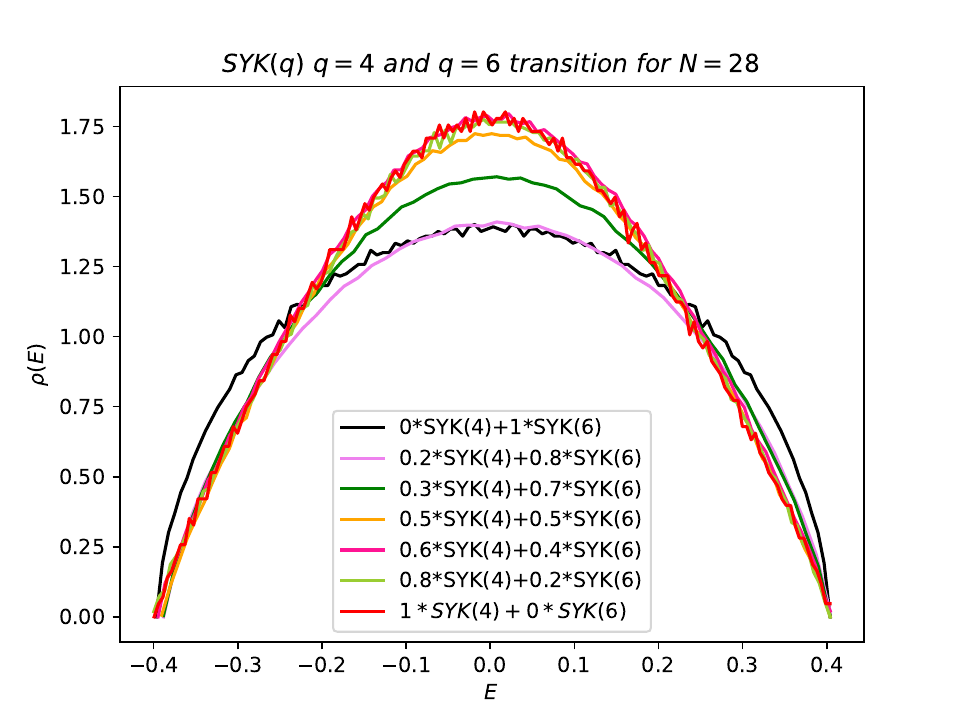}}
\caption{(a) Density of states with varying $\lambda$, eq.(\ref{fH}),  for $N=28$ SYK model (b) Density of states with varying $\lambda$, after rescaling the axis appropiately. }
		\label{SYKq4q6transition}
	\end{figure}
	
	
	The discussion in this section is  indicative of the rich set of behaviours one can get for various Gaussian ensembles, with varying degrees of randomness. 
	A more definitive analysis is clearly called for, especially in the large $N$ limit,  and we hope to return to this in subsequent work, as was also mentioned above. 

\section{ Summary and Discussion}
\label{sec:Discussion}

The conventional SYK model, with $\psi^q$ couplings, is referred to as the non-local SYK model in this paper, due to the  all-to-all nature of its fermionic couplings. 
This model and the Gaussian Unitary Ensemble (GUE) are both random systems, with  some similarities. Their spectral form factors are  similar and the level statistics, of nearest neighbour energy levels, in both follow the same  Wigner distribution. However, there are some important difference as well. In the large $N$ limit, the non-local SYK model has interesting  low-energy behaviour characterised by the breaking of   time reparametrisation invariance which  gives rise  to a Schwarzian action.  This is not true in  the GUE. 

The pattern of symmetry breaking for the SYK model   determines much of its low energy behaviour. Its low-energy thermodynamics is governed by a saddle point resulting in a linear specific heat, and a big entropy, $S_0$, scaling like $N$, eq.~(\ref{entSYK}), eq.~(\ref{csc}),   close to ground state. And the OTOCs exhibit chaotic behaviour with a Lyapunov exponent saturating the chaos bound. These features are absent in the GUE. Many  properties of the GUE, for   $L\times L$  matrices,  are tied to the $U(L)$ invariance  of the resulting  probability distribution. This invariance results in hyperfast scrambling for operators  and also  for  the OTOCs, which 
exhibit a saturation time of order the thermalisation time, see appendix \ref{rmt4pt}.  The low-energy thermodynamics for the GUE is  also quite different, see section \ref{revsec}. 


The behaviour of the non-local SYK model is of considerable interest from the point of view of gravity, agreeing in particular  with JT gravity and being akin,   quite universally, to near-extremal black holes with an $AdS_2$ factor in their near horizon geometry,{\cite{Maldacena:2016hyu,Nayak:2018qej}}. 
The differences in behaviour with the GUE prompt one to ask why these differences arise and more generally when can  behaviour of interest from the point of view of gravity arise, 
as the randomness is varied\footnote{Hyperfast scrambling might  also be of interest from the point of view of gravity in dS space and more generally, cosmology, see \cite{Susskind:2021esx}}. Such questions 
were some of the central motivations   behind this paper. 

Both the GUE and the SYK model correspond to Gaussian Ensembles, see section \ref{rmtassyk}. In the GUE, with $L\times L$   Hermitian matrices, $H$, all  $L^2$ matrix elements are random variables drawn from independent Gaussian distributions with the same variance and no mean. Equivalently, one can expand a general matrix $H$ in the basis of products of  fermionic operators, eq.~(\ref{sumh}).  The number of fermion flavours, $N$, needed is related to $L$ by $L=2^{N\over 2}$. The total number of coupling constants in this  fermionic basis is $2^N=L^2$, which is of course the same as the   number of independent matrix elements in $H$,   and  for the GUE these couplings in the fermionic basis are also  independent Gaussian random variables with equal variance and vanishing mean. 
In contrast, the non-local SYK model has fewer random variables.  For the SYK$_q$ model, only operators with products of $q$ fermions are present in the Hamiltonian, giving rise to ${}_NC_q\sim \order(N^q)$ random variables, which is much smaller than 
$L^2$. 

It is quite clear that some of the difference between the two models arise because of this reduction in randomness. To investigate this further, we  studied two kinds of models  in this paper. First, starting with the GUE we reduced the number of random variables by setting some number of matrix elements to vanish, at random, and choosing the remaining non-vanishing elements from the same kind of Gaussian distribution as in the GUE. This gave rise to the sparse random matrix model studied in section \ref{sprmt}. As the number of non-vanishing matrix elements, $n$, is reduced, we  move away from the GUE and one can study  the effects of reduced randomness. 
Second, at the other end with much reduced randomness, we considered the case of the local SYK model where only $4$ adjacent fermions were coupled to each other through $\psi^4$ type couplings, unlike the  non-local SYK$_4$ model where all fermions are coupled to each other in $\psi^4$ type couplings, section \ref{sykmat}. This model has only $\order(N)$ random couplings, which is even smaller than the $\order(N^4)$ couplings in the corresponding non-local SYK$_4$ model.

For the sparse random matrix model we found that as the number of random variables was reduced there was a marked change of behaviour when $n$ reached a value $n_c$, which was numerically found to scale like ${n_c\over L^2} \sim L^{-0.48}$, for large $L$. 
This change of behaviour is tied to the reduction in eigenvalue repulsion, resulting in energy eigenvalues clumping in an increasingly pronounced manner near the origin ($E=0)$, fig.~\ref{finiteeffectforp}. 
At values of $n<n_c$  our analysis showed that the behaviour becomes  quite different  from the GUE and  several  properties begin to differ. The spectral form factor now showed evidence of attraction between eigenvalues; the level statistics began to depart from the Wigner statistics as $n$ decreases and become more Poisson-like distributed; { the ramp in the SFF is flattened} and the OTOCs indicated a slower growth of the correlators. A key limitation of our study is that the numerical work could only be carried out at finite $N$ and many of  our conclusions could not be reliably extrapolated to the infinite $L,N,$  limit. In particular, we could not investigate the change of behaviour near the transition at $n_c$, in the $L\rightarrow \infty$ limit. It would also be very interesting to know if for $n<n_c$  a Schwarzian mode  arises in this limit.

For the local SYK model, studied in section \ref{sykmat},  we found  that the behaviour was quite different from the non-local SYK case. These differences could be seen in the spectral form factor, which has some similarities in fact with the sparse random matrix model  case (for {$1\ll n\ll n_c$}); in the level statistics, which was Poisson and not Wigner; and also, in  a qualitative sense, in the OTOCs. 

 It would be interesting to understand our results in greater depth, augmenting the numerical work presented here with additional  analytical insights. 
As mentioned above, the GUE and the SYK models both belong to  the class of Gaussian ensembles, where the Hamiltonian is determined by couplings drawn from Gaussian random distributions. Our investigation reveals that this class is in fact quite rich and such a detailed understanding, we think, will prove quite rewarding.

Some analytical results along the lines of our investigation  are already known. It was shown in \cite{Swingletalk,Xu:2020shn} that starting with the non-local SYK$_q$ model,  if we set some of the $\psi^q$ couplings to vanish at random, the behaviour would continue to be of the SYK$_q$ type,  as long as the total number of non-vanishing couplings scaled like $N^\alpha, \alpha>1$, in the large $N$ limit. The local model we have studied has $\order(N)$ couplings, which is smaller and lies outside this range. In contrast,  the sparse random matrix model of section \ref{sprmt} exhibits a change of behaviour when the number of couplings is $n_c \sim L^{1.52}$ - this  is much bigger than any power of $N$. 
The results of \cite{Xu:2020shn} can be extended to the case where $q$ is not fixed and also diverges when  $N \rightarrow \infty$. One finds, as discussed in appendix \ref{quadbosnmodel},  that a $G -\Sigma$ type of saddle point which arises in the non-local model, continues to arise here,  with the same breaking of time reparametrisation invariance etc, as long as $q<\sqrt{N}$ and the total number of non-vanishing terms scales  like $N^\alpha$, with $\alpha>1$. If we think of  the GUE  in the fermionic basis, eq.~(\ref{psiaspaul}), when  $N\rightarrow \infty$, its Hamiltonian  will be determined by the $\psi^q$ terms with $q={N\over 2}$, since these terms 
out-number all others. Thus we see that the 
sparse random matrix model  of section \ref{sprmt} is   related to the fixed $q$ model of \cite{Xu:2020shn},  with $q$ now scaling linearly with $N$. 

More generally, one could consider many other models in the  class of Gaussian ensembles. 
For example, we can include a mean value as well as  a variance in the normal distributions, see section \ref{rmtassyk} and \cite{Arka2023}. Varying their ratio allows us to explore what happens as we go from ordered to disordered systems.
 We could also study such Gaussian models  from the perspective  of the breaking of the $U(L)$ symmetry present in the GUE, see section \ref{rmtassyk}. 
 For example, one can have blocks in the random Hamiltonian of GUE type, and  couple these blocks  to each other with intermediate blocks that have  reduced randomness, including of the  SYK type.  The resulting behaviour would then be quite rich. As far as the spread of entanglement goes for example,  which could be measured through OTOCs, or the growth of operators, one would expect GUE like blocks to exhibit hyperfast scrambling, with $t_{scrambling} \sim \order(1)$,  this would become somewhat slower $t_{scrambling} \sim \order(\ln N)$,
  in the non-local SYK blocks of the random matrix and then even slower,  $t_{scrambling} \sim \order(N)$,   in  the blocks where the couplings preserve   spatial locality. 
{ Bottle neck} for the spread of entanglement, may perhaps  arise in highly ordered blocks. A detailed study along these lines is also left for the future.

\section*{Acknowledgement} 
We thank Kedar Damle, Abhijit Gadde, Nilmani Mathur,  Yoshinori Matsuo,  Shiraz Minwalla, Upamanyu Moitra, Onkar Parrikar,  Vikram Tripathy, Nandini Trivedi and particularly Subir Sachdev and Masaki Tezuka for valuable discussions.
The work of TA and NI were supported in part by JSPS KAKENHI Grant Number 21J20906(TA), 18K03619(NI). 
The work of NI and SS was also supported by MEXT KAKENHI Grant-in-Aid for Transformative Research Areas A ``Extreme Universe'' No.\ 21H05184.
 AM, SS and ST acknowledge support from the Department of Atomic Energy, Government of India, under  Project Identification No. RTI 4002, and acknowledge support from the Quantum Space-Time Endowment of the  Infosys Science Foundation. AM, SS and ST  acknowledge the  use  of the DTP, TIFR, computing cluster and especially thank Kapil Ghadiali and Ajay Salve for their help. AM thanks Nandini Trivedi and Ilya Gruzberg for valuable discussion. AM also thanks the Department of Physics, The Ohio State University for the University Fellowship.


\appendix

\section{Sparse SYK}
\label{quadbosnmodel}

In this { appendix}, we discuss the sparse version of SYK following  \cite{Xu:2020shn}. 
The sparse SYK model considered in \cite{Xu:2020shn} is a variant of the original SYK model where each of the $q$-fermi interaction are assigned a probability. The Hamiltonian for the model is given by 
\begin{align}
	H=\sum_{Q} j_Q x_Q \psi_Q, \qquad Q=\{i_1,\dots i_q\}, \quad i_1<i_2<\dots i_q\label{qspsyk}
\end{align}
where the variables $j_Q$ and $x_Q$ are  independent random variables, with $j_Q$ being drawn from a Gaussian distribution and $x_Q$ is a boolean variable with a binary distribution. Note that in the notation above $Q$ denotes a specific set of $q$  flavour indices, $\{i_1, i_2, \cdots i_q\}$, and $\psi_Q$ denotes the product of the 
corresponding $q$ fermions, $\psi_{i_1}\psi_{i_2}\cdots \psi_{i_q}$. 
By the probability distribution we mean more precisely 
\begin{align}
	p(j_Q)=\frac{1}{\sqrt{2\pi J^2\sigma}}e^{-\frac{j_Q^2}{2J^2\sigma}}	\implies \langle j_{Q}j_Q'\rangle =J^2 \sigma\,\, \delta_{Q,Q'}\label{jvar}
\end{align}
and 
\begin{align}
	&	p(x_Q=1)=p,\nonumber\\
	& p(x_Q=0)=1-p\label{pxprop}
\end{align}
{ The case of $p=1, \sigma=\frac{(q-1)!}{N^{q-1}}$} 
would correspond to the usual SYK model, where $N$ is the total number of fermion flavours. The variable $x_Q$ being a binary variable will be unity for some bonds and zero for others and so only some of the bonds are turned on in any instantiation. The total number of terms in the Hamiltonian, {$n$, is when $p=1$ } 
\begin{align}
	n={}_NC_q \,. \label{totterm}
\end{align}
For $p<1$,  the number of terms in any instantiation of the model is given by 
\begin{align}
	n=p\,\,{}_NC_q\label{npl1}
\end{align}
The disorder averaged path integral for the model is given by 
\begin{align}
	Z=\int [D\psi] e^{i\int dt( \psi_ii \del_t \psi_i)}\prod_Q[dJ_Q][dx_Q] p(J_Q)p(x_Q)e^{-i\int { dt} j_Qx_Q\psi_Q}\label{pidef}
\end{align}
Doing the integral over $j_Q$ first, we get
\begin{align}
	Z\propto\int [D\psi]e^{i\int dt( \psi_ii \del_t \psi_i)}\prod_Q[dx_Q] p(x_Q)e^{-\frac{J^2\sigma}{2} \int { dt} \int { dt'} x_Q^2 \psi_Q(t)\psi_Q(t') }\label{pijin}
\end{align}

Now doing the average over the sparseness variable $x_Q$, we get
\begin{align}
	Z\propto\int [D\psi]e^{i\int dt( \psi_ii \del_t \psi_i)}\prod_Q\left(pe^{- \frac{J^2\sigma}{2}\int { dt} \int { dt'} \psi_Q(t)\psi_Q(t') }+1-p\right)\label{pixin}
\end{align}

Now, if $p$ is such that the average number of bonds turned on  is 
\begin{align} 
n=N^\alpha
\end{align} 
for some $\alpha>0$, we have
\begin{align}
	p=\frac{\,\, N^\alpha}{_NC_q\,\,}\label{pval}
\end{align}

\subsection{$q$ = fixed case}

Let us first consider where $q$ is fixed with large $N$ limit. Then we get 
\begin{align}
	p\sim \frac{q!}{N^{q-\alpha}}\label{plan} \,.
\end{align}
Let us consider the case where both $p$ and $\sigma$ small, {\it i.e.}, 
\begin{align}
\label{consistent1}
p \to 0 \,, \quad \sigma \to 0 \,.
\end{align}
For $p \to 0$ in the large $N$ limit with $q$ fixed, we require
\begin{align}
	q>\alpha \,. \label{alq}
\end{align}
Then we can exponentiate the term in the brackets in \eqref{pixin} and so we get
\begin{align}
	Z &\sim \int [D\psi]e^{\frac{i}{2}\int dt( \psi_ii \del_t \psi_i)}\prod_Q\exp\left({p \lbrace  e^{- \frac{J^2\sigma}{2}\int { dt} \int { dt'}  \psi_Q(t)\psi_Q(t') }-1 \rbrace  }\right) \\
	& \sim \int [D\psi]e^{\frac{i}{2}\int dt( \psi_ii \del_t \psi_i)} \prod_Q \exp\left({  {- \frac{p J^2\sigma}{2}\int { dt} \int { dt'}  \psi_Q(t)\psi_Q(t') }  }\right) 
	\label{expzp}
\end{align}
In the second line, we assume $\sigma \to 0$ and also that $\langle \psi_Q\psi_Q\rangle \sim \order{(1)}$. We will see the self-consistency of these assumption soon. 

We now introduce the flavour dependent bi-local fields $\Sigma_Q(t,t'), G_Q(t,t')$ and follow the same manipulations as in the case of usual SYK. The path integral then becomes
\begin{align}
	Z&\propto \int [D\psi][D\Sigma_i][DG_i]\Bigg[e^{\frac{i}{2}\int dt( \psi_ii \del_t \psi_i)+\frac{1}{2}\int { dt} \int { dt'} \Sigma_i(t,t')(G_i(t,t')-\psi_i(t)\psi_i(t'))}\nonumber\\
&~~~~~~~~~~~~~~~~~~~~~~~~~~~~~~~~~~~~~~~~    \prod_Q \exp\left({  {- \frac{p J^2\sigma}{2}\int { dt} \int { dt'}  \psi_Q(t)\psi_Q(t') }  }\right)  \Bigg]\nonumber\\
	&= \int [D\psi][D\Sigma_i][DG_i]e^{\left[\frac{1}{2}\int dt( \psi_i(- \del_t-\Sigma_i) \psi_i)+\half\int { dt} \int { dt'} \Sigma_i(t,t')G_i(t,t')+\sum_Q  {- \frac{p J^2\sigma}{2}\int { dt} \int { dt'} G_Q(t,t') }  \right]}\nonumber\\
	&= \int [D\Sigma_i][DG_i]e^{\left[\int \int \frac{1}{2}\log\det(-\delta(t-t')\del_t-\Sigma_i)+\half\int { dt} \int { dt'} \Sigma_iG_i+\sum_Q {- \frac{p J^2\sigma}{2}\int { dt} \int  { dt'} G_Q(t,t') }
	\right]}\label{pathmap}
\end{align}
%
where  the symbol $\sum_Q$ is the summation symbol with sum being taken over all sets $Q$ of the form in eq.~\eqref{qspsyk} and  in the second line above we have used the notation that 
\begin{align}
	G_Q(t,t')=G_{i_1}(t,t')\dots G_{i_q}(t,t')\label{gqpro}
\end{align}
For a moment let us assume that there exists a solution where all flavour two point functions are same, {\it i.e.},
\begin{align}
	\Sigma_{i_1}(t,t')=\Sigma_{i_2}(t,t')\equiv\Sigma(t,t')\nonumber\\
	G_{i_1}(t,t')=G_{i_2}(t,t')\equiv G(t,t')\label{allgssm}
\end{align}
then the path integral becomes
\begin{align}
	Z\propto \int [D\Sigma][DG]e^{\left[\int \int \frac{N}{2}\log\det(-\delta(t-t')\del_t-\Sigma)+\frac{N}{2}\int\int\Sigma G - {}_NC_q   { \left( \frac{p J^2\sigma}{2} \right) \int \int  G^q }  \right]}\label{allsmapar}
\end{align}
The equation of motion obtained by varying with respect to $\Sigma$ and  $G$ are then given by 
\begin{align}
		G(t)&=(- \del_t-\Sigma(t))^{-1}\nonumber\\
	\Sigma&= q \, p \,  \sigma \frac{  {}_NC_q }{N} \, J^2 G^{q-1} 
		\label{eomsip}
\end{align}
First, the way it works in the conventional SYK model is as follows. The parameters for the conventional SYK model are
\begin{align}
	p=1\,, \quad \sigma=\frac{(q-1)!}{N^{q-1}} \quad \Rightarrow  \quad  q \, p \, \sigma \frac{  {}_NC_q }{N} = 1
	\label{convsykpar}
\end{align}
Then it is clear that eq.~\eqref{eomsip} becomes
\begin{align}
	\Sigma 
	&=  {J^2}G^{q-1} \,,  
		\label{consyksg}
\end{align}
which gives exactly the conventional SYK saddle point, where $G\sim O(1)$ in terms of $N$ counting. 

However it must be clear now that the way to obtain conventional SYK saddle point, therefore  $G\sim O(1)$, 
is not only \eqref{convsykpar}, but more generically possible as long as  
\begin{align}
\quad  q \, p \, \sigma \frac{  {}_NC_q }{N} = 1 \,, \quad \mbox{with} \quad p \to0 \,, \quad \sigma \to 0 
\end{align}
is satisfied. 
The condition of $p \to 0$ and $\sigma \to 0$ is due to the consistency with \eqref{consistent1}. 
With \eqref{plan}, this is equivalent to 
\begin{align}
	p \sim \frac{q!}{N^{q-\alpha}} \,, \quad \sigma \sim \frac{N^{1-\alpha}}{q}  \,, \quad \mbox{with \, $\alpha > 1$} \label{sigmordn}
\end{align}
in the large $N$ limit. In this case, the number of random variables, $n$, is 
$n=  {N^\alpha}$ with $\alpha > 1$. 
Thus, we conclude that even in sparse case, as long as  \eqref{sigmordn} is satisfied, 
we end up with the same saddle point equation and therefore solution as in the conventional SYK, \eqref{consyksg}. 
It is also worth mentioning,  as we see   from eq.(\ref{sigmordn}), that as   the sparseness increases, i.e.,  as $\alpha$ becomes smaller, the variance $\sigma$ increases. 

On the other hand, for $\alpha=1$ case, the model has become sufficiently sparse that the effective number of terms is of $\order{(N)}$. In this case, the approximation we used in \eqref{expzp} breaks down.The action then becomes 
\begin{align}
	Z\propto \int [D\Sigma][DG]e^{\left[\int \int \frac{N}{2}\log\det(-\delta(t-t')\del_t-\Sigma)+\frac{N}{2}\int\int\Sigma G+{}_NC_q \left(\,pe^{- \frac{J^2\sigma}{2}\int \int  G^q }-p\right)\right]}\label{allsmapar1}
\end{align}
 It is straightforward to see the modified equation of motion obtained by varying $G$ in this case
\begin{align}
	\Sigma&=\left(-{J^2\sigma q}\right)\frac{{}_NC_q \,p}{N}\left(e^{- \frac{J^2\sigma}{2}\int \int  G^q }\right)G^{q-1}
	\label{n1modeqs}
\end{align}

Another way to see the conclusion in eq.~\eqref{sigmordn} is to compute $\langle \tr(H^2)\rangle$, which is given by 
\begin{align}
	\langle \tr(H^2)\rangle =  {}_NC_q \,\, p \,J^2\,\sigma=J^2 \, N^\alpha \sigma\label{trh2}
\end{align}
For the saddle point solutions to exist we require $\langle \tr(H^2)\rangle\sim J^2N$ and so we find
\begin{align}
	J^2N\sim J^2 N^\alpha \sigma\implies \sigma\sim \order \left({N^{1-\alpha}} \right) \label{sialptr}
\end{align}
Thus, we reach the conclusion that $\alpha > 1$ in this viewpoint. 

Note that all the manipulations above are done for a fixed $q$. Another interesting limit to consider is when $q$ also becomes large. This can be done in two ways. First, we can take $N\rightarrow\infty$ followed by $q\rightarrow\infty$. Alternatively, we can consider $q\sim N^a,\,\,1>a >0$,  so that $q^{\frac{1}{a}}/ N$ is fixed as $N\rightarrow\infty$. In both these cases, the above manipulations have to be redone carefully. 

\subsection{Large $q$  limit, No sparseness }
\label{largeq}
{\underline{\bf{$N\rightarrow\infty$ with $q$ large but fixed: }}}

In the previous subsection, we focussed on the case of finite $q$ and large $N$. Another interesting regime to consider is when both $q$ and $N$ are large. 
Let us first  analyze the case of large, fixed $q$ but with $N$ taken to infinity without introducing any sparseness. This part will be a review of \cite{Maldacena:2016hyu}.The path integral in this case in terms of $G,\Sigma$ fields is given by 
\begin{align}
	Z\propto \int [D\Sigma][DG]e^{\frac{N}{2}\left[\int \int \log\det(-\delta(t-t')\del_t-\Sigma)+\int\int\Sigma G - {}_NC_q   { \left( \frac{ J^2\sigma}{N} \right) \int \int  G^q }  \right]}\label{allsmapara}
\end{align}
The equations of motion for $G$ and $\Sigma$ fields are
\begin{align}
	G(t)&=(-\del_t-\Sigma(t))^{-1}\nonumber\\
	\Sigma(t)&= (J^2\sigma q)\,\, \frac{{}_NC_q}{N} G^{q-1}
	\label{gsig}
\end{align}
In momentum space this becomes
\begin{align}
	{G(\omega)}=(-i\omega-\Sigma(\omega))^{-1}\label{gom}
\end{align}
In the large $q$ limit, we can take the ansatz for $G$ to be given by 
\begin{align}
	G(t)=\frac{1}{2}{\text{sgn}}(t)e^{\frac{g(t)}{q-1}}\label{gt}
\end{align}
so that we can do an expansion in $\frac{1}{q}$ as
\begin{align}
	G(t)\sim\frac{1}{2}\text{sgn}(t)\left(1+\frac{g(t)}{q}+\dots\right)\label{1qwex}
\end{align}
The fourier transform of the above is given by
\begin{align}
	G(\omega)\simeq \frac{i}{\omega}+\frac{1}{q}[\text{sgn}\times g](\omega)\label{gw}
\end{align}
The crucial point here is that, usually in the analysis at fixed $q$, low energies correspond to small $\omega$ and so the explicit $\omega$ dependence in eq.~\eqref{gom} is dropped in solving the saddle point equation. However, in the large $q$ limit, even in the regime of low energies, we should not drop this term but instead do a perturbation around the free theory result as shown in eq.~\eqref{1qwex}.

Using eq.~\eqref{gw} in eq.~\eqref{gom}, it is easy to see that 
\begin{align}
	\Sigma(\omega)\sim-\frac{\omega^2}{q}[\text{sgn}\times g](\omega) \,.
	\label{sigom}
\end{align}
We can self-consistently assume that in terms of $q, N$ counting $g(t)\sim \order{(1)}$ so that we get $\Sigma(t)\sim\order{(q^{-1})}$. On the other hand, using eq.~\eqref{gt} in the second equation in  eq.~\eqref{gsig} gives
\begin{align}
	\Sigma(t)=(J^2\sigma q) {{}_NC_q\over 2^{q-1}  { N}}\text{sgn}(t)e^{g(t)}\label{stelq}
\end{align}
Comparing eq.~\eqref{sigom} and eq.~\eqref{stelq},   we get
\begin{align}
	\del_t^2(\text{sgn}(t)g(t))=(J^2\sigma q^2) {{}_NC_q\over 2^{q-1} N}\text{sgn}(t)e^{g(t)}\label{largqgeq}
\end{align}
We can choose $\sigma$ to cancel all factors of $N$ and $q$ above, so that $g(t)$ can be obtained to be of order unity. We take, 
\begin{align}
	\sigma=\frac{2^{q-1}N}{q^2\, {}_NC_q} \label{sigmainlq}
\end{align}
{
The general solution for $g(t)$ is then given by 
\begin{align}
	e^{g(t)}=\frac{2c^2}{J^2}\frac{1}{\sin(c(|\tau|+\tau_0)^2)}\label{gtval}
\end{align}
where $c, \tau_0$ are the integration constants. 
Let us now turn to the analysis when $q$ scales as a power of $N$.
	
{	\underline{\bf{ $q\rightarrow \infty$ as $q\sim N^a$, \, $a>0$ : }}}
\par
The above discussion of the saddle point for the large $q$ limit can be easily generalized to the case when $q$ scales as a power of $N$. In this case, the path integral in terms of the fields $G,\Sigma$ is still given by eq.\eqref{allsmapara}. The last term in the exponent $G^q$ has an exponent which scales as a power of $N$. However, one can still self-consistently find a saddle point along the same lines as before by choosing $\sigma$ judiciously. In fact, the saddle point equations are still given by eq.\eqref{gsig} and the corresponding saddle point solutions are given by eq.\eqref{gom}, \eqref{sigom} with $g(t),\sigma$ still given by eq.\eqref{gtval},\eqref{sigmainlq} respectively. However, as we shall show below, the saddle point is a good approximation only for $a$ satisfying
\begin{align}
	a<\frac{1}{2}\label{abdy}
\end{align}
For $a>\half$, the saddle point analysis breaks down. 



\subsection{Large $q$ limit with Sparseness}
\label{larqwisp}
Below we examine the effects of sparseness for large values of $q$. For $q$ large and fixed as $N\rightarrow \infty$  we show that  a consistent saddle point can be found  in a manner similar to the one discussed above. 
For  the case when $q$ scales as a power of $N$, with sparseness present, we also carry out a saddle point analysis. 
Following the  saddle point analysis,  we then turn to a discussion of when the saddle point we have is a good one.  We find that the saddle point we have found is valid only for $a<1/2$, both with and without sparseness being present.  The arguments for the break down of the saddle point when $a>1/2$  is similar for the cases with and without sparseness, and we present a general analysis applicable to both cases below. 
}

 The steps leading to eq.~\eqref{allsmapar} are valid even when $q$ is large as long as 
	\begin{align}
	\bigg\vert	{J^2\sigma\int\int G^q}\bigg\vert\ll 1\label{condssma}
	\end{align}
	 and so are the equations eq.~\eqref{eomsip}. If $\sigma$ is not small, this step is not valid and we should derive the appropriate equations from eq.~\eqref{allsmapar1}.  Let us for now consider the case where $\sigma\rightarrow 0$. We use the ansatz eq.~\eqref{gt} for $G(t)$. The first equation in eq.~\eqref{eomsip} then leads to eq.~\eqref{sigom} as before where as the second equation now becomes
\begin{align}
		\Sigma(t)=(J^2\sigma q) {{}_NC_q \,p\over 2^{q-1}}\text{sgn}(t)e^{g(t)}\label{stelqp}
\end{align}
As before,  equating eq.~\eqref{stelqp} and eq.~\eqref{sigom}  we get
\begin{align}
		\del_t^2(\text{sgn}(t)g(t))=(J^2\sigma q^2) {{}_NC_q\,\, p\over 2^{q-1} N}\text{sgn}(t)e^{g(t)}\label{largqgeq}
\end{align}
Thus, for a good large $N$ and large $q$ limit to exist, we now have to take $\sigma$ to be given by 
\begin{align}
		\sigma=\frac{2^{q-1}N}{q^2\,\,{}_NC_q\,\,p} =\frac{2^{q-1}N}{q^2 N^\alpha}
		\label{signcp}
\end{align}
where we used ${}_NC_q\,p=N^\alpha$.  Naively it may appear that $\sigma$ is large in the large $q$ and large $N$ if $q\sim N^a,a>0$ due to the factor of $2^{q-1}$. However this factor is cancelled by a similar factor coming from $G^q$ in eq.~\eqref{condssma}.
From eq.~\eqref{signcp} and for $q\sim N^a$, we see that eq.~\eqref{condssma} is satisfied as long as
\begin{align}
	\alpha>1-2a\label{alphaa}
\end{align}
 Thus, we see that eq.~\eqref{condssma} is satisfied even if the total effective number of terms is linear in $N$, {\it i.e.,} $\alpha=1$, if $q$ is large. This is to be contrasted to the case of finite $q$ where we required $\alpha>1$, see eq.~\eqref{sigmordn}.
%
\subsection{Validity of Saddle point analysis}
\label{valsadpt}
Here we  argue that, both without and with sparseness  we cannot make $q$ arbitrarily large as the saddle point approximation will break down for $q>\sim \order{(\sqrt{N})}$. 
One way to understand this is to compute the fluctuations around the saddle point and see when they become important. To see this consider the expansion of the fields around the saddle 
\begin{align}
	\Sigma=\Sigma_0+\Sigma,\quad G=G_0+\delta G \,. \label{pertars}
\end{align}
Expanding the action in eq.~\eqref{allsmapar} by considering the fluctuations above, and noting the expansions
\begin{align}
	&(G_0+\delta G)^q\simeq G_0^q+G_0^{q-1}\delta G+{}_qC_2G_0^{q-2}\delta G^2+{}_qC_3G_0^{q-3}\delta G^2\nonumber\\
&	\ln\det(-\del_t-(\Sigma_0+\delta\Sigma))= \ln\det(G_0^{-1})+\ln\det(1+G_0\delta \Sigma)\nonumber\\ 
&\det(1+\epsilon A)\simeq 1+\epsilon\Tr(A)+\epsilon^2(\Tr A^2+(\Tr(A))^2)\nonumber\\
&	\qquad\qquad +\epsilon^3 (\Tr A^3+(\Tr(A))^3+\Tr A^2\Tr A+(\Tr(A))^2\Tr A)
	\label{expgsig}
\end{align}
where $A$ is any matrix in the last line above and $\epsilon$ is a small parameter in which the expansion is done. In our case $\Tr$ corresponds to the trace in time coordinates $t,t'$, the arguments of the functions $G,\Sigma$.
We find that the quadratic and cubic terms in fluctuations have the schematic form (ignoring the time integrals and traces)
\begin{align}
	-S\simeq&N(a_1(G_0\delta \Sigma)^2+a_2(G_0\delta \Sigma)^3)+N\delta\Sigma\delta G\nonumber\\
	&-{}_NC_q \frac{p J^2\sigma}{2}\left(\frac{q^2}{2}G_0^{q-2}\delta G^2+\frac{q^3}{6}G_0^{q-3}\delta G^3\right)
\end{align}
where $a_1,a_2$ are some $\order(1)$ coefficients. 
Noting that $\Sigma_0\sim \order(q^{-1}),G_0\sim \order(1)$ in terms of $N$ and $q$ counting given by eq.~\eqref{gt},\eqref{stelq} and that $\sigma$ is given by eq.~\eqref{signcp}, we immediately see from the quadratic terms in the fluctuations that 
\begin{align}
	\delta \Sigma\sim\order\left(\frac{1}{\sqrt{N}}\right) \,, \quad \delta G\sim \order\left(\frac{1}{\sqrt{N}}\right) \,. \label{sigord}
\end{align}
Using these estimates in the cubic terms, we find that the cubic terms are suppressed compared to the quadratic terms if and only if 
\begin{align}
	\frac{q}{\sqrt{N}}\ll 1. \label{dbsc}
\end{align}
{So, precisely when $q^2\gtrsim \order(N)$, the saddle point estimates fail}. In these cases there are other methods such as Chord diagram techniques \cite{Berkooz:2018jqr} that can used to solve the model in such a limit. We shall not delve further into these in this appendix. 

\section{Four Point function in RMT}
\label{rmt4pt}
In this section, we shall elaborate on the four point function calculations in GUE. To begin with, we illustrate the calculation in detail for the case of 2 point function which can then be straightforwardly generalized to four point function. Let us consider the averaged two point function of two operators $A,B$, with the averaging done over the Haar ensemble. This is given by 
\begin{align}
	\langle A(t)B(0)\rangle_{GUE}&=\int dH \Tr(A(t)B(0))\nonumber\\
	&=\int dH \Tr(e^{-i Ht}A(0)e^{iHt}B(0))\nonumber\\
	&=\frac{1}{\text{Vol}(U)}\int dU dH \Tr(Ue^{-i Ht}U^\dagger A(0)Ue^{iHt}U^\dagger B(0))\label{2ptsim}
\end{align}
where in obtaining the third line we used the fact that the Hamiltonian $H$ and its measure are invariant under $H\rightarrow U H U^\dagger$.
We now use the identity
\begin{align}
	\int dU (U^\dagger)^{a}_b U^c_d (U^\dagger)^{e}_f U^g_h=\frac{1}{L^2-1}\left(\delta^a_d\delta^c_b\delta^e_h\delta^g_f+\delta^a_h\delta^g_b\delta^e_d\delta^c_f-\frac{1}{L}\left(\delta^c_b\delta^e_d\delta^g_f\delta^a_h+\delta^a_d\delta^c_f\delta^e_h\delta^g_b\right)\right)\label{2ptui}
\end{align}
where $L$ is the rank of the matrix. Using this in eq.~\eqref{2ptsim} it is easy to see that we get
\begin{align}
	\langle A(t)B(0)\rangle_{GUE}=&\frac{\langle Z \bar{Z}\rangle_H}{L^2-1}\left(\Tr(A(0)B(0))-\frac{1}{L}\Tr(A(0))\Tr(B(0))\right)\nonumber\\
	&+\frac{L}{L^2-1}\left(\Tr(A(0))\Tr(B(0))-\frac{1}{L}\Tr(A(0)B(0))\right)\label{atbo}
\end{align}
where the traces on the rhs above involve only operators at time $t=0$ and  
\begin{align}
	Z=\Tr(e^{-i Ht}),\,\,	\bar{Z}=\Tr(e^{i Ht}),\nonumber\\
	\langle Z\bar{Z}\rangle_H=\int dH\, \Tr(e^{-i Ht})\,\Tr(e^{i Ht})\label{zzbdef}
\end{align}
Further, for simplicity, if we consider operators such that $\Tr(A(0))=0=\Tr(B(0))$, then we get
\begin{align}
		\langle A(t)B(0)\rangle_{GUE}=\left(\frac{	\langle Z\bar{Z}\rangle_H-1}{L^2-1}\right)\Tr(A(0)B(0))\label{1ptzer}
\end{align}
To leading order in large $L$, the quantity $	\langle Z\bar{Z}\rangle_H$ goes like $\order(L^2)$ and thus gives the dominant contribution to the two point function above.  So, in the large $L$ limit, we have
\begin{align}
		\langle A(t)B(0)\rangle_{GUE}\simeq\frac{	\langle Z\bar{Z}\rangle_H}{L^2}\Tr(A(0)B(0))
\end{align}
We now turn to the 4-pt function. Since we are mainly interested in the OTOCs, we consider a 4-pt function of the form
\begin{align}
	\langle A(t)B(0)C(t)D(0)\rangle_{GUE}&=\int  dH\, \Tr(A(t)B(0)C(t)D(0))\nonumber\\
	&=\int  dH\,\Tr(e^{-i H t}A(0)e^{i H t}B(0)e^{-i H t}C(0)e^{i H t}D(0))\nonumber\\
	=\frac{1}{\text{Vol}(U)}&\int dU dH\,\Tr(Ue^{-i H t}U^\dagger A(0)Ue^{i H t}U^\dagger B(0)Ue^{-i H t}U^\dagger C(0)Ue^{i H t}U^\dagger D(0))\label{4ptgueexp}
\end{align}
The analog of the identity eq.~\eqref{2ptui} we now need invoves four $U$s and same number of $U^\dagger$s. This identity is very cumbersome and so we shall not write it out explicitly here. Again, assuming all one point functions are zero, we find that, in the large $L$ limit, the leading term in the four point function is given by 
\begin{align}
		\langle A(t)B(0)C(t)D(0)\rangle_{GUE}\simeq {\langle Z \bar{Z} Z \bar{Z}\rangle_H\over L^4}\Tr(A(0)B(0)C(0)D(0))\nonumber\\
		\simeq \left(\frac{J_1(2t)}{t}\right)^4		\Tr(A(0)B(0)C(0)D(0))
		\label{4ptlarl}
\end{align}
The discussion so far pertained to the case of $\beta=0$. We can generalize these results for the case of finite $\beta$ as follows. The appropriate 4-pt function we for the finite temperature case is of the form 
\begin{align}
	\langle y A(t)y B(0) y C(t) y D(0)\rangle ,\quad y=e^{-\frac{\beta H}{4}}\label{yh}
\end{align}
The analog of eq.~\eqref{4ptgueexp} now becomes
\begin{align}
		&\langle y A(t)y B(0) y C(t) y D(0)\rangle = \nonumber\\
		&\frac{1}{\text{Vol}(U)}\int dU dH\,\Tr(Ue^{-i H \tilde{t}}U^\dagger A(0)Ue^{i H \tilde{t}^*}U^\dagger B(0)Ue^{-i H \tilde{t}}U^\dagger C(0)Ue^{i H \tilde{t}^*}U^\dagger D(0))\label{fint4pt}
\end{align}
where 
\begin{align}
	\tilde{t}=t-i\frac{\beta}{4},\quad \tilde{t}^*=t+i\frac{\beta}{4}\label{ttdde}
\end{align}

Thus, in the limit of large $L$, the 4-pt function essentially is given by eq.~\eqref{4ptlarl} with the simple modification that now 
\begin{align}
	Z=\Tr(e^{-iH\tilde{t}}),\,\bar{Z}=\Tr(e^{iH\tilde{t}^*})\label{ztzbt}
\end{align}
with $\tilde{t},\tilde{t}^*$ as defined in eq.~\eqref{ttdde}. Thus, the 4-pt function in the limit of large $L$ is given by 
\begin{align}
		\langle A(t)B(0)C(t)D(0)\rangle_{GUE}
	\simeq \left |\frac{J_1(2\tilde{t})}{\tilde{t}}\right |^4		\Tr(A(0)B(0)C(0)D(0))
	\label{4ptlarlfb}
\end{align}

{
\section{Symmetries of Sparse random matrices}
\label{symsparuni}
In this appendix, we shall show that the sparse random matrices we constructed in section \ref{sprmt} do not preserve any of the Unitary symmetries of the full random matrix. For random matrices constructed as outlined in  section \ref{sprmt}, the partition function can be written as
\begin{align}
	Z_{\text{sparse-RMT}}=\sum_{x_{ij}={0,1}} p({x_{ij}})\int \left(\prod_{ij}dH_{ij}\right)e^{-\tr  M M^\dagger},\quad M_{ij}=x_{ij}H_{ij}
	\label{zsprminbol}
\end{align}
where $x_{ij}$ for $i>j$ is a binary distributed random variable with the distribution
\begin{align}
	p(x_{ij})=\begin{cases}
		p,\quad & x_{ij}=1\\
		1-p,\quad &x_{ij}=0
	\end{cases}
\label{pxij}
\end{align}
and 
\begin{align}
	&x_{ji}=x_{ij}, \quad \text{for} \,\, i>j\nonumber\\
	&x_{ii}=0
\end{align}
The full random matrix corresponds to $p=1$ for all the random variables $x_{ij}$. Let us consider the class of unitary ensemble in which case the matrix $H$ is a hermitian matrix. The action in eq.\eqref{zsprminbol} written explicitly reads
\begin{align}
	\tr MM^\dagger &=\sum_{ij} M_{ij}(M^\dagger)_{ji}=\sum_{ij} M_{ij}(M^*)_{ij}\nonumber\\
	&=\sum_{ij} x_{ij}H_{ij}x^*_{ij}H_{ij}^*\nonumber\\
	&=\sum_{ij} x_{ij}^2H_{ij}H_{ji}\label{bfuact}
\end{align}
where in obtaining the last line we used the fact that $H$ is hermitian and so $H^*=H^T$.
The unitary symmetry of the full random matrix corresponds to the invariance of the action as well as the measure for the matrices $H$ under 
\begin{align}
	H\rightarrow U H U^\dagger\label{husy}
\end{align}
when all $x_{ij}=1$.  We would like to show that this invariance is broken when the binary variables are unity with a probability $p<1$. To show this, let us do a unitary transformation for $H$ and define the new matrix $\tilde{M}$ as
\begin{align}
	\tilde{M}_{ij}=x_{ij}\tilde{H}_{ij},\quad \tilde{H}_{ij}=(UHU^\dagger)_{ij}
\end{align}
Since, the measure for the matrices $H$ is invariant under the unitary transformation of $H$, the path integral for the sparse random matrix changes only due to the action. Let the path integral for the transformed $H$ be denoted bu $UZ$
\begin{align}
	UZ_{\text{sparse-RMT}}=\sum_{x_{ij}={0,1}} p({x_{ij}})\int \left(\prod_{ij}dH_{ij}\right)e^{-\tr  \tilde{M} \tilde{M}^\dagger},\quad \tilde{M}_{ij}=x_{ij}\tilde{H}_{ij}\label{uspaz}
\end{align}
The action, explicitly, is given by 
\begin{align}
	\tr \tilde{M}\tilde{M}^\dagger &=\sum_{ij}\tilde{M}_{ij}(\tilde{M}^\dagger)_{j,i}=\sum_{ij}\tilde{M}_{ij}\tilde{M}_{ij}^*\nonumber\\
&=	\sum_{ij} x_{ij}^2 (UHU^\dagger)_{ij} (UHU^\dagger)_{ji}\label{mmda}
\end{align}
It is easy to see from the above that unless $x_{ij}=1$ for all $i,j$, the dependence on the unitary matrices will not be trivial. Also, the above action is not equivalent to eq.\eqref{bfuact} under any transformation of the variables of $x_{ij}$. So, the action is not invariant under the unitary symmetries for general value of $x_{ij}$ and hence the unitary symmetry is broken. One way to intuitively understand this result is as follows. For any fixed values of $x_{ij}$, some subgroup of the full $U(L)$ symmetry can be preserved. However, after averaging over the binary variables, we have shown above that the $U(L)$ symmetry is broken. 
}
\section{Systematic Sparseness}
\label{sysspa}
In this appendix, we consider a different way of sparsing a random matrix. 
In this version of constructing a sparse random matrix, we start by picking a positive integer $p\in [1,L]$. We then 
take $M_{ij}$ to have the following form,
%
\begin{align}
	M_{ij} = \begin{cases}
		0 \qquad \mbox{when} \quad i - j \neq 0  \quad (\mbox{mod $p$}) \,\\
		\neq 0 \quad \mbox{when} \quad i - j = 0  \quad (\mbox{mod $p$})
	\end{cases}
	\label{syssprdef}
\end{align}

More precisely: the GUE, eq.(\ref{GUE}), corresponds to the partition function
\be
\label{gdis}
Z=\frac{1}{Z_0} \int \prod_i dM_{ii} \ \prod_{i<j}  d\text{R}(M_{ij}) d\text{I}(M_{ij} ) \ e^{-\bigl ( {L \over 2} \bigl [\sum_i M_{ii}^2 + \sum_{i<j} 2 \text{R}(M_{ij})^2+ 2 \text{I}(M_{ij})^2\bigr] \bigr) }
\ee
where $Z_0$ is some normalization factor. 
This  shows that  the  matrix elements  are independent Gaussian random variables with the diagonal  elements $M_{ii}$ having variance ${1\over L} $; among the   off-diagonal elements  $M_{ij}$, with $ i<j$,  can be taken to be the independent variables, with $M_{ij}$ for $i>j$ being determined by the condition $M^\dagger=M$.  $\text{R}(M_{ij}),\text{ I}(M_{ij})$ denote the real and imaginary parts of $M_{ij}$ and these have variance ${1\over  2 L }$.

Now in the systematic randomness case the matrix elements which are non-zero,  $M_{ij}$, with $i-j=0\ {\rm mod} \ p$, eq.~(\ref{syssprdef})
are drawn in the same manner as independent Gaussian Random variables; {\it i.e.},  
the diagonal elements $M_{ii}$ are taken to have variance ${1\over L}$ and among the off diagonal elements, $M_{ij}$ with $i-j=0 \ {\rm mod}  \ p$ that are non-zero, with $M_{ij}, i<j$, being the independent ones and the  variance for $\text{R}(M_{ij}), \text{I}(M_{ij})$ being  ${1\over 2 L}$.


Let us now consider a few special cases. For $p=1$, the random matrix is fully random, {\it i.e.}, all the elements of the matrix are non-zero random elements. For $p=L$,  the only non-zero elements are the  diagonal elements.  More generally, let $L=p \, q$, then the matrix breaks up into $p$ block diagonal sub-matrices\footnote{We thank Yoshinori Matsuo for pointing this out to us.}, each of rank $q$. The first block has the matrix elements $M_{ij}$ with $i$ and $j$ both being equal to $1 \ {\rm mod} \ p$. The second block has the matrix elements $M_{ij}$, with $ i, j = 2  \ {\rm mod}  \ p$, etc. 
The total number of (real) random elements is  therefore equal to  $q^2 p =L^2/p$. 
It is also easy to see that each block is a GUE of $q\times q$ matrices with variance being $L$.

Thus we see that 
the amount of sparseness can be controlled by varying the value of $p$ with only a fraction $1/p$ of the $L^2$ total matrix elements being  present.  

Due to the   block diagonal nature, with each block being a GUE,  most of  the properties of this ensemble  easily follow from those of the GUE as we will see below in Fig \ref{srmtsydosf} and \ref{srmtsydosfT}. 
	\begin{figure}[H]
\hspace{-10mm}
\subfigure[]{\includegraphics[width=7.5cm,height=5cm]{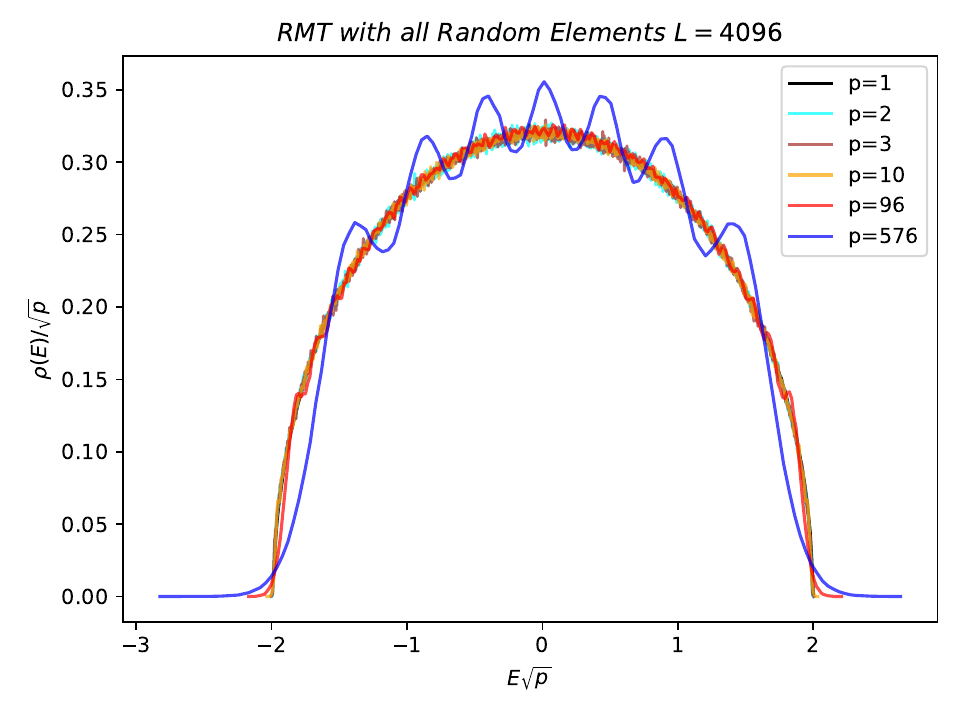}}
\subfigure[]{\includegraphics[width=7.5cm,height=5.3cm]{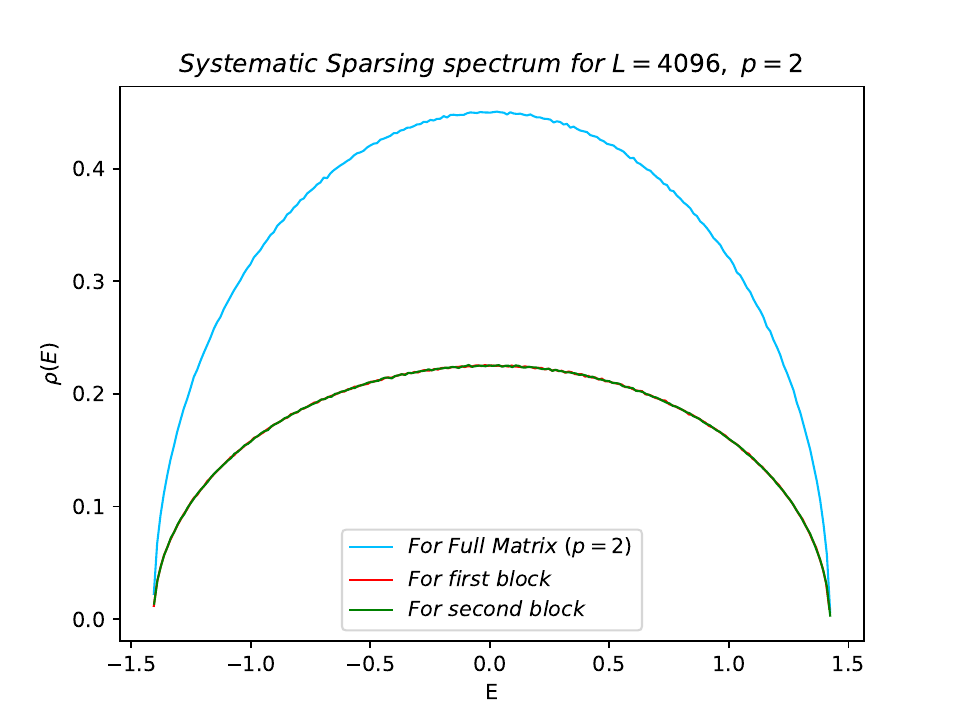}}
		\caption{(a) Spectrum for the systematic sparseness with $L=4096$ and various values of parameter $p$. (b) For $p=2$ case both blocks follows Wigner semicircle.}  
		\label{srmtsydosf}
	\end{figure}

The above is a plot of the rescaled density of states for various values of $p$. It is evident from the above plot that the except for highly sparse matrix, the case of $p=576$ in the plot above when only $\frac{L^2}{p}$ random elements are kept, the rest of the cases fit well with a Wigner distribution. For the case of $p=576$, the size of each block is $\frac{L}{p}\sim 7$ and hence the wiggles around the Wigner distribution. 
The case of $p=2$ is shown explicitly in the right panel where it can be seen that the two blocks have identical distribution, each contributing half to the total distribution, as expected. 

		\begin{figure}[H]
\hspace{-10mm}
\subfigure[]{\includegraphics[width=7.5cm,height=5.5cm]{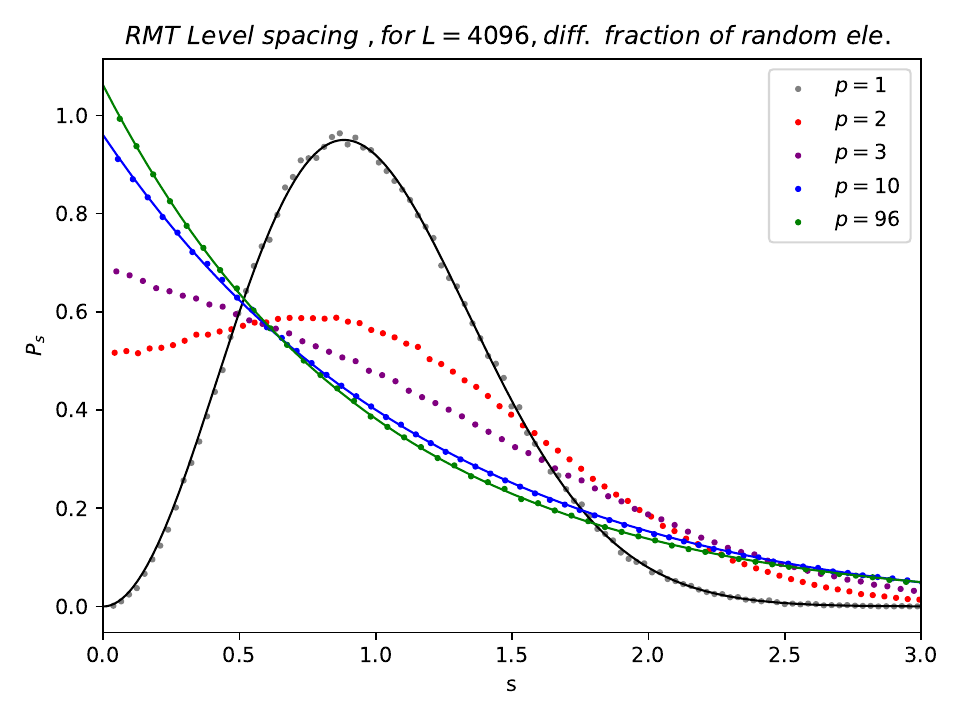}}
\hspace{-3mm}
\subfigure[]{\includegraphics[width=7.5cm,height=5.5cm]{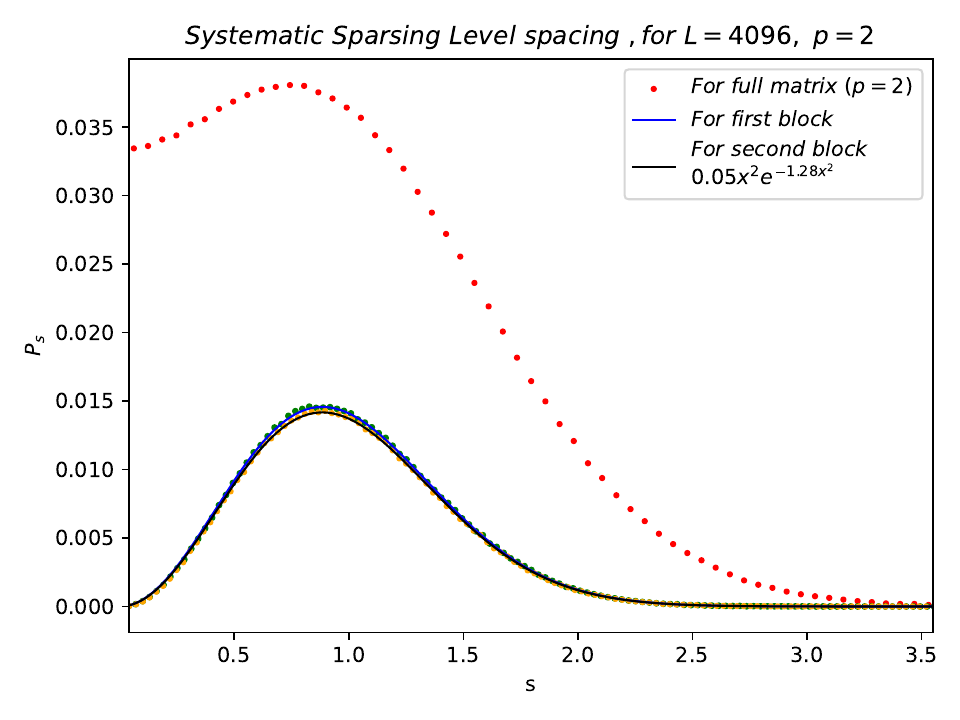}}
\caption{(a) Nearest neighbour level spacing after unfolding the spectrum for different $p$ values
(b) Level spacing for $p=2$ blocks. Each blocks satisfy Wigner surmise individually.}
\label{srmtsydosfT}
\end{figure}
The above plot shows the level spacing distribution for the systematically sparsed random matrix. The left panel is the level spacing distribution for unfolded eigenvalues for the full random matrix and for varying values of $p$. As can be seen from the plots, for $p>1$, when the full matrix splits in to blocks, the level spacing distribution is no longer follows Wigner surmise. As $p$ increases, the distribution moves from Wigner Surmise to a Poisson distribution. This is just because of the fact that as the value of $p$ increases, the full matrix splits into more number of blocks each of which are uncorrelated. So, increasing $p$ corresponds to decreasing correlations between the eigenvalues of the full matrix. Each of the sub blocks in each matrix are still completely correlated as can be seen from the right panel. The mixing of the blocks leads to the distribution becoming more Poisson-like. 

We shall now calculate the 2-pt function for the case of sparse random matrix corresponding to systematic sparseness. 
As already mentioned above the hamiltonian for the systematically sparse random matrix becomes a block diagonal matrix. In particular, the unitary matrices which preserve this block diagonal form the symmetry group for such Hamiltonians. So, we shall compute the averaged 2-pt function in a manner similar to the previous section \ref{rmt4pt} but with the understanding that the corresponding unitary matrices are also block diagonal to preserve block diagonal nature of the Hamiltonian. Let the Hamiltonian be a block diagonal matrix with $p$ blocks each of size $q$. Any matrix element of a block diagonal unitary matrix will be of the form
\begin{align}
	U^{a}_{a'}\neq 0 \quad \text{iff }\,\,\,a=\alpha q+i, \, a'=\alpha q+i', \quad 1\leq \alpha\leq p, \quad 1\leq i,i'\leq q.\label{ublck}
\end{align}

The expression for the two point function is still given by the expression in eq.~\eqref{2ptsim}. The operators $A$ and $B$ need not be block diagonal. The measure $dH dU$ in eq.~\eqref{2ptsim} means that
\begin{align}
	dH dU=\prod_{\gamma}dU_\gamma\prod_\kappa dH_{\kappa}\label{mesblck}
\end{align}
where the quantities with subscripts $\rho,\kappa$ indicate the corresponding blocks.  Also, since we are assuming $A,B$ are non block diagonal, the product $U^\dagger A U$ will be of the form
\begin{align}
	(U^\dagger A U)^a_b&=(U^{\dagger})^a_{a'}A^{a'}_{b'}U^{b'}_b\nonumber\\
	&=(U^\dagger_\alpha)^{i}_{i'}(A_{\alpha\beta})^{i'}_{j'}(U_{\beta})^{j}_{j'}\label{uausy}
\end{align}
where in the second line the quantities with the subscripts corresponds to the appropriate blocks. A similar expression can be written for operator involving $B$. The two point function eq.~\eqref{2ptsim} written explicity in terms of blocks reads
\begin{align}
		\langle A(t)B(0)\rangle_{Sys}&=\int dU d H \left(e^{-i H t}\right)^{d}_a\left(e^{i H t}\right)^{b}_c \left(U^\dagger A(0) U\right)^{a}_b(U^\dagger B(0) U)^{c}_d\nonumber\\
		&=\int  d H  \left(e^{-i H t}\right)^{d}_a\left(e^{i H t}\right)^{b}_c (A_{\alpha\beta}(0))^{i'}_{j'}(B_{\rho\sigma}(0))^{k'}_{l'}\nonumber\\
&~~~~~~~~~~~~~~~~~~~~~~~~\times\int \prod_{\gamma}dU_\gamma (U^\dagger_\alpha)^{i}_{i'}(U_{\beta})^{j'}_{j}(U^\dagger_\rho)^{k}_{k'}(U_{\sigma})^{l'}_{l}\nonumber\\
		=\int  \prod_\kappa d H_{\kappa}&  \left(e^{-i H_\alpha t}\right)^{l}_i\left(e^{i H_\rho t}\right)^{j}_k \delta_{\alpha\sigma}\delta_{\beta\rho} (A_{\alpha\beta}(0))^{i'}_{j'}(B_{\rho\sigma}(0))^{k'}_{l'}\nonumber\\
&~~~~~~~~~~~~~~~~~~~~~~~~\times\int \prod_{\gamma}dU_\gamma (U^\dagger_\alpha)^{i}_{i'}(U_{\beta})^{j'}_{j}(U^\dagger_\rho)^{k}_{k'}(U_{\sigma})^{l'}_{l}\label{atbo2ptsys}
\end{align}
where we used 
\begin{align}
	a=\alpha q+i,\, b=\beta q+j,\, c=\rho q+k,\, d=\sigma q+l\nonumber\\
	\quad \text{with} \,\,1\leq \alpha,\beta,\rho,\sigma\leq p,\,\quad 1\leq i,j,k,l\leq q\label{varind}
\end{align}
 We now need the analog of the identity eq.~\eqref{2ptui}. It is given by 
 \begin{align}
 	\int \prod_{\gamma} dU_\gamma \,&(U^\dagger_\alpha)^{i}_{i'}(U_{\beta})^{j}_{j'}(U^\dagger_\rho)^{k}_{k'}(U_{\sigma})^{l}_{l'}=\frac{\delta_{\alpha\beta\rho\sigma}}{q^2-1}\Big(\delta^{i}_j \delta^{j'}_{i'}\delta^k_l\delta^{l'}_{k'}+\delta^{i}_l\delta^k_j\delta^{l'}_{i'}\delta^{j'}_{k'}\nonumber\\
&~~~~~~~~~~~~~~~~~~~~~~~~~~~~~~~~~~~~~~~-\frac{1}{q}(\delta^i_j\delta^{j'}_{k'}\delta^k_l\delta^{l'}_{i'}+\delta^{j'}_{i'}\delta^k_j\delta^{l'}_{k
 	}\delta^i_l)   \Big)\nonumber\\
 	&\quad+\frac{1-\delta_{\alpha\beta\rho\delta}}{q^2}\left(\delta_{\alpha\beta}\delta_{\rho\sigma}\delta^i_j\delta^{j'}_{i'}\delta^k_l\delta^{l'}_{k'}+\delta_{\alpha\sigma}\delta_{\rho\beta}\delta^i_l\delta^{l'}_{i'}\delta^{j'}_{k'}\delta^k_j\right)
 	\label{usysid}
 \end{align}
The delta function with four indices means that it gives unity only if  block diagonal indices are  such that $\alpha=\beta=\rho=\sigma$ and  zero otherwise. 
Using the identity eq.~\eqref{usysid} in eq.~\eqref{atbo2ptsys} and contracting the indices appropriately, we get
\begin{align}
		\langle A(t)B(0)\rangle_{Sys}&=\sum_\alpha\frac{1}{q^2-1}\Big[\left(q-\frac{1}{q}Z_\alpha\bar{Z}_\alpha\right)\Tr(A_{\alpha\alpha}(0))\Tr(B_{\alpha\alpha}(0))\nonumber\\
&~~~~~~~~~~~~~~~~~~~~~~~~~~+(Z_\alpha\bar{Z}_\alpha-1)\Tr(A_{\alpha\alpha }(0)B_{\alpha\alpha}(0))\Big]\nonumber\\
		&\,~~~~~~~~~~~\,+\sum_{\alpha,\beta}\frac{(1-\delta_{\alpha\beta})}{q^2}Z_\alpha\bar{Z}_\beta\Tr(A_{\alpha\beta}(0)B_{\beta\alpha}(0))\label{finsys2pt}
\end{align}
where we have explicitly written the summation to make it clear and where $Z_\alpha=\Tr(e^{-iH_\alpha t}),\bar{Z}_\alpha=\Tr(e^{iH_\alpha t})$. The first term is the usual term that one get if the operators $A,B$ are also block diagonal. The second term is an extra contribution that arises due to the off diagonal blocks in $A,B$. 

The four-point can also be generalized in a manner analogous to the two point function but  involves many more terms. We do not explicitly write it here except to note that the leading term in the OTOC will also be given by eq.~\eqref{4ptlarl}.

\section{Wigner surmise}
\label{Plnrder}
In this appendix we outline the derivation of the distribution of the ratio of nearest neighbour eigenvalues, eq.~\eqref{prb}.
For Wigner Surmise  $p(r)$ is given by
	 \begin{align}
	 	&p(r)=K\frac{(r+r^2)^\beta}{(1+r+r^2)^{1+\frac{3\beta}{2}}},\nonumber\\
	 	\text{where}\quad &\text{GOE}: \beta=1,\,\, K=\frac{27}{8}\nonumber\\
	 	\qquad\quad\,\, &\text{GUE}: \beta=2,\,\, K=\frac{81\sqrt{3}}{4\pi}\nonumber\\
	 	\qquad\quad\,\,&\text{GSE}: \, \beta=4,\,\,  K=\frac{729\sqrt{3}}{4\pi}\label{wiprn}
	 \end{align}
 A quick derivation of the above result for the case of $3\times 3 $ matrix is shown below.

\subsection{$3 \times 3$ matrix}
Consider  $3 \times 3$ matrix $H$. 
We consider both GOE ($\beta = 1$) and GUE $(\beta =2)$ here. 
One can diagonalize this matrix by orthogonal  matrix for $\beta =1$ and unitary matrix for $\beta=2$ and obtain the three eigenvalues, where
\be
\lambda_1 < \lambda_2 < \lambda_3
\label{inequalitylambda}
\ee
Then the eigenvalue ratio $r$ is given by 
\be
r \equiv \frac{\lambda_3 - \lambda_2}{\lambda_2 - \lambda_1} \ge 0
\ee 
and the probability for finding eigenvalue ratio  $r$ is given by  
\begin{align}
	p(r) &\propto \int_{-\infty}^{\lambda_2} d \lambda_1  \int_{-\infty}^{+\infty} d \lambda_2  \int_{\lambda_2}^{+\infty} d \lambda_3 \, \Delta(\lambda_i) \, \delta \left(r - \frac{\lambda_3 - \lambda_2}{\lambda_2 - \lambda_1} \right) e^{- \frac{\beta}{4} \mbox{Tr}\left( \sum_i \lambda_i^2\right)} 
\end{align}
Here we introduce Vandermonde determinant; 
\be
\Delta(\lambda_i) = \Pi_{i < j}^3 \left( \lambda_j - \lambda_i \right)^\beta \,.
\ee 
To simplify the calculation we change variables $(\lambda_1 , \lambda_3) \to (x, y)$ where 
\be
x \equiv \lambda_2 - \lambda_1 \,, \quad  y \equiv \lambda_3 - \lambda_2 \,.
\ee
Then, inequality eq.~\eqref{inequalitylambda} gives 
\be
x \ge 0 \,, \quad  y \ge 0 \,, 
\ee
and 
\be
\Delta(\lambda_i) = x^\beta y^\beta (x+y)^\beta \,.
\ee
With this, $P(r)$ becomes 
\begin{align}
	p(r) &\propto \int_{0}^{+\infty} d x  \int_{-\infty}^{+\infty} d \lambda_2  \int_{0}^{+\infty} d y \, x^\beta y^\beta (x+y)^\beta \, \delta \left(r - \frac{y}{x} \right) \nonumber\\
	&\quad e^{- \frac{\beta}{4} \left( 3 (\lambda_2  -\frac{x-y}{3})^2 +x^2 + y^2 - \frac{1}{3} \left(x-y \right)^2 \right)} \nonumber\\
	&\propto \int_{0}^{+\infty} d x   \int_{0}^{+\infty} d y \, x^{\beta+1} y^\beta (x+y)^\beta \, \delta \left(x r - {y} \right)  e^{- \frac{\beta}{4} \left(x^2 + y^2 - \frac{1}{3} \left(x-y \right)^2 \right)} \nonumber \\
	&\propto \int_{0}^{+\infty} d x   \,r^\beta (1+r)^\beta \, x^{3 \beta+1}  \,   e^{- \frac{\beta x^2}{4} \left(1 + r^2 - \frac{1}{3} \left(1-r  \right)^2 \right)}\nonumber  \\
	&\propto \frac{r^\beta (1+r)^\beta}{\left( 1 + r + r^2 \right)^{1 + \frac{3 \beta}{2}}}
\end{align}
where in the third line, we use 
\be
\delta\left(r - \frac{y}{x}\right) = x \delta(x r -y )
\ee
and in the fourth line, integration over $y$ is conducted. 
Therefore we obtain 
\be
p(r) = K_\beta  \frac{r^\beta (1+r)^\beta}{\left( 1 + r + r^2 \right)^{1 + \frac{3 \beta}{2}}} \,,
\label{rhorexpression}
\ee
where $K_\beta$ is a normalization factor, which we can easily obtain from 
$
\int_0^\infty dr\, p(r) = 1 
$
as 
\begin{align}
	K_{\beta=1} &= \frac{27}{8} \, \, \quad \quad  \mbox{for GOE $(\beta =1)$}  \,, \nonumber\\
	K_{\beta=2} &= \frac{81 \sqrt{3}}{4 \pi} \, \quad \mbox{for GUE $(\beta =2)$} \,.
\end{align}
As is seen in this example, eq.~\eqref{rhorexpression} is correct expression only for $3 \times 3$ matrix. In higher $L \times L$ case, the expression deviates from eq.~\eqref{rhorexpression}.  See for example, \cite{atas2013distribution}.

\section{Error Analysis for non-local SYK OTOC fit}
\label{eranl}

In this appendix we present a detailed error analysis we have done for the case of OTOCs. We have taken non-local SYK OTOC as the set up to perform the error analysis since we already know the Lyapunov exponent is the limit of large $N$ saturates the Chaos bound value of $2\pi T$.  

The main point of doing the error analysis is to estimate the maximum time till which the growth of the OTOC can be approximated by an exponential. Starting with the smallest values of time, the exponential fit will be good till a certain time beyond which the exponential approximation to the OTOC will not be good and we are interested in estimating this critical time. We will  denote this critical time as $t_c$ below. 

The error analysis is carried out using the Monte Carlo method.
 Given that our numerical data is a discrete set as a function of time, we first fit our data to the function
\begin{align}
	F_{th}=A+B e^{\lambda t},\quad 0<t<t_{\text{max}}\label{theoretic}
\end{align}
for various values of $t_{\text{max}}$. This is done using the best fit package of Python ($scipy.optimize.curve\_fit$). The python output will be the values of $A,B,\lambda$ for the best fit case  along with the associated covariance matrix $\sigma_{\alpha\beta}$, of which we only show the diagonal elements corresponding to $\sigma_{AA},\sigma_{BB},\sigma_{\lambda\lambda}$ denoted by $\sigma_A,\sigma_B,\sigma_\lambda$ in the table. This is the starting point for 
the analysis. 
We explain the methods for a single value of $\beta$. We start by choosing a value of $t_{\text{max}}$, the maximum time till which we fit the exponential. Let $y=\{A,B,\lambda\}$ and $\sigma_y=\{\sigma_A,\sigma_B,\sigma_\lambda\}$.

\subsection{Monte Carlo Simulation}
\label{montecarlo}
This method is described in \cite{Press2007} Chapter 15.6.1, `Monte Carlo Simulation of Synthetic Data Sets'.
We will also explain it, to some extent, below. 

 From the  quantities $ y=\{A,B,\lambda\}$ and $\sigma_y=\{\sigma_A,\sigma_B,\sigma_\lambda\}$, mentioned above,  we can build three independent normal distribution of 5000 elements, around mean $y_{i}$ with variance $\sigma_{y_{i}}^{2}$. So in total we have $5000^{3}$ elements of type $\{A_{i},B_{j},\lambda_{k}\}$ where $A_{i}\in N(A,\sigma_{A}),~B_{j}\in N(B,\sigma_{B}),~\lambda_{k}\in N(\lambda,\sigma_{\lambda})$. We retain only those data points which are within $1\sigma$ deviation from the mean value. \\
Next we fix a time point $(t_{n})$ and generate the corresponding OTOC value using eq. \eqref{theoretic} for $F_{th}(t_{n},A_{i},B_{j},\lambda_{k})$. Around each $t_{n}$ we now have a total of $\approx 5000^{3}$ such points. We take this distribution of $\approx 5000^{3}$ points and compute the $\sigma$ for that distribution. This $\sigma$ can be denoted by $\sigma_{t_{n}}$, we use $\sigma_{n}$ . \\
Now we use this to compute the value of $\chi^2$ using the formula (6.14.38), \& (15.1.6) in \cite{Press2007}
\begin{align}
\displaystyle
\chi_0^2(t_{max})=\sum_{t_{n}\leq t_{max}}\frac{(F_{numerical}(t_{n})-F_{th}(t_{n},A,B,\lambda))^{2}}{\sigma_{n}}
	\label{choi2}
\end{align}
then compute $P(\chi^2>\chi_0^2)$ given by 
\begin{align}
	P(\chi^2>\chi_0^2)=\int_{\chi_0^2}^{\chi^2}\frac{1}{2^{\frac{\nu}{2}}\Gamma{\frac{\nu}{2}}}(\chi^{2})^{\frac{\nu}{2}-1}e^{-\frac{\chi^{2}}{2}}d\chi^{2}
	\label{pchio2}
\end{align}
We repeat this for various values of $t_{\text{max}}$. The one with lowest value of $P$ will give  the best fit  and we set $t_c$ equal to this value of $t_{\text{max}}$. 
Once we have fixed $t_c$ then our best estimate for the Lyapunov exponent is the corresponding fitted parameter($\lambda$) from Python. 

In fact this more systematic way of obtaining $t_c$ and  $\lambda$ agrees with the more naive way of fitting ``with the eye". 

This justify our choosing of $t_{max}$ for fitting the data naively with “the eye”. The error in $\lambda$ for the value of $\beta$ under consideration corresponds to the standard deviation obtained from the best fit till $t_{\text{max}}$. From the data shown below in fig.\ref{otoc1nonr}, we see that for $\beta=20,30,50$, correspondingly we get $t_{\text{max}}=4,5.5,7$. Thus for smaller $\beta$, the exponential fit is only good till small value of $t$. 
\begin{figure}[H]
	\begin{minipage}[b]{\linewidth}
	\hspace{-13mm}
\subfigure[OTOC vs t plot different $\beta$ ]{		\includegraphics[width=7.7cm,height=7cm]{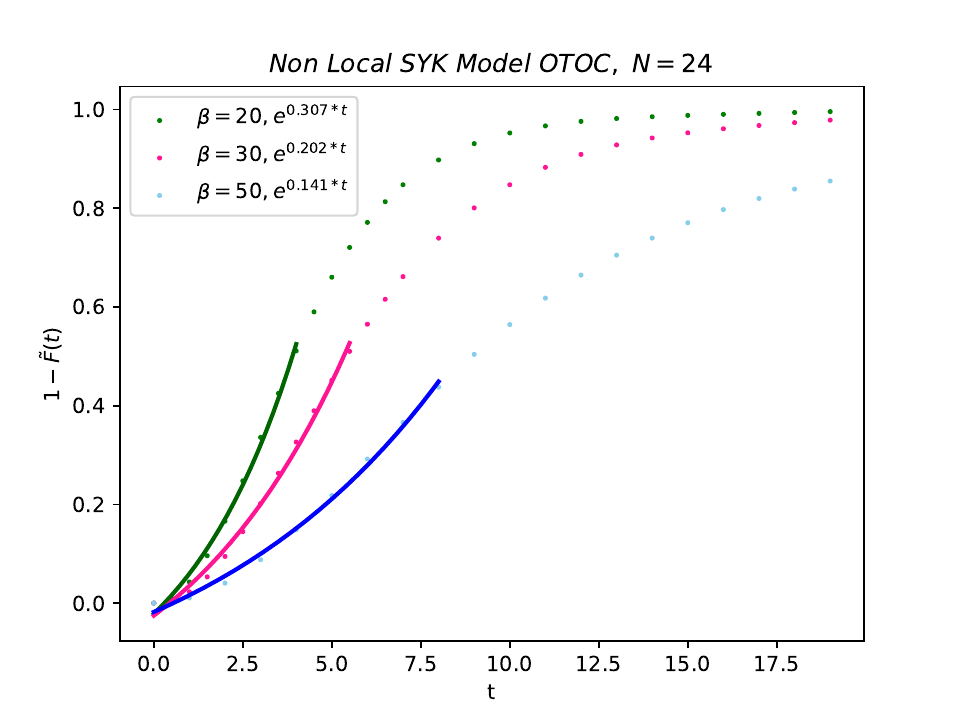}}
          \hspace{-10mm}
\subfigure[$\beta=20,30,50$ Chi-square fitting]{		\includegraphics[width=8.5cm,height=7cm]{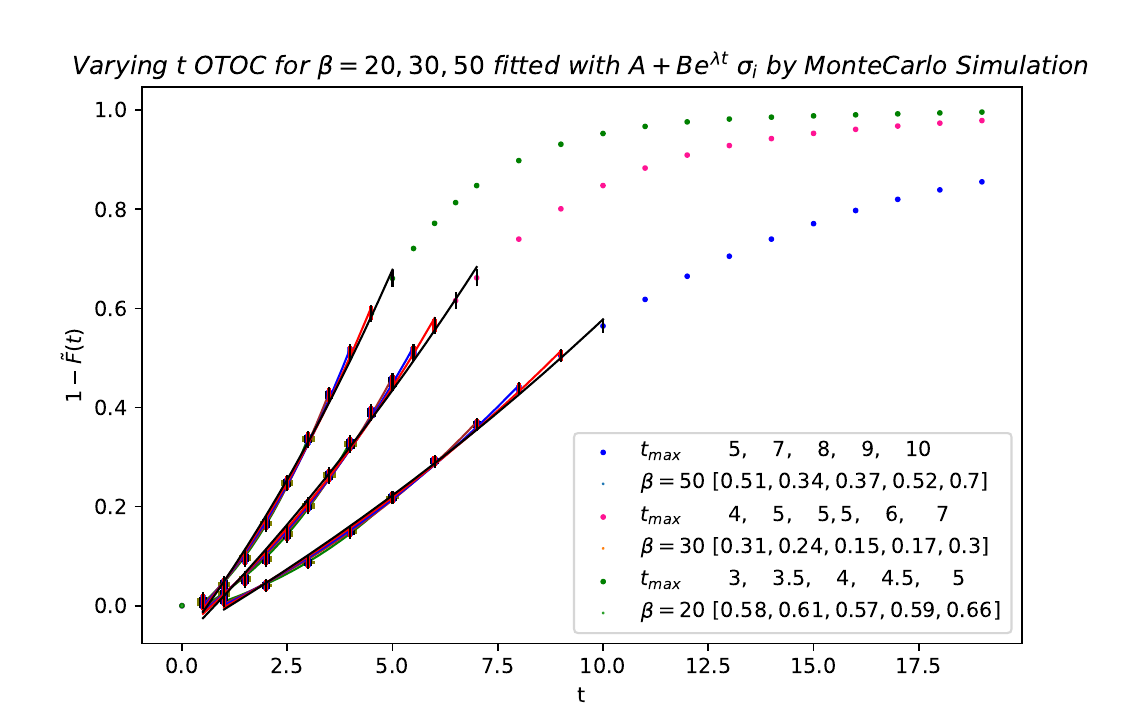}}
\end{minipage}
\caption{Non-local SYK model, $G(t) = 1 - {\tilde F}(t)$ as a function of time and Chi-Square fit value for different $\beta=20,30,50$. In (b) the values under $[~~]$ denotes the corresponding $P(\chi^2>\chi_0^2)$ up to mentioned $t_{max}$ values in pervious line.}
\label{otoc1nonr}
\end{figure}

\begin{figure}[H]
	\begin{minipage}[b]{\linewidth}
	\hspace{-13mm}
\subfigure[$\beta=30$ Chi-square fitting]{	\includegraphics[width=8cm,height=7cm]{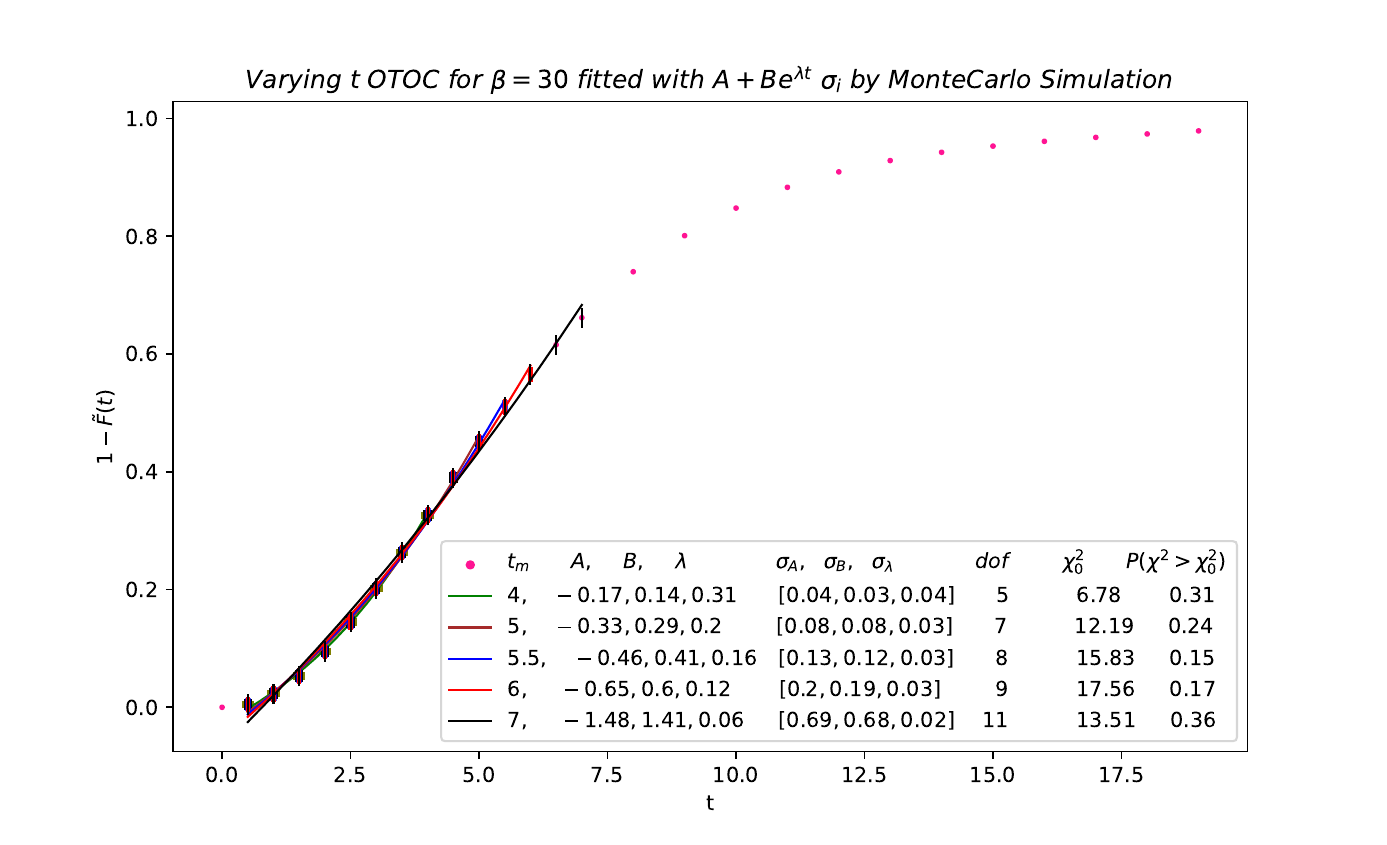}}
\hspace{-10mm}
\subfigure[$\beta=50$ Chi-square fitting]{	\includegraphics[width=8cm,height=7cm]{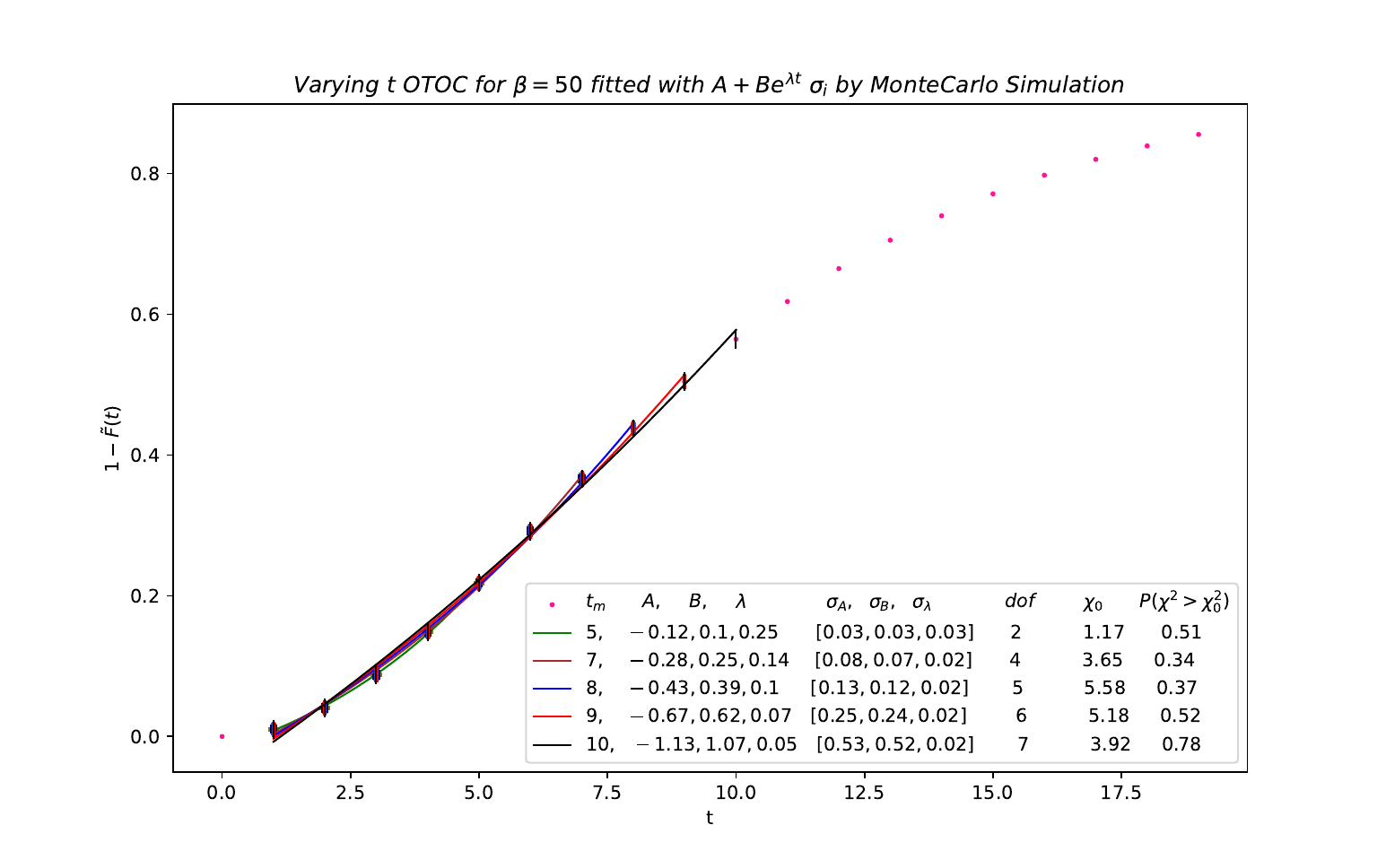}}
	\end{minipage}            \hspace{2mm}
\caption{Non-local SYK model, $G(t) = 1 - {\tilde F}(t)$ as a function of time and Chi-Square fit value for different $\beta$, with fitting parameters $(A,B,\lambda)$, $\sigma_{A},\sigma_{B},\sigma_{\lambda}$, degrees of freedom, $\chi_{0}^{2}$ and $P(\chi^{2}>\chi_{0}^{2})$}
\label{otoc1nonra}
\end{figure}

\section{Lyapunov Exponent for Local SYK }
\label{ExtraLyapunov}
In the local SYK section \ref{sykmat}, we have discussed about OTOCs in Local SYK model for $\bar{G}_{i,i+1}$ and $\bar{G}_{i,i+2}$, see fig.~\ref{lotoc1}. We had argued there that there is no good large $N$ limit that suppresses the connected part compared to the disconnected part. However, the plots for OTOCs are qualitatively similar to the case of Non-local SYK and so we can try to extract a Lyapunov exponent. This is shown in the plots below. 
\begin{figure}[H]
\hspace{-10mm}
\subfigure[]{\includegraphics[width=8cm,height=7cm]{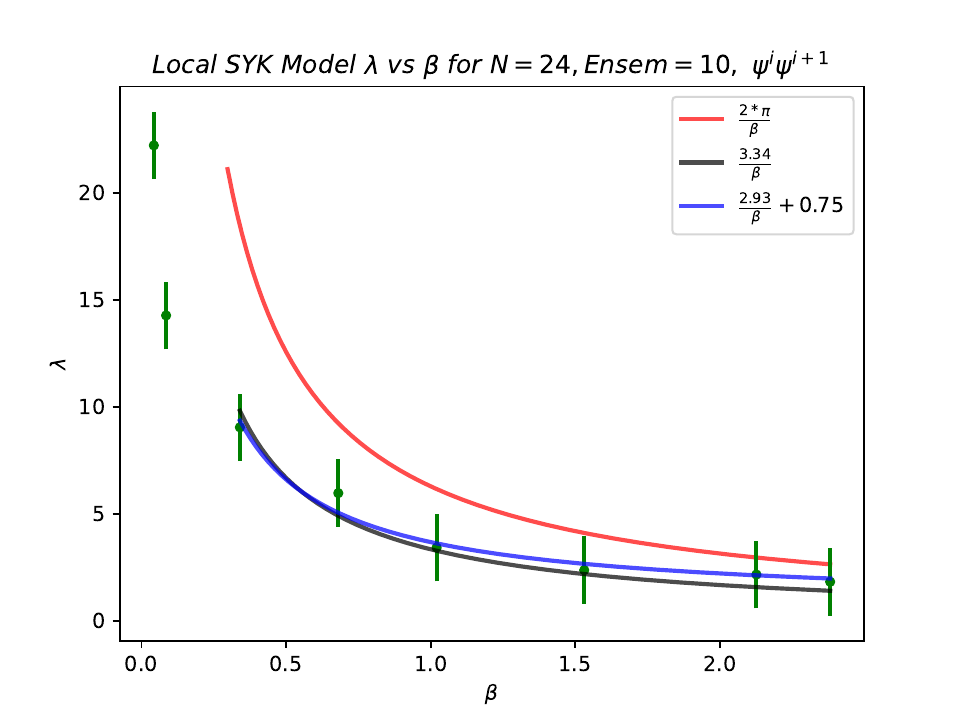}}
\hspace{-7mm}
\subfigure[]{\includegraphics[width=8cm,height=7cm]{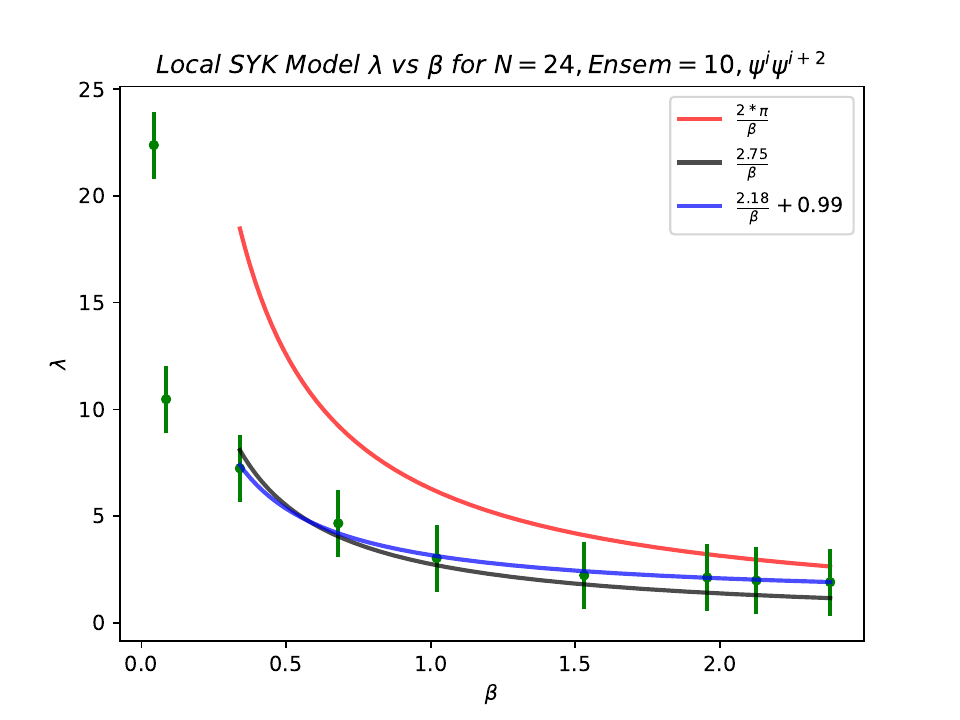}}
\caption{Lyapunov exponent behavior at different termperature for Local SYK model. (a) $\lambda$ vs $\beta$ for ${\bar G}_{i,i+1}$ (b) $\lambda$ vs $\beta$ for ${\bar G}_{i,i+2}$.
}
\label{otoc2lya}
\end{figure}
In fig.~\ref{otoc2lya}, the blue curves are for fitting the Lyapunov exponents at different $\beta$ with a functional form
$\lambda=\frac{A}{\beta}+B$. Black curves are for fitting with $\lambda=\frac{A}{\beta}$. The green error bars correspond to $1\sigma$ error obtained from fitting the OTOCs with eq.~(\ref{theoretic}). 
Within our numerical limitations we see that both fitting gives value of $A$ to be lesser than $2\pi$.
%

\newpage


\end{document}